\documentclass[poms,final,nonblindrev]{poms1_V1} 

\OneAndAHalfSpacedXI

\usepackage{appendix}

\usepackage{bm}
\usepackage{amsmath}
\usepackage{amssymb}
\usepackage{graphicx}

\usepackage{amsmath}
\usepackage{enumerate}

\usepackage{url}
\usepackage[hidelinks]{hyperref}
\hypersetup{breaklinks=true}
\usepackage[utf8]{inputenc}
\usepackage{mathtools}
\usepackage{amsbsy}
\usepackage{amssymb}
\usepackage{color}
\usepackage{ulem}
\usepackage{placeins}
\usepackage{booktabs}
\usepackage{multirow}
\usepackage{hyperref}
 \usepackage{booktabs,subcaption,amsfonts,dcolumn}
\usepackage{xcolor,colortbl}
\usepackage{subcaption}
\usepackage{verbatim}
\usepackage{chngcntr}
\usepackage{apptools}
\usepackage{movie15}
\usepackage{epstopdf}
\usepackage{endnotes}
\usepackage{tabularx}
\usepackage{comment}
\newcolumntype{Y}{>{\centering\arraybackslash}X}
\let\footnote=\endnote

\usepackage{natbib}
 \bibpunct[, ]{(}{)}{,}{a}{}{,}%
 \def\bibfont{\small}%
\TheoremsNumberedThrough     
\ECRepeatTheorems

\EquationsNumberedThrough    

\begin{document}



\RUNTITLE{Sriram, Sinha and Choudhary}

\TITLE{Predictive Hotspot Mapping for Data-driven Crime Prediction}

\ARTICLEAUTHORS{%
\AUTHOR{Karthik Sriram}
\AFF{Operations and Decision Sciences\\ Indian Institute of Management Ahmedabad, 380015, Gujarat, India, \EMAIL{karthiks@iima.ac.in}} 
\AUTHOR{Ankur Sinha}
\AFF{Operations and Decision Sciences\\ Indian Institute of Management Ahmedabad, 380015 Gujarat, India, \EMAIL{asinha@iima.ac.in}} 
\AUTHOR{Suvashis Choudhary}
\AFF{Indian Police Service (Retired), India, \EMAIL{suvindia@gmail.com}}
} 

\ABSTRACT{%
Predictive hotspot mapping is an important problem in crime prediction and control. An accurate hotspot mapping helps in  appropriately targeting the available resources to manage crime in cities. With an aim to make data-driven decisions and automate policing and patrolling operations, police departments across the world are moving towards predictive approaches relying on historical data. In this paper, we create a non-parametric model using a spatio-temporal kernel density formulation for the purpose of crime prediction based on historical data. The proposed approach is also able to incorporate expert inputs coming from humans through alternate sources. The approach has been extensively evaluated in a real-world setting by collaborating with the Delhi police department to make crime predictions that would help in effective assignment of patrol vehicles to control street crime. The results obtained in the paper are promising and can be easily applied in other settings. We release the algorithm and the dataset (masked) used in our study to support future research that will be useful in achieving further improvements.
}%


 \HISTORY{Received: December 2023; Accepted: February 2026 by Jayashankar Swaminathan, after two revisions.}

\maketitle

%


\section{Introduction}
\label{S:intro}

Predicting and preventing crime is a persistent problem for law enforcement agencies
who are often working with limited amount of resources with the aim to reduce such
incidents. A number of studies have looked at the use of a variety of predictive algorithms
that rely on historical data for crime prediction and patrolling decisions \citep{brunsdon2007visualising,gerber2014predicting,hart2014kernel,hu2018spatio,brandtetal2022}. The governments
across the world are looking to automate operations in several contexts by capitalizing on
technology and data. Process automation based on historical data is required, but law enforcement agencies often intend to have the flexibility to incorporate real-time intelligence arriving from alternate sources as well. 
In this paper, we propose an algorithm based on a non-parametric model for crime prediction
to automate surveillance and patrolling; the approach is flexible such that it is able to account for past data as well as allows human/expert inputs while making predictions.

Currently, in most parts of the world patrolling is done through police vehicles moving on
roads in urban areas and through aerial vehicles to monitor large outdoor areas, such as
forests, coastlines, borders, and crowded locations. With the advancement of drone
technology, one is likely to see the use of a swarm of drones for patrolling urban areas with
the benefit that the drones would provide a bird’s eye view of the surroundings, and can
capture high-resolution images and videos that would be useful for surveillance and
investigation. While the drone technology will not completely replace police cars because of
various limitations like its inability to operate under extreme weather and the immediate
need of a human officer in a variety of incidents, it is pertinent that the use of multiple sub-systems
to operationalize the task of patrolling and surveillance will require a much greater
level of automation. Any kind of automation will rely on predictive algorithms that can
quickly analyze past data and immediately accept expert inputs to chart out a plan for
effective utilization of the available resources to meet the desired objectives.
The smart city initiatives across the world promise a variety of benefits, like increased
efficiency, sustainable development, better quality of life and safety of citizens. Among all
these benefits, safety of the citizens is at the core of the development and implementation
of smart city initiatives. Smart technologies, such as artificial intelligence, sensors and cameras
promise prevention of crimes, rapid response to emergencies and improved safety. Real-time
availability of data in these initiatives can be capitalized by quickly converting it into
informed and timely decisions for improved efficiency and effectiveness of city operations.
Certain activities like patrolling and surveillance may require expert inputs and alternative
sources of information to be used in conjugation with historical data.
{\color{black}New information, for instance, new construction sites, newly created bus stops, areas with changed lighting conditions, places with modified traffic flow, etc., may not get captured readily based on historical data until a few crimes happen in those locations. Other kinds of information like offenders out on parole or after completion of sentence may also impose added risk in localities. These kinds of information are best highlighted by expert inputs and not past data alone.}
Unfortunately, most of the data-driven prescriptive analytics approaches for urban policing
do not provide a mechanism to incorporate such expert information arriving from alternate
sources while making predictions and suggesting an operational plan. We recognize this gap
in the literature on patrolling operations and propose an algorithm to create spatio-temporal
hotspot maps based on past data and potential expert inputs to help in patrolling decisions.

This paper emerged as part of a collaborative project of the authors with the Delhi police department to develop strategies driven by data analytics for reducing street crimes. Street crime involves snatching and robberies, where snatching typically involves forceful grabbing of property such as a necklace or a wallet from an unsuspecting victim, whereas robbery additionally involves a threat of bodily harm to the victim who may refuse to part with the property. While these are petty crimes, they can cause serious damage to the morale of the public, their perception of safety and their confidence in the police. Further, it is a known concern that street crime offenders, if not checked and deterred, will likely commit more serious crimes in the future \citep{singh2018delhi}.  Unlike burglaries or other property thefts that get reported much later after the event has occurred, street crimes get reported almost immediately. So, with adequate planning of patrolling resources, the police can either preempt the crime or apprehend the offenders. The Delhi police adopt different approaches to tackle street crime. For example,  they may assign special patrolling resources at key locations to deter street crime, as they did with  the ``Prakhar anti-street-crime-vans" \citep{pti2019delhi}, or assign patrolling resources to carry out  random checks \citep{manral2021delhi} on vehicles at specific  locations and  timings with an aim to apprehend the offenders. Clearly, effectiveness of such patrolling resources requires good knowledge of timing and locations prone to street crime. Our objective in this paper is to propose an effective method to obtain hotspot maps to predict the locations prone to street crime at different times of the day, and discuss its implications for operational decisions on patrolling. 
Our approach for hotspot maps uses kernel density estimation. As a general technique, kernel density estimate (kde) is a well known  estimate for probability density function (pdf) based on univariate as well as multivariate data (see  \citealt{silverman1986}). In particular,  bivariate kde is naturally applicable to spatial hotspot mapping since it involves modeling of two dimensional (longitude, latitude) coordinates of the event locations. Further, the kde can be extended to a trivariate setting to include a temporal component to jointly model the timing of the events, and  has therefore been used for generating spatio-temporal hotspot maps in many applications including crime management (see \citealt{brunsdon2007visualising}, \citealt{porter_reich2012}, \citealt{hu2018spatio}, \citealt{yuanetal2019}). In this paper, we make a number of methodological contributions to how kde is used in the crime prediction context and how expert input is incorporated in the algorithm to generate crime hotspots. The important methodological contributions and contextual insights from this study are highlighted below:
\\
\textcolor{black}{{\bf (1) Model contributions.}}  We construct a non-parametric model with an aim to estimate the spatio-temporal probability density for crime. Specifically, we use a spatio-temporal block-weighted adaptive kernel density formulation for hotspot prediction that incorporates timing during the day using a circular distribution kernel, as well as temporal differences across days using a temporal-block-weighting approach. To our best knowledge, this is the first study that tackles multiple real-world complexities arising in the context of crime prediction using hotspots in a single framework.
For instance, \citet{wang2023takde} use temporally block-weighted kde but in a univariate context not requiring modeling of timing during the day or adaptive kde, \citet{brunsdon2007visualising} give a spatio-temporal formulation incorporating timing during the day but without temporal block-weighting or adaptive kde, and
\citet{backlinetal2018} use an adaptive kde formulation but in a univariate context without incorporating timing during the day. 
Additionally, none of the methods available in the literature discuss about incorporating human judgment in this context. We formulate our model by bringing together multiple ideas in Section \ref{S:main_model} in a Bayesian framework that also provides the flexibility to incorporate expert inputs. The model is inherently non-linear for which we provide an estimation procedure. We find that the predictive accuracy of the modeling approach is consistent across all weeks of forecast in our data.
\\
{\bf (2) Contextual Insights.} The hotspot mapping model developed in this paper that allows expert inputs to be incorporated is a first of its kind approach being applied in the context of street crime in Delhi, and India, in general. The decision makers are averse to using predictive models in case they are not involved in the process or they feel that the crucial inputs available with them cannot be incorporated in the model. Our proposed approach gives a strong control to the decision maker and with the right balance between the historical data and expert inputs the results are also significantly better than relying on past data alone. Our methodology  applied to this context leads to some useful insights:~\\
 (i)  Our spatio-temporal approach of jointly modeling the location with the timing in the day,  achieves better accuracy in predicting street crime locations than a purely spatial formulation of modeling data restricted to a particular time interval. This means that the street crime event locations during a particular time interval, say 8 pm to 12 am, depend not only on the historical event locations during the same time interval but also on the historical event locations at other times of the day. ~\\
 (ii) We have learnt from the Delhi police that their patrolling allocations are generally stable, not changing much from week to week with only some cruicial inputs incorporated.  However, our analysis presents a useful insight that the prone locations can change every week, requiring regular changes to patrolling allocations depending on the time during the day. ~\\
 (iii) Key locations such as metro stations, famous temples, and busy markets are considered by the police to be prone to street crime events and hence attract more patrolling. Our analysis highlights two important aspects: first, while most of these key locations are prone to street crime as expected, there are some that are not. Second, there are several locations other than these key locations that are prone to street crime and may require more patrolling.~\\
 (iv) Finally, the expert inputs to the model may play an important role in leading to a better predictive outcome, justifying the need for a procedure that can capitalize on both historical data and expert inputs. Our procedure allows such an integration leading to a collaborative approach to making predictions. Currently, there is no systematic mechanism to capture and utilize expert intelligence from alternative sources in the context of street crimes in Delhi. Therefore, we demonstrate the working of the algorithm with the help of simulated expert inputs.

{\color{black}
Our formulation for predicting likely crime locations is based on kernel density estimation that has its own advantages, especially for integrating expert inputs, and for other reasons discussed later. For spatio-temporal predictions in various other contexts researchers have used a variety of methods, for instance, \cite{faghih2016incorporating} model hourly arrival and departure rates using a panel data model with Gaussian spatio-temporal correlated errors, in the presence of several explanatory variables. Such a model may not be applicable in a crime prediction situation where the response variable is binary (i.e. 1 implies event happened at a location at a given time, and 0 implies event did not happen).}

{\color{black}Typically, point-process approaches are used in applications, similar to ours, and work by modeling the number of events indexed by space and time as a stochastic process, while introducing spatio-temporal correlations.} For example, \cite{flaxmanetal2019}  use a log-Gaussian Cox process (LGCP) formulation, where the centering of the Gaussian process is modeled using a weighted combination of kernel densities. \citet{shirota_gelfand2017} also use a log-Gaussian Cox process formulation where they incorporate a circular correlation structure for timing of events during the day, in addition to spatial correlations. Other approaches include self-exciting point processes (\citealt{mohleretal2011}),  adaptation of Hawkes processes (\citealt{zhuang_mateu2019}), and auto-regressive mixture models (\citealt{taddy2010}).  These alternate approaches inherently imply some weighted average of past experience and the regression type of framework is conducive for including relevant explanatory variables. While we explored the use of these approaches, we preferred the kde approach due to the following considerations. First, these methods are more useful in the presence of explanatory variables that may have good predictive power, which in the context of street crime in Delhi are not readily known or available. Second, the estimation and implementation of these models turn out to be computationally challenging in our context. For example, a LGCP formulation in our case for any given week, would need to look at event incidences within 36,263 locations in Delhi every day, which over a past window of 52 weeks amounts to over 12 million records. So, estimation and implementation of spatio-temporal model formulations based on  LGCP on a regular basis would be computationally prohibitive. Third, the number of locations with event incidences, on any given day, is relatively very less (only few hundreds per day) compared to the number of total locations, leading to a highly imbalanced dependent variable, which presents its own challenges. On the other hand, the kde approach requires only to  focus on the locations with events, thus circumventing the aforementioned challenges with data size. {\color{black}Despite these advantages, we acknowledge that the kde approach has certain limitations, for instance; the kernel density approach treats space uniformly ignoring land use types, road networks, barriers or rivers; interdependence of crime events like retaliatory crimes cannot be learned; densities near the boundaries can be underestimated due to lack of neighbouring data points beyond the edges; etc.}

The paper is structured as follows. We describe our problem and data in Section \ref{S:data}. In Section \ref{S:main_model},  we formulate our proposed model, and provide an approach for estimation. In Section \ref{S:results}, we compare results from different model scenarios, and in Section \ref{S:discussion} we discuss the implication of the results for operational decisions in the Delhi street crime context. We conclude the paper in Section \ref{S:conclusion}. The dataset used in this study has been masked without any loss of integrity and is made publicly available for further research at: {\url{https://www.kaggle.com/datasets/ankurzing/spatio-temporal-crime-incidents}.

\section{Problem and data description}
\label{S:data}

The city of Delhi reported the highest crime rate among 19 Indian metropolitan cities with population larger than 2 million as per the National Crime Records Bureau report of 2020. The Delhi police department has implemented several measures to prevent and counter crime in the city. The use of data and technology has increased over the years in crime prevention, but a predictive approach for crime control has still not been fully exploited by the Delhi police department. Usually, crime data is obtained from the list of formal complaints registered at the police stations. However, a large number of crimes in Delhi are street crimes, where victims typically call 100 (or 112) emergency-helpline-number to report a street crime event to the `police control room' (PCR). When a victim in distress calls  the PCR, some details pertaining to the caller as well as the event are recorded by a call attendant, who then locates a patrol vehicle nearest to the incident and intimates via the police wireless communication system.  The patrol vehicle reaches the spot of the caller and assesses whether a crime has indeed been committed or not. In case of a crime, the patrol vehicle intimates the nearest police station of jurisdiction and hands over the case for further processing.  If the crime has happened and if the victim is willing, then the police station registers a `first information report', which is a formal complaint necessary to pursue further investigation. Sometimes it is possible that the victim may choose not to file a formal complaint, considering the time and subsequent effort that may be needed while the case gets investigated. There could also be some situations where the patrol vehicle may not be able to trace the victim, e.g.  if the victim changes mind and leaves the spot. Based on police sources, less  than 15\% of PCR calls on street crime get registered as formal complaints. 
In any case, the volume of PCR calls are considered by the Delhi police as a more accurate representation of the level of street crimes than the formally registered complaints. In this paper, we use data provided to us by the Delhi police on all PCR calls related to street crimes in Delhi from October 2019 to March 2021, captured in the Delhi police database (DPD). The fields available in DPD are provided through Table~\ref{T:data_desc}, from which we use the date, time and event location information. Since the data is digitally captured in DPD, the approach developed in this paper can be integrated with the system to generate hotspot maps on a regular basis to help in patrolling decisions. The data used in this study is not publicly available, and was procured following due process with permission from the Delhi police department. {\color{black}We performed a number of steps to clean and prepare the data for our analysis. The data that we obtained contained verified calls (i.e. prank calls were excluded); however, it required some cleaning in terms of removal of duplicate entries or erroneous latitude/longitude entries. For a detailed discussion on data preparation and descriptive analysis, we refer the reader to Section~\ref{S:data_prep}.} Through this study we are also releasing a masked version of the cleaned dataset that would be useful for future research.

\begin{table}
\centering
\caption{Fields in Delhi police database (DPD) }
\label{T:data_desc}
\renewcommand{\arraystretch}{1.3}
\resizebox{\textwidth}{!}{
\begin{tabular}{l p{13cm}}
\toprule
 Variable& Description\\
\hline
Event id& Identification number for the street crime event.\\
\hline
 Date & Date of the street crime event (day, month and year). This is automatically captured by the system during the call.\\
 \hline
 Time& Time of the street crime event during the day (hour, minutes, seconds). This is automatically captured by the system during the call.\\
 \hline
 Location description &  Text describing the location (address) of the event. This is entered manually into the system by the PCR call attendant during or immediately after the emergency call. This is  typed in English but may contain some words of the Hindi language.\\
 \hline
 Police station & This is the name of the police station, in whose jurisdiction the street crime event has occurred. This is manually entered into the system by the PCR attendant based on the information received from the patrol vehicle after it that visits the event location.\\
 \hline
 Police station (geo coordinates) & Latitude and longitude, which we (manually) looked up for the police station, in whose jurisdiction the street crime event has occurred.\\
 \hline
 Event location (geo coordinates) &  Latitude and longitude of the event location entered in the system by the PCR call attendant, or if not entered, geocoded based on the text address descriptions.\\
\bottomrule
\end{tabular}}
\end{table}


Using the crime data, we formulate a `spatio-temporal block-weighted adaptive kernel density model' to jointly model the location and timing of street crimes in Delhi. It is expected that the likely locations for street crime depend on the times of the day. Also, the event locations in the historical temporal blocks (e.g. past weeks) tend to be indicative of the event locations in a future temporal block (e.g. upcoming week), although the relative weights to be attached to the historical temporal blocks for an accurate prediction are a priori unclear. Our proposed method generates hotspot maps for different times of the day in an upcoming temporal block, by determining suitable weights for the historical temporal blocks along with other parameters required for the kde. In our application, each temporal block consists of a set of consecutive days, i.e., a week.
Since `week' is a balanced time frame for strategic planning of resources without being too frequently demanding of the valuable time of senior officers, or without being too delayed in utilizing recent data, we use it as the temporal block. However, our approach can be easily adapted to accommodate temporal blocks of other sizes (e.g. `1 day' or `1 month') as well.  The model formulation and estimation are described in detail in Section \ref{S:main_model} and were implemented using the R package (\citealt{cran}). We also note that the background maps used for plotting the spatial distribution in this paper use the leaflet package in R (\citealt{leaflet}).

\begin{figure}[ht]
\begin{subfigure}[t]{.48\textwidth}
\center
\includegraphics[width=8cm]{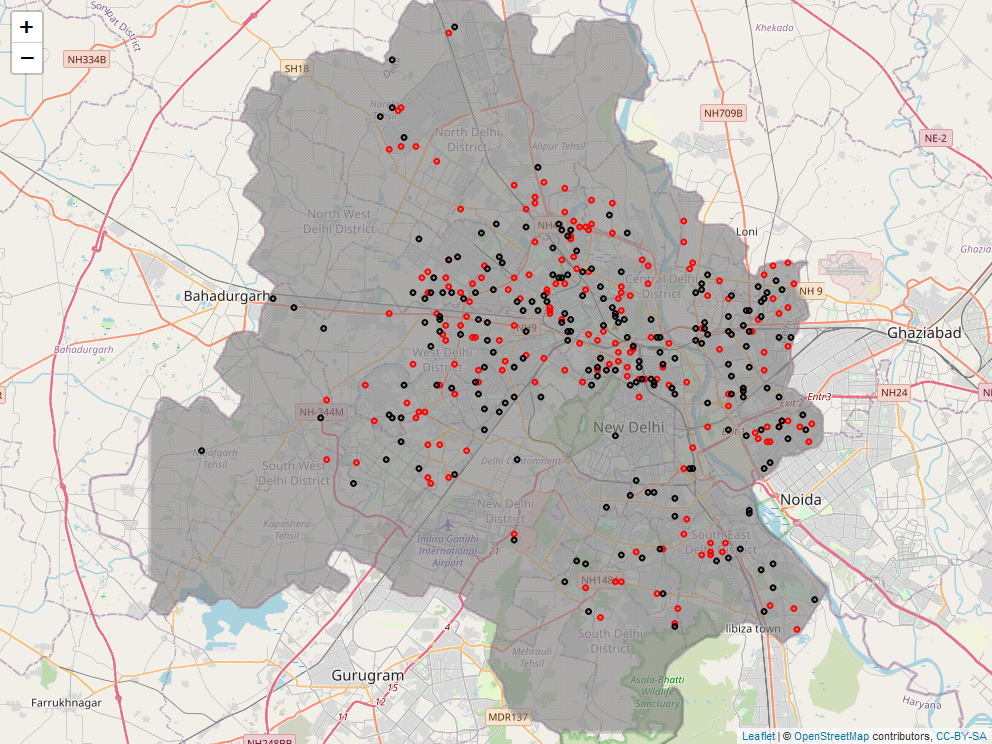}
\caption{\textcolor{black}{Actual locations (black dots) predicted using as-is locations of  the previous  week (red dots).}}
\end{subfigure}\hfill
\begin{subfigure}[t]{.48\textwidth}
\center
\includegraphics[width=8cm]{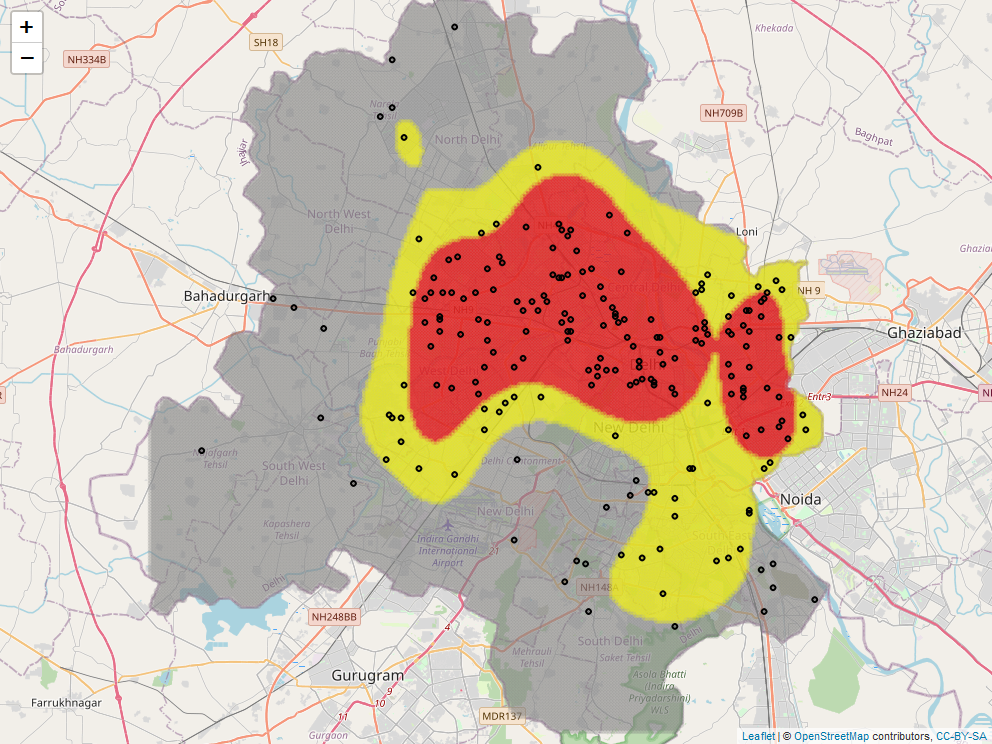}
\caption{Actual locations (black dots) predicted using spatial kde on past 1 week locations. Top 20\% area (red) captures 60\% events and next 20\% area (yellow) captures an additional 27\% events)}
\end{subfigure}

\begin{subfigure}[t]{.48\textwidth}
\includegraphics[width=8cm]{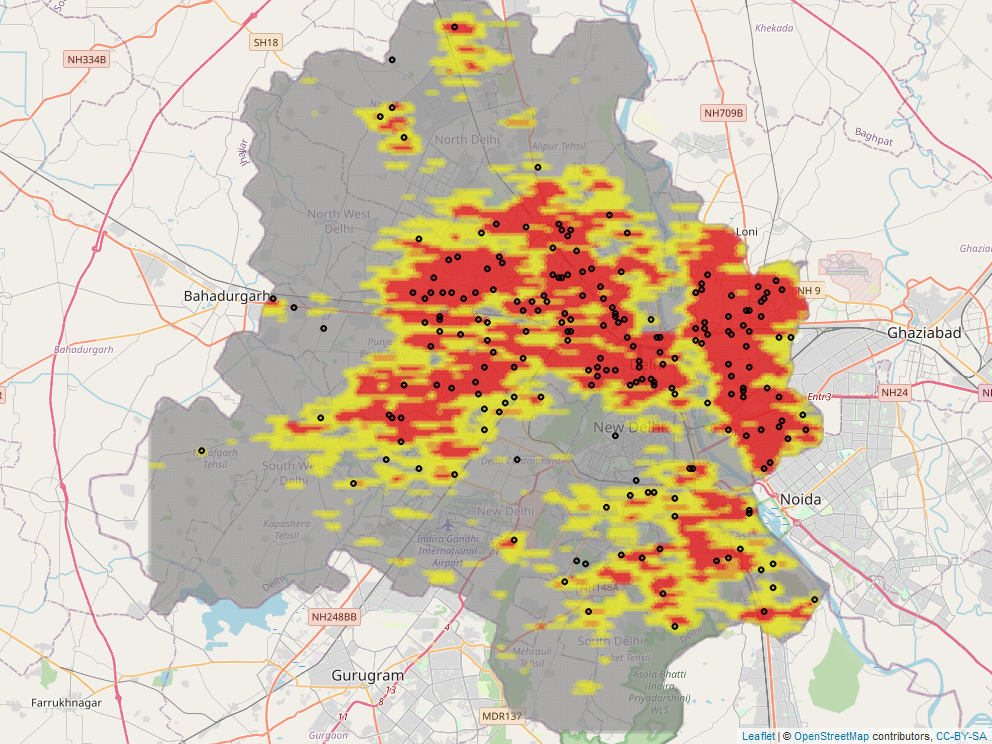}
\caption{Actual locations (black dots) predicted using spatial kde on past 52 weeks. Top 20\% area (red) captures 76\% events and next 20\% are (yellow) captures an additional 18\% events)}
\end{subfigure}\hfill
\begin{subfigure}[t]{.48\textwidth}
\center
\includegraphics[width=6.8cm]{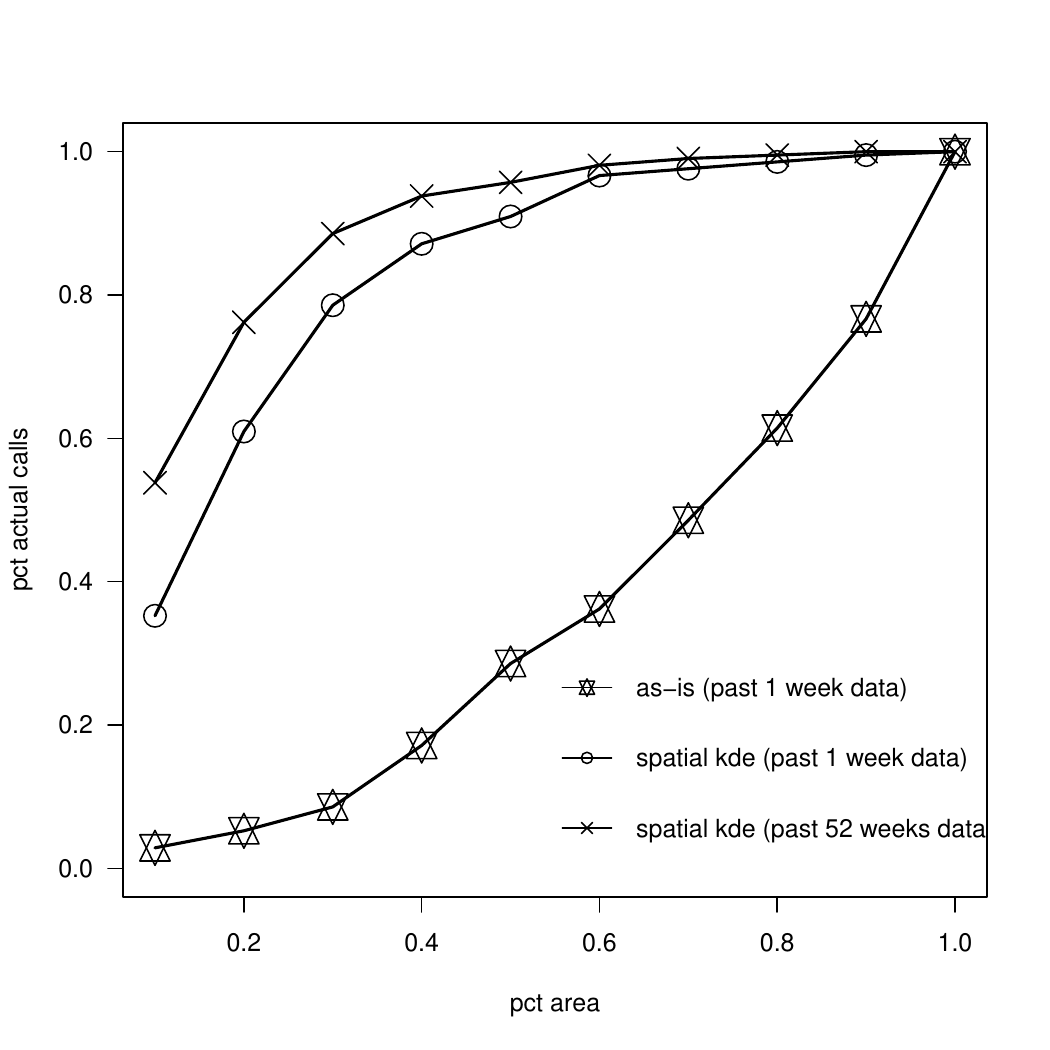}
\label{F:callsarea}
\caption{`Event-area curve': \% events captured versus \%  area monitored (in descending order of  locations predicted as prone to events).  }
\end{subfigure}
\vspace{2mm}
\caption{Preliminary predictive approaches to predict street crime locations in Delhi during `8 pm to 12 am' for the week of  21 March 2021 (The grey shaded area is the approximate map of Delhi).}
\label{F:calllocs21mar}
\end{figure}

Next, we discuss a few preliminary predictive approaches that rely only on past data and do not take expert inputs. Let us start with an ad hoc approach,  which is to consider the `as-is' event locations of a current week as predictive of street crime locations for an upcoming week. On the map of Delhi, Part (a) of Figure \ref{F:calllocs21mar} shows the actual locations (black spots) of  street crime events that occurred during the time interval `8 pm to 12 am' in the week starting  21 March 2021, along with the predicted locations (red spots) given by the as-is event locations of its previous  week.  It is clear  that actual event locations in the week starting 21 March 2021 rarely overlap with the `as-is' locations of its previous week due to spatial variations. Therefore, some smoothing of the past experience, such as that made possible by kde, is desirable while predicting prone locations. In Parts (b) and (c) of Figure \ref{F:calllocs21mar}, we compute a spatial kde based on all street crime events that occurred in the past 1 week and past 52 weeks prior to 21 March 2021, respectively, but within the time interval 8 pm to 12 am.  As locations predicted to be prone to street crime events, we mark the top 20\% locations in decreasing order of kde in red and the next 20\% locations in yellow color. More the number of actual events that the red and yellow areas succeed in capturing better is the prediction. In Part (b) of Figure \ref{F:calllocs21mar}, the red and yellow spots cumulatively capture 60\% and 87\% actual events, respectively,  and in Part (c) they capture 76\% and 94\% actual events, respectively. Further, as shown in Part (d) of Figure \ref{F:calllocs21mar}, the predictive performance can be summarized using the `event-area-curve', a plot of the percentage actual events captured versus the percentage  area monitored (in descending order of  locations predicted as prone). We see in the figure that the spatial kde estimated on past 52 weeks of data is  the best, capturing the largest percentage of actual events by monitored area. The illustrated models in the figure are based on purely spatial formulation of kde. In Section \ref{S:main_model}, we formulate a more general spatio-temporal block-weighted kde model enabling better predictive accuracy leading to useful operational insights in the context of street crime in Delhi.


 

\section{Proposed model for hotspot mapping}
\label{S:main_model}
In this section, we formulate our proposed model, and provide the estimation procedure. We first describe the modeling approach for estimating the distribution based on historical data alone and then discuss the process used to incorporate expert input in the distribution that would be used to create the hotspot mapping.
\subsection{Model formulation}
The kde is used to estimate univariate and multivariate probability densities. Bivariate kde can be used to, therefore, model spatial distribution of events.   In its simple form, bivariate kde at a point $(s_x, s_y)$ based on sample data points $\{(x_i, y_i), i=1,2,\ldots, N\}$ can be written as
\begin{eqnarray}
{f}(s_x, s_y) & = &\frac{1}{N}\sum_{i=1}^{N} \kappa_{h_1, h_2}(s_x-x_i, s_y-y_i) ,\label{eq:bivkde}
\end{eqnarray}
where $\kappa_{h_1, h_2}(\cdot, \cdot)$ is a suitably chosen bivariate kernel function, typically a symmetric pdf, centered at $(0,0$) with bandwidths $(h_1,h_2)$. 
Taking $\kappa_{h_1, h_2}(\cdot, \cdot)$ in Equation (\ref{eq:bivkde}) to be the standard Gaussian kernel, we can compute `spatial' kde at any given grid location, with (longitude, latitude) coordinates  $(s_x, s_y)$, based on $N$ past event locations $\{(x_i, y_i), i=1,2,\ldots, N)\}$, as
\begin{eqnarray}
{f}(s_x, s_y) & = &\frac{1}{N}\sum_{i=1}^{N} \left\{  \frac{1}{h_1}\kappa\left(\frac{s_x-x_i}{h_1}\right)\cdot \frac{1}{h_2}\kappa\left(\frac{s_y-y_i}{h_2} \right) \right\}, \label{eq:KDE_pred}
\end{eqnarray}
where 
\begin{equation}
\kappa(u)=\frac{1}{\sqrt{2\pi}}e^{-u^2/2}. \label{eq:gaussiankernel}
\end{equation}

Our starting point is  Equation \ref{eq:KDE_pred}, which shows kde for the bivariate case based on standard Gaussian kernel and fixed bandwidth. Large values of the bandwidths $(h_1,h_2)$ will lead to extensive smoothing, whereas small values lead to overfitting the data distribution. Several rules have been proposed for the choice of bandwidth parameter in kde (e.g. \citealt{silverman1986}, \citealt{sheather2004}). It is also known that the estimation accuracy can be potentially improved by using `adaptive kde' , where  the bandwidth choice can be made to either depend on the point of estimation (i.e., depending on $(s_x, s_y)$), or the sample data points (i.e., depending on $(x_i, y_i)$). An early estimator given by  \citet{Loftsgaarden_Quesenberry1965} is equivalent to  kde based on a uniform kernel and variable bandwidth. The variable bandwidth is chosen based on the distance of the point of estimation to $k$th nearest sample data point (for some suitable value of $k$).  In contrast, \cite{breimanetal1977} makes the bandwidth depend on each sample data point, specifically its distance to the $k$th nearest sample data point.
A refinement given by \cite{abramson1982} is to take $h(x_i, y_i)\propto \left(\hat{f}_p(x_i,y_i)\right)^{-1/2}$, where $\hat{f}_p$ is a preliminary kde based on fixed bandwidth, and \cite{vankerm2003} gives its implementation  in the SPSS package.  \cite{terrell_scott1992} give further insights on  the implications of kernel and bandwidth choices in the univariate as well as multivariate setting.   \cite{Filippone_Sanguinetti2011} consider estimation of the fixed bandwidth kde using a Bayesian approach. \citet{backlinetal2018}  suggest a Bayesian adaptive kde with  $h(x_i, y_i)=  \alpha\left( \hat{f}_p(x_i,y_i)\right)^{-\beta}$, where $\hat{f}_p$ is a preliminary kde with fixed bandwidth, and give an approximate computation for the posterior distribution of $(\alpha, \beta)$. 

Bivariate (adaptive) kde has been used for spatial hotspot mapping in many applications (e.g. \citealt{yuanetal2019}). Following \citet{brunsdon2007visualising}, Equation (\ref{eq:KDE_pred}) can be extended to a spatio-temporal kde, by choosing an appropriate kernel for the time dimension $\kappa_\tau(\cdot)$ along with a bandwidth $\tau$, as follows
\begin{eqnarray}
\widehat{f}(s_x, s_y,t) & = &\frac{1}{n}\sum_{i=1}^{n} \left\{  \frac{1}{h_1}\kappa\left(\frac{s_x-x_i}{h_1}\right)\cdot \frac{1}{h_2}\kappa\left(\frac{s_y-y_i}{h_2} \right)\cdot \kappa_{\tau}\left(t-t_i\right) \right\}, \label{eq:KDEspt_pred}
\end{eqnarray}
where $\kappa(\cdot)$ is as in (\ref{eq:gaussiankernel}), and $t_i$ is the timing during the day of the event in the location $(x_i, y_i)$.  If the objective is to jointly model the timing of events during the day, then the kernel needs to be periodic in a $24$ hour interval. So, as in \citet{brunsdon2007visualising}, one can use a circular pdf such as the von Mises pdf  as the kernel, scaled appropriately to be supported on $0$ to $24$ hours, as shown in Equation (\ref{eq:vonmises}).
\begin{eqnarray}
 \kappa_\tau(u)=\frac{1}{24 I_0(\tau)}exp(\tau \cos(\pi u/12)),  ~ u\in [0, 24) ,\label{eq:vonmises}
 \end{eqnarray}
 where $I_0(\tau)=\frac{1}{2\pi}\int_{0}^{2\pi}e^{\tau \cos(\theta)}d\theta$ is the modified Bessel function of the first kind of order $0$.  The parameter $\tau$ is the concentration parameter, which plays the role of bandwidth in the circular case, with $\tau=0$ corresponding to a uniform pdf and higher values of $\tau$ leading to more concentration of the kernel pdf around $0$.  If the periodicity property for time differences is not required, then one may use a suitable kernel as  done by \citet{porter_reich2012} or  \citet{hu2018spatio}. 
 
 We now describe our proposed model. In our spatio-temporal kde formulation, we will incorporate timing during the day as well as temporal differences across days, by using a block-weighting approach. To help describe our kde formulation, we will organize the data in temporal blocks of time and use the notations as shown in Table \ref{T:blockdata}. Our proposed `spatio-temporal block-weighted adaptive kde' approach is as follows. We will model the events in a temporal block $b$, conditional on $B$ historical temporal blocks of data as
 \begin{eqnarray}
 (s_{x1}, s_{y1}, t_1), (s_{x2}, s_{y2}, t_2), \ldots, (s_{xn}, s_{yn}, t_n) \stackrel{ iid}{\sim} f(\cdot, \cdot, \cdot), \label{eq:model}
 \end{eqnarray}
where
 \begin{eqnarray}
f(s_x, s_y,t) =\sum_{i=1}^B \frac{w_i}{n_i}\sum_{j=1}^{n_i} \left\{  \frac{1}{h_{1ij}}\kappa\left(\frac{s_x-x_{ij}}{h_{1ij}}\right)\cdot \frac{1}{h_{2ij}}\kappa\left(\frac{s_y-y_{ij}}{h_{2ij}} \right) \cdot \kappa_{\tau_{ij}}\left(t-t_{ij}\right)\right\}, \label{eq:KDE_proposed}
\end{eqnarray}
and $\kappa(\cdot)$ is as in (\ref{eq:gaussiankernel}) and $\kappa_{\tau}(\cdot)$ is as in (\ref{eq:vonmises}). Motivated by the approach of \citet{backlinetal2018}, we define the adaptive bandwidths as follows:
\begin{eqnarray}
h_{1ij}=\frac{1}{\alpha_1A_{ij}^{\beta_1}}, ~h_{2ij}=\frac{1}{\alpha_2A_{ij}^{\beta_2} }\mbox{ and } \tau_{ij}=\frac{1}{h_{3ij}^2} , \mbox{ with } h_{3ij}=\frac{1}{\alpha_3 A_{ij}^{\beta_3}},  \label{eq:adaptbw}
\end{eqnarray}
where 
\begin{eqnarray}
A_{ij}= \hat{f}_p(x_{ij},y_{ij},t_{ij})/G,  ~G=\left(\prod_{i=1}^B\prod_{j=1}^{n_i} \hat{f}_p(x_{ij},y_{ij},t_{ij}) \right)^{1/n_{tot}},~n_{tot}=n_1+n_2+\cdots+n_{B}, \label{eq:A}
\end{eqnarray}
and $\hat{f}_p$ is a preliminary kde with fixed bandwidths (i.e. constraining $h_{1ij}=h_1, h_{2ij}=h_2, \tau_{ij}=\tau$, for all $i,j$). Note that G is just the geometric mean of the values of $\hat{f}_p$ computed over all the data points in the temporal blocks $1,2,\ldots, B$.  
We utilize an adaptive bandwidth formulation (i.e. $\tau_{ij}$) for the circular density used to model timing as part of the multivariate kde.
Our formulation for $\tau_{ij}$ is guided by the fact that von Mises pdf $\kappa_{\tau}(\cdot)$, for high values of the concentration parameter $\tau$, approximately matches the normal distribution with variance $1/\tau$. Hence, we take the idea used for $h_{1ij}$ and $h_{2ij}$ in Equation (\ref{eq:adaptbw}) and apply it to $h_{3ij}$. {\color{black}Note that the {\it iid}-assumption does not imply time independence in our overall modeling framework. Our model incorporates time dependence in two ways: (i) The spatio-temporal probability density in the current week is conditioned on the observed locations and times from the past 52 weeks, capturing the dependence between the current week and previous weeks. (ii) A circular probability density kernel accounts for events occurring at similar times across different days. Our simplifying assumption is that, conditional on past 52 weeks of observations, the observations within a time-block are independent. It is common and reasonable to assume that observations are {\it iid} in a short time-block.}

In our formulation (\ref{eq:KDE_proposed}), the terms in the summation for historical event locations recieve different weights ($w_i$) depending on the temporal block ($i$) to which they belong.  Kde with differently weighted terms have arisen in other problems, e.g. \citet{harvey_orysh2012} use time-weighting to estimate kde for time-series data,   \cite{Gao_Zhong2010} study randomly weighted kde where the weights are assumed to be uniformly distributed, and \cite{kim_scott2012} derive a weighted kde solution to an M-estimation problem with a robust loss function in the presence of contaminated data.  To handle streaming data that are gathered in temporal blocks \citet{wang2023takde} use temporally block-weighted kde in the univariate case, with a fixed bandwidth, but their problem context does not require modeling of timing within the day. Our formulation is in the multivariate (spatio-temporal) setting and incorporates modeling of timing of the events during the day along with spatio-temporally adaptive bandwidths. In the context of street crime in Delhi, the crime events do not form a  time series as they are not linked to a common offender.  So, events that are close temporally need not be close spatially, but the circular distribution pattern of timing of events during the day is worth capturing, which we do by using a circular pdf as the kernel for timing.  Also,  locations of past temporal blocks (e.g. past weeks) can be indicative of locations prone to street crime during an upcoming temporal block (e.g. upcoming week). We do not assume that a recent historical temporal block is necessarily more predictive of an upcoming temporal block. Instead, we leave the weights for the temporal blocks flexible and dynamically estimate them whenever a new temporal block of data becomes available.

\begin{table}
\centering
\caption{Data organised for block-weighted adaptive kde }
\label{T:blockdata}
\renewcommand{\arraystretch}{1.2} 
\resizebox{\textwidth}{!}{
\begin{tabular}{|l|c|c|c|c|c|}
\hline
~& \multicolumn{4}{c|}{Historical blocks}& Latest block  \\
\hline
 Temporal block & $b-B$ & $b-B+1$& $\ldots$ & $b-1$ & $b$ \\
\hline
Block index $i$ & $1$ & $2$& $\ldots$ & $B$ & $~$  \\
\hline
Lon. coordinates& $x_{11},x_{12},\ldots,x_{1n_1}$ & $x_{21},x_{22},\ldots,x_{2n_2}$& $\ldots$ & $x_{B1},x_{B2},\ldots,x_{Bn_B}$& $s_{x1},s_{x2},\ldots,s_{xn}$  \\
\hline
Lat. coordinates& $y_{11},y_{12},\ldots,y_{1n_1}$ & $y_{21},y_{22},\ldots,y_{2n_2}$& $\ldots$ & $y_{B1},y_{B2},\ldots,y_{Bn_B}$& $s_{y1},s_{y2},\ldots,s_{yn}$  \\
\hline
Event time (in hours)& $t_{11},t_{12},\ldots,t_{1n_1}$ & $t_{21},t_{22},\ldots,t_{2n_2}$& $\ldots$ & $t_{B1},t_{B2},\ldots,t_{Bn_B}$& $t_{1},t_{2},\ldots,t_{n}$ \\
\hline
Block weight& $w_1$ & $w_2$& $\ldots$ & $w_B$ & $~$ \\
\hline
\end{tabular}}\\
\end{table}

\subsection{Model estimation}
\label{S:estimation}
 The likelihood for the data on the spatio-temporal coordinates of street crime events in the training temporal-block $b$, i.e. $(\boldsymbol{s}_x, \boldsymbol{s}_y, \boldsymbol{t}):=\left\{ (s_{xl},s_{yl},t_l), ~l=1,2,\ldots, n\right\}$, based on the model Equations (\ref{eq:model}) and (\ref{eq:KDE_proposed}) is given by
 \begin{eqnarray}
 L\left(  (\boldsymbol{s}_x, \boldsymbol{s}_y, \boldsymbol{t}) \big\vert \alpha_1, \beta_1, \alpha_2, \beta_2, \alpha_3, \beta_3\right)= \prod_{l=1}^nf(s_{xl},s_{yl}, t_l).\label{eq:lik}
 \end{eqnarray}
 We use a Bayesian approach to estimate the parameters $(\alpha_1, \beta_1, \alpha_2, \beta_2, \alpha_3, \beta_3, w_1, \ldots, w_B)$ by assuming the following independent non-informative priors.
 \begin{eqnarray}
 &&\Pi(\alpha_1, \alpha_2, \alpha_3) \propto 1,~ \alpha_1, \alpha_2, \alpha_3\in (0,\infty), \label{eq:pr_alpha}\\
  &&\Pi(\beta_1, \beta_2, \beta_3) \propto 1,~ \beta_1, \beta_2, \beta_3\in [0,1], \label{eq:pr_beta}\\
  &&\Pi(w_1, \ldots, w_B)=\mbox{Dirichlet}(1,1,\ldots,1),~w_i\geq0, ~\sum_{i=1}^Bw_i=1. \label{eq:pr_w}
 \end{eqnarray}
 Our approach is flexible in allowing for a suitable value between $0$ and $1$ to be estimated for the parameters $\beta_1, ~\beta_2$ and $\beta_3,$ as against other rules of thumb, e.g. \citet{breimanetal1977} suggest $1/d$ (when dealing with $d$ dimensions), and  \citet{abramson1982} suggests 1/2 irrespective of the dimension.  To help the estimation of parameters via Gibbs sampling, we use the standard data augmentation technique. For each data point $l\in \{1,2,\ldots, n\}$ in the training temporal-block $b$, we define the random vector of data augmentation variables $(I_l, J_l)$, whose possible values are given by $\mathcal{S}_{IJ}:=\left\{(i,j),~i\in\{1,2,\ldots, B\} , ~j\in\{1,2,\ldots, n_i\}\right\}$, and the probabilities are given by
 \begin{eqnarray}
 \Pi(I_l=i, J_l=j)=w_i/n_i, ~(i,j)\in \mathcal{S}_{IJ}.
 \end{eqnarray} 
 Accordingly, the likelihood of the data after including the data augmentation variables, is given by
 \begin{eqnarray}
 &&L\left( (\boldsymbol{s}_x, \boldsymbol{s}_y, \boldsymbol{t}, (I_1,J_1),\ldots, (I_n,J_n)\big\vert \alpha_1, \beta_1, \alpha_2, \beta_2, \alpha_3, \beta_3,w_1, \ldots, w_B\right)\nonumber\\
 &&=\prod_{l=1}^n\left\{ \frac{1}{h_{1I_lJ_l}}e^{ -\frac{\left(s_{xl}-x_{I_lJ_l} \right)^2}{2h^2_{1I_lJ_l}}} \cdot  \frac{1}{h_{2I_lJ_l}}e^{ -\frac{\left(s_{yl}-y_{I_lJ_l} \right)^2}{2h^2_{2I_lJ_l}}} \cdot \frac{e^{\tau_{I_lJ_l} \cos\left( (t_l-t_{I_lJ_l})\pi/12\right)}}{24I_{0}(\tau_{I_lJ_l})}\cdot \frac{w_{I_l}}{n_{I_l}}\right\}.\label{eq:lik2}
 \end{eqnarray}
 It is easy to check that marginalizing the likelihood (\ref{eq:lik2}) over the variables $\left\{ (I_l,J_l), l\in\{1,2,\ldots,n\}\right\}$ gives the likelihood (\ref{eq:lik}). Then, after incorporating the prior distributions in Equations (\ref{eq:pr_alpha}, \ref{eq:pr_beta}, \ref{eq:pr_w}), the idea is to simulate sample observations from the posterior distribution given by
 \begin{eqnarray}
 &&\Pi\left( \alpha_1, \beta_1, \alpha_2, \beta_2, \alpha_3, \beta_3, w_1, \ldots,w_B, (I_1, J_1),\ldots, (I_n,J_n)\big\vert \boldsymbol{s}_x, \boldsymbol{s}_y, \boldsymbol{t}\right)\nonumber\\
&& \propto \prod_{l=1}^n\left\{ \alpha_1 A^{\beta_1}_{I_lJ_l}e^{ -\frac{\left(s_{xl}-x_{I_lJ_l} \right)^2\alpha^2_1 A^{2\beta_1}_{I_lJ_l}}{2}} \cdot  \alpha_2A^{\beta_2}_{I_lJ_l}e^{ -\frac{\left(s_{yl}-y_{I_lJ_l} \right)^2\alpha^2_2 A^{2\beta_2}_{I_lJ_l}}{2}} \cdot \frac{e^{\alpha_3^2A^{2\beta_3}_{I_lJ_l}\cos\left( (t_l-t_{I_lJ_l})\pi/12\right)}}{24I_{0}(\alpha_3^2 A^{2\beta_3}_{I_lJ_l})}\cdot \frac{w_{I_l}}{n_{I_l}}\right\}.\label{eq:posterior}
 \end{eqnarray}
 We use Gibbs sampling to generate the simulations, the details of which are given in Section~\ref{S:gibbs}. In particular, the simulations can be used to obtain summaries from the distribution of $(\alpha_1,\beta_1, \alpha_2, \beta_2, \alpha_3, \beta_3, w_1, \ldots, w_B)$, which would be of main interest. For the preliminary kde required in Equation \ref{eq:A},  we use kde with a fixed bandwidth, obtained by going through the same estimation procedure described earlier in this section but by taking $A_{ij}=1$ for all $i,~j$.   
 
 \subsection{Incorporating expert input and prediction}
 \label{S:prediction}
 Our objective is be to predict the event locations for an upcoming temporal block $(b+1)$ based on the data in the historical $B$ temporal blocks $b, b-1, \ldots, b-B+1$.  Let us denote the set of parameters of the kde model (\ref{eq:KDE_proposed}) by 
 \begin{equation}
 \Theta=(\alpha_1,\beta_1, \alpha_2, \beta_2, \alpha_3, \beta_3, w_1, \ldots, w_B). \label{eq:Theta}
 \end{equation}
  Using the approach in Section \ref{S:estimation} and modeling the data in block $b$ as a function of data in blocks $b-1,\ldots, b-B$,  we obtain the posterior distribution $\Pi\left( \Theta \vert \boldsymbol{s}_x, \boldsymbol{s}_y, \boldsymbol{t}\right) $. 

{\color{black}
To enhance the predictive capabilities of our model, we can extend the proposed block-weighted procedure to incorporate expert inputs in the model. As a case in point, the Delhi Police has implemented a mobile application tool, ``eBeat Book'' designed to interface with various data systems, such as motor vehicle registries, criminal databases, and PCR calls. This tool aids on-ground patrolling by allowing officers to, for example, check the background of a vehicle by entering its registration number or take a photo of a suspected individual for face recognition matching. Additionally, patrol officers can log the location (latitude-longitude and description) of critical places within their area (beat), which can inform deployment decisions. Building on this existing framework, we propose an adaptation of the ``eBeat Book'' tool to systematically collect expert inputs on potential street crime locations. Officers on duty could be instructed to identify and mark on a map areas where they predict, based on their professional judgment, that crimes such as snatching are likely to occur in the upcoming week. The tool would then record the latitude and longitude of these locations, providing essential data for the predictive model. New information, for instance, areas where new construction has started, places which are dimly lit because of failure of traffic lights, make shift (temporary) or new bus stops, etc., are potential locations that can be highlighted by experts as these may not get captured by historical data. To substantiate this proposal, we conducted in-depth interviews with the head of Delhi PCR and multiple patrol officers, focusing on the feasibility of integrating expert inputs and their experiences with the mobile application. The findings from these interviews are detailed in Section~\ref{S:interview_findings} of the paper.}

In the presence of expert inputs, the posterior distribution is learned conditional on historical data fom previous $B$ blocks and expert inputs from $E$ block (refer to Table~\ref{T:blockdata2}). We refer to this posterior distribution as $\Pi_H\left( \Theta \vert \boldsymbol{s}_x, \boldsymbol{s}_y, \boldsymbol{t}\right): \Theta=(\alpha_1,\beta_1, \alpha_2, \beta_2, \alpha_3, \beta_3, w_1, \ldots, w_B, w_E)$, and the steps for estimating this posterior distribution remains the same as discussed in Section \ref{S:estimation} apart from the following modifications. The events in a temporal block $b$ are modeled, conditional on $B$ historical temporal blocks and $E$ expert block as 
\begin{eqnarray}(s_{x1}, s_{y1}, t_1), (s_{x2}, s_{y2}, t_2), \ldots, (s_{xn}, s_{yn}, t_n) \stackrel{ iid}{\sim} f(\cdot, \cdot, \cdot), \label{eq:model2}\end{eqnarray}
where
 \begin{eqnarray}
f(s_x, s_y,t) =\sum_{i=1}^B \frac{w_i}{n_i}\sum_{j=1}^{n_i} \left\{  \frac{1}{h_{1ij}}\kappa\left(\frac{s_x-x_{ij}}{h_{1ij}}\right)\cdot \frac{1}{h_{2ij}}\kappa\left(\frac{s_y-y_{ij}}{h_{2ij}} \right) \cdot \kappa_{\tau_{ij}}\left(t-t_{ij}\right)\right\} + \notag \\
\frac{w_E}{n_E}\sum_{j=1}^{n_E} \left\{  \frac{1}{h_{1Ej}}\kappa\left(\frac{s_x-x_{Ej}}{h_{1Ej}}\right)\cdot \frac{1}{h_{2Ej}}\kappa\left(\frac{s_y-y_{Ej}}{h_{2Ej}} \right) \cdot \kappa_{\tau_{Ej}}\left(t-t_{Ej}\right)\right\}, \label{eq:KDE_proposed2}
\end{eqnarray}
and $n_{tot}=n_1+n_2+\cdots+n_{B-1}+n_{B}+n_{E}$. Note that there is an extra term that gets added with weight $w_E$ in $f(s_x, s_y,t)$. In the absense of expert data block, the weight $w_E$ is zero, therefore, one obtains $\Pi_H\left( \Theta \vert \boldsymbol{s}_x, \boldsymbol{s}_y, \boldsymbol{t}\right) = \Pi\left( \Theta \vert \boldsymbol{s}_x, \boldsymbol{s}_y, \boldsymbol{t}\right)$.  Hereafter, we will use the new posterior distribution of parameters, $\Pi_H\left( \Theta \vert \boldsymbol{s}_x, \boldsymbol{s}_y, \boldsymbol{t}\right)$, to make predictions for $b+1$ as a function of data in the preceding $B$ blocks, viz. $b, b-1, \ldots, b-B+1$, and expert inputs for $b+1$ from $E$ block.

\begin{table}
\centering
\caption{Data organised for block-weighted adaptive kde with a block containing expert inputs}
\label{T:blockdata2}
\renewcommand{\arraystretch}{1.3} 
\resizebox{\textwidth}{!}{
\begin{tabular}{|l|c|c|c|c|c|c|}
\hline
~& \multicolumn{4}{c|}{Historical blocks}& Expert block & Latest block \\
\hline
Temporal block & $b-B$ & $b-B+1$& $\ldots$ & $b-1$ & Expert inputs for $b$ & $b$ \\
\hline
Block index $i$ & $1$ & $2$& $\ldots$ & $B$ & $E$ & $~$  \\
\hline
Lon. coordinates& $x_{11},x_{12},\ldots,x_{1n_1}$ & $x_{21},x_{22},\ldots,x_{2n_2}$& $\ldots$ & $x_{B1},x_{B2},\ldots,x_{Bn_B}$& $x_{E1},x_{E2},\ldots,x_{En_E}$ & $s_{x1},s_{x2},\ldots,s_{xn}$  \\
\hline
Lat. coordinates& $y_{11},y_{12},\ldots,y_{1n_1}$ & $y_{21},y_{22},\ldots,y_{2n_2}$& $\ldots$ & $y_{B1},y_{B2},\ldots,y_{Bn_B}$& $y_{E1},y_{E2},\ldots,y_{En_E}$ & $s_{y1},s_{y2},\ldots,s_{yn}$  \\
\hline
Event time (in hours)& $t_{11},t_{12},\ldots,t_{1n_1}$ & $t_{21},t_{22},\ldots,t_{2n_2}$& $\ldots$ & $t_{B1},t_{B2},\ldots,t_{Bn_B}$& $t_{E1},t_{E2},\ldots,t_{En_E}$ & $t_{1},t_{2},\ldots,t_{n}$ \\
\hline
Block weight& $w_1$ & $w_2$& $\ldots$ & $w_B$ & $w_E$ & $~$ \\
\hline
\end{tabular}}\\
\end{table}

Using similar notations as in Table \ref{T:blockdata2}, let us denote the data in the $B$ historical temporal blocks preceding block $(b+1)$ and expert inputs in block $E$ as follows
$$\{(x^\prime_{ij}, y^\prime_{ij}, t^\prime_{ij}), ~ i=1,2,\ldots, B-1,B,E, ~~ j=1,2,\ldots, n^\prime_{i}\}.$$
For simplifying the notations in the later part of the paper, let us denote $E=0$, which reduces the above representation to
$$\{(x^\prime_{ij}, y^\prime_{ij}, t^\prime_{ij}), ~ i=0,1,2,\ldots,B, ~~ j=1,2,\ldots, n^\prime_{i}\}.$$
The adaptive bandwidths corresponding to the above coordinates are given by
\begin{eqnarray}
h^\prime_{1ij}=\frac{1}{\alpha_1A_{ij}^{\prime\beta_1}}, ~h_{2ij}=\frac{1}{\alpha_2A_{ij}^{\prime\beta_2} }\mbox{ and } \tau^\prime_{ij}=\frac{1}{h^{\prime 2 }_{3ij}} , \mbox{ with } h^\prime_{3ij}=\frac{1}{\alpha_3 A^{\prime\beta_3}},  \label{eq:adaptbw2}
\end{eqnarray}
where 
\begin{eqnarray}
A^\prime_{ij}= \hat{f}_p(x^\prime_{ij},y^\prime_{ij},t^\prime_{ij})/G,  ~G=\left(\prod_{i=1}^B\prod_{j=1}^{n^\prime_i} \hat{f}_p(x^\prime_{ij},y^\prime_{ij},t^\prime_{ij}) \right)^{1/n^\prime_{tr}},~n^{\prime}_{tr}=n^{\prime}_{0}+n^{\prime}_1+n^{\prime}_2+\cdots+n^{\prime}_{B}, \label{eq:A2}
\end{eqnarray}
where $\hat{f}_p$ is the same preliminary kde, constructed using data in Table \ref{T:blockdata},  with a fixed bandwidth, by going through the same estimation procedure described in Section \ref{S:estimation},  by taking $A_{ij}=1$ for all $i,~j$. The predictive spatio-temporal pdf at any point $(s_x^\prime, s_y^\prime, t^\prime)$,  for the temporal block $(b+1)$ is given by 
\begin{eqnarray}
&&\hat{f}(s^\prime_x, s^\prime_y,t^\prime) = \int \sum_{i = 0}^{B} \frac{w_i}{n^\prime_i}\sum_{j=1}^{n^\prime_i} \left\{  \frac{1}{h^\prime_{1ij}}\kappa\left(\frac{s_x-x_{ij}}{h^\prime_{1ij}}\right)\cdot \frac{1}{h^\prime_{2ij}}\kappa\left(\frac{s_y-y_{ij}}{h^\prime_{2ij}} \right) \cdot \kappa_{\tau^\prime_{ij}}\left(t-t_{ij}\right)\right\} d\Pi_H\left( \Theta \vert \boldsymbol{s}_x, \boldsymbol{s}_y, \boldsymbol{t}\right),
\label{eq:KDE_postpredict}
\end{eqnarray}
where $\Theta$ is the set of parameters and the integral in the above expression is over the parameter space with respect to the posterior probability measure.  Numerical evaluation of the integral  (\ref{eq:KDE_postpredict}) requires repeated computation of kde in the integrand several times. This computation can be very time consuming and based on our assessment takes more than 12 hours for a single round of computation for the city of Delhi on a standard PC with an Intel(R) Core(TM) i7-7700 CPU@3.6GHz and 16 GB RAM. When the runs are done on an HPC server having 24 clusters with configuration Intel(R) Xeon(R)
Gold 6140 CPU@2.30 GHz and 64 GB RAM, the task gets completed in less than 1 hour.  
In this paper, we propose to use the following approximate predictive density, which is obtained by computing (\ref{eq:KDE_proposed}) at the posterior mean value of the parameters. This serves our purpose  in terms of a reasonably good predictive accuracy as shown in the Section \ref{S:results}. Specifically, we compute
\begin{eqnarray}
&&\tilde{f}(s^\prime_x, s^\prime_y,t^\prime) = \sum_{i=0}^B \frac{\tilde{w}_i}{n^\prime_i}\sum_{j=1}^{n^\prime_i} \left\{  \frac{1}{\tilde{h}_{1ij}}\kappa\left(\frac{s^\prime_x-x^\prime_{ij}}{\tilde{h}_{1ij}}\right)\cdot \frac{1}{\tilde{h}_{2ij}}\kappa\left(\frac{s^\prime_y-y^\prime_{ij}}{\tilde{h}_{2ij}} \right) \cdot \kappa_{\tilde{\tau}_{ij}}\left(t^\prime-t^\prime_{ij}\right)\right\},
\label{eq:KDE_postpredictapprox}
\end{eqnarray}
where 
\begin{eqnarray}
\tilde{h}_{1ij}=\frac{1}{\tilde{\alpha}_1A_{ij}^{\prime\tilde{\beta}_1}}, ~\tilde{h}_{2ij}=\frac{1}{\tilde{\alpha}_2A_{ij}^{\prime\tilde{\beta}_2} }\mbox{ and } \tau^\prime_{ij}=\frac{1}{\tilde{h}^{ 2 }_{3ij}} , \mbox{ with } \tilde{h}_{3ij}=\frac{1}{\tilde{\alpha}_3 A^{\prime\tilde{\beta}_3}},  \label{eq:adaptbwapprox}
\end{eqnarray}
with $\tilde{\Theta}=(\tilde{\alpha}_1,\tilde{\beta}_1, \tilde{\alpha}_2, \tilde{\beta}_2, \tilde{\alpha}_3, \tilde{\beta}_3, \tilde{w}_0, \ldots, \tilde{w}_B)$ being the mean of the vector of parameters $\Theta$ from the posterior distribution $\Pi_H\left( \Theta \vert \boldsymbol{s}_x, \boldsymbol{s}_y, \boldsymbol{t}\right)$.  
Based on (\ref{eq:KDE_postpredictapprox}),  for the temporal block $(b+1)$, we compute the (approximate) predictive spatial density in a given time interval $[t_1, t_2]$ of the day, as 
\begin{eqnarray}
&&\tilde{f}(s^\prime_x, s^\prime_y) = \frac{\sum_{i=0}^B \frac{\tilde{w}_i}{n^\prime_i}\sum_{j=1}^{n^\prime_i} \left\{  \frac{1}{\tilde{h}_{1ij}}\kappa\left(\frac{s^\prime_x-x^\prime_{ij}}{\tilde{h}_{1ij}}\right)\cdot \frac{1}{\tilde{h}_{2ij}}\kappa\left(\frac{s^\prime_y-y^\prime_{ij}}{\tilde{h}_{2ij}} \right) \cdot \int_{t_1}^{t_2}\kappa_{\tilde{\tau}_{ij}}\left(t^\prime-t^\prime_{ij}\right)dt^\prime\right\} }{ \sum_{i=0}^B \frac{\tilde{w}_i}{n^\prime_i}\sum_{j=1}^{n^\prime_i} \left\{ \int_{t_1}^{t_2}\kappa_{\tilde{\tau}_{ij}}\left(t^\prime-t^\prime_{ij}\right)dt^\prime\right\} }.
\label{eq:KDE_postpredictapprox_sp}
\end{eqnarray}
For practical purposes, it helps to divide the city into a grid of 36,263  boxes, each of approximate dimension $200m\times 200m$, and identify any location with the coordinates of the center of the grid box containing it. Accordingly, we compute the prediction (\ref{eq:KDE_postpredictapprox_sp}) only at the centers of the 36,263 grid boxes. Based on the discussions in this section, we create multiple models and discuss the results in the next section. We also propose a metric to evaluate the predictive accuracy of the various models.

{\color{black}
The proposed method offers a flexible framework that can be extended to incorporate additional auxiliary information in the form of covariates, enhancing the model's predictive capabilities. This adaptability ensures that the formulation remains relevant and robust across varied operational contexts. A brief discussion on integrating such covariates is provided in Section~\ref{sec:covariates}, highlighting the potential for more informed and context-sensitive density predictions. 
However, in the data shared by Delhi police, such covariates have not been captured and therefore not studied in our implementation. As a part of future work, we intend to look at the importance of additional information, like, population density, socio-economic indicators, weather conditions and past patrol vechicle locations for producing better estimates. Interestingly, dynamic covariates are likely to be more useful as compared to static covariates that are endogenous in the spatio-temporal crime data.
}

\section{Results}
\label{S:results}
We consider different model formulations based on kde in this section, and identify the best performing model. To begin with, we consider all models without any expert inputs. After having identified the best performing model on historical data, we do a further analysis by incorporating expert inputs with varying levels of accuracy to evaluate the performance of the chosen model. As mentioned earlier, we use `week' as a temporal block in our analysis, however, one can choose smaller or larger temporal blocks as the situation demands. Next, we define the models (without expert inputs) that have been considered in this section. ~\\
 {\bf Model 1:} This is a purely spatial model, as in Equation (\ref{eq:KDE_pred}), but based on only past 1 week (i.e. one past temporal block) of event locations during a given daily time interval. Since this model formulation does not incorporate timing of the events during the day for estimating the kde for a given time interval (such as 8 pm to 12 am), we use only those events of the past temporal block that occurred during the given time interval.~\\
 {\bf Model 2:} This is a spatio-temporal model, as in Equation (\ref{eq:KDEspt_pred}), but based on only past 1 week (i.e. one past temporal block) of event locations and timings. Since this model incorporates timing of events during the day, unlike Model 1, the estimation is based on all event locations along with their timing in the previous week, rather than restricting only to events within a given time interval.~\\
 The fixed-bandwidth and adaptive bandwidth cases of Models 1 and 2 are estimated using the Bayesian approach given in Section \ref{S:estimation}, of course suitably modified in the case of Model 1 to exclude the parameters related to timing. Our preliminary analysis (as shown in Figure \ref{F:calllocs21mar}) suggests that using a longer time frame of data might give better predictive performance. The rest of the model scenarios allow for  past 52 weeks of data (i.e. approximately past 1 year).~\\
 {\bf Model 3:} This is a spatio-temporal model, as in Equation (\ref{eq:KDEspt_pred}), but  we use Silverman's rule of thumb (SROT) for the bandwidth parameters (e.g. \citealt{silverman1986}, \citealt{sheather2004}), and use equal weights for the temporal-blocks. Since SROT gives a simple approach to obtaining the fixed bandwidth parameters, this serves as a benchmark on whether the computational effort of using a Bayesian estimation is worthwhile.  Specifically, we take
 \begin{eqnarray}
 \begin{cases} \tilde{h}_1 = 0.9 \times min(sd(\tilde{x}), IQR(\tilde{x})/1.34)\times n_{tot}^{-1/5},\\
 \tilde{h}_2= 0.9 \times min(sd(\tilde{y}), IQR(\tilde{y})/1.34)\times n_{tot}^{-1/5},\\
 \tilde{h}_3 = 0.9 \times 24/(2\pi)\times circularsd(\tilde{2\pi t/24})\times n_{tot}^{-1/5}, ~~\tilde{\tau}= 1/\tilde{h}_3^2,
 \end{cases} \label{eq:h_srot}
 \end{eqnarray}
 where $\tilde{x}=\{x_{ij}\},$ $\tilde{y}=\{y_{ij}\}$ and $\tilde{t}=\{t_{ij}\}$, are collection of data points from the historical temporal blocks for the longitude, latitude and timing, respectively, for $ i=1,2,\ldots,B, ~j=1,2,\ldots, n_i$. Note that $sd$ and $IQR$ denote standard deviation and inter-quartile range, respectively. For $h_3$, we use circular standard deviation given the circular nature of the timing variable. {\color{black}The adaptive bandwidths in this case are obtained following \citet{abramson1982} using Equation (\ref{eq:adaptbw}) with $\alpha_1=\tilde{h}_1$, $\alpha_2=\tilde{h}_2$,  $\alpha_3=\tilde{h}_3$ and $\beta_1 = \beta_2= \beta_3 =0.5$.}\\
 {\bf Model 4:} This is a purely spatial model, as in Equation (\ref{eq:KDE_pred}), and based on past 52 weeks of event locations. Similar to Model 1, since this model formulation does not incorporate timing of the events during the day for estimating the kde for a given time interval (such as 8 pm to 12 am), we use only those events of the past weeks that occurred during the given time interval.~\\
{\bf Model 5:} This is a spatio-temporal model, as in Equation (\ref{eq:KDEspt_pred}),  based on past 52 weeks of event locations and timings. Since this model incorporates timing of events during the day, unlike Model 4, the estimation is based on all event locations along with their timing in the previous week, rather than restricting only to events within a given time interval.

The fixed bandwidth and adaptive bandwidth cases of Models 4 and 5 are estimated using the Bayesian approach given in Section \ref{S:estimation}, where the procedure for Model 4 is suitably modified to exclude the estimation of parameters related to timing. Table~\ref{tab:rev_model_comp} provides a comparison of the models considered in this study with respect to their characteristics.

\begin{table}[h!]
\centering
\caption{\textcolor{black}{Comparison of models with respect to their characteristics considered in the experiments.}}
\begin{tabular}{lccccc}
\toprule
Model   & Past periods considered & Spatial & Temporal & Adaptive Bandwidth & Adaptive Weights \\ \midrule
Model 1 & 1 week                           & $\checkmark$     & $\times$                    & $\checkmark$                & $\times$                     \\ 
Model 2 & 1 week                           & $\checkmark$     & $\checkmark$             & $\checkmark$            & $\times$                     \\ 
Model 3 & 52 weeks                         & $\checkmark$     & $\checkmark$            & $\checkmark$     & $\times$                     \\ 
Model 4 & 52 weeks                         & $\checkmark$     & $\times$                   & $\checkmark$                & $\checkmark$                     \\ 
Model 5 & 52 weeks                         & $\checkmark$     & $\checkmark$            & $\checkmark$            & $\checkmark$                 \\ 
\bottomrule
\end{tabular}
\label{tab:rev_model_comp}
\end{table}
\vspace{-2mm}


\subsection{Predictive accuracy with historical data}
\label{S:accuracy}
We propose `area under the event-area curve' (AUC) metric based on `predictive accuracy index' (PAI) to evaluate the predictive performance of various models. PAI is the ratio of the percentage of actual events captured to a chosen monitoring area (see e.g. \citealt{flaxmanetal2019}, \citealt{hu2018spatio}). In line with the idea of PAI, to obtain a single summary metric for a given model, we look at the ``event-area'' curve, which we define as the plot of the percentage of actual events captured versus percentage monitored area in descending order of  the predicted likelihood of street crime events as per the model. We then compute the AUC as our summary metric to measure the accuracy of the model. An illustration of the event-area curve is shown in Part (d) of Figure \ref{F:calllocs21mar} in Section \ref{S:intro}, where the spatial kde estimated on past 52 weeks  is seen to capture the largest percentage of actual events by area and hence has the highest AUC. We will use AUC to compare different model scenarios in the next section.

To obtain predictions for a temporal block $(b+1)$ we model the events in a temporal block $b$, based on its $B$ historical temporal blocks (see Table \ref{T:blockdata}). Then, as described in Section \ref{S:accuracy}, to check the performance of the model in predicting the event locations of the (out of sample) block $(b+1)$, we compute the event-area curve and the corresponding area under the curve (AUC). So, we obtain the predicted kde for each of the 25 upcoming weeks using the approach in Section \ref{S:prediction}, starting with the week of 4 October 2020 up to the week of 21 March 2021,  based on each of the Models 1 to 5. 

Figure \ref{F:AUCoverall} shows the over-all average AUC for the different models with respect to actual event locations during different time intervals in the day. It is clear from the figure that Models 3 to 5, which are based on past 52 weeks of data, outperform Models 1 and 2, which are based only on the recent past 1 week of data. We  see that there is benefit to a spatio-temporal kde formulation (\ref{eq:KDEspt_pred}), as is done in Model 5, over a purely spatial kde formulation  (\ref{eq:KDE_pred}), as is done in Model 4; although the latter is still estimated based on events that occurred within a particular time interval. Further, Model 5 obtains the required bandwidth parameters via Bayesian estimation, which clearly outperforms Model 3, which uses Silverman's rule of thumb. 

To get insights into the performance of the models week-wise, Figure \ref{F:AUCweekwise} shows the AUC by week, for different time-intervals in the day. It is clear that the week-wise AUC performance is in line with the observations made in the previous paragraph based on the overall average across weeks. Model 5 consistently outperforms all the other model scenarios.  In practice, the model predicted density will be used to mark the top areas most likely to experience crime. Table \ref{T:top2040pct} shows a summary of the `percentage of actual events captured for any upcoming week', in the top 20\% and top 40\% monitored areas in different time intervals of the day, based on model predictions for each of the 25 upcoming weeks (i.e. week of 4 October 2020 up to week of  March 21 2021). Again, we see that Model 5 captures the highest percentage of events, across all time intervals. In particular, Model 5 predictions capture 75-80\% of actual events in the top 20\% monitored areas, and 93-97\% events in the top 40\% monitored areas, on an average. Further, we can see that the standard deviations for the different time groups is lower for Model 5, compared to the other models, which indicates more certainty in the predictions than Models 1 to 4. The posterior summaries for the parameters of Model 5 has been relegated to Section~\ref{S:additional_results}.

\vspace{-5mm}
\begin{figure}[ht]
\center
\includegraphics[height=7cm]{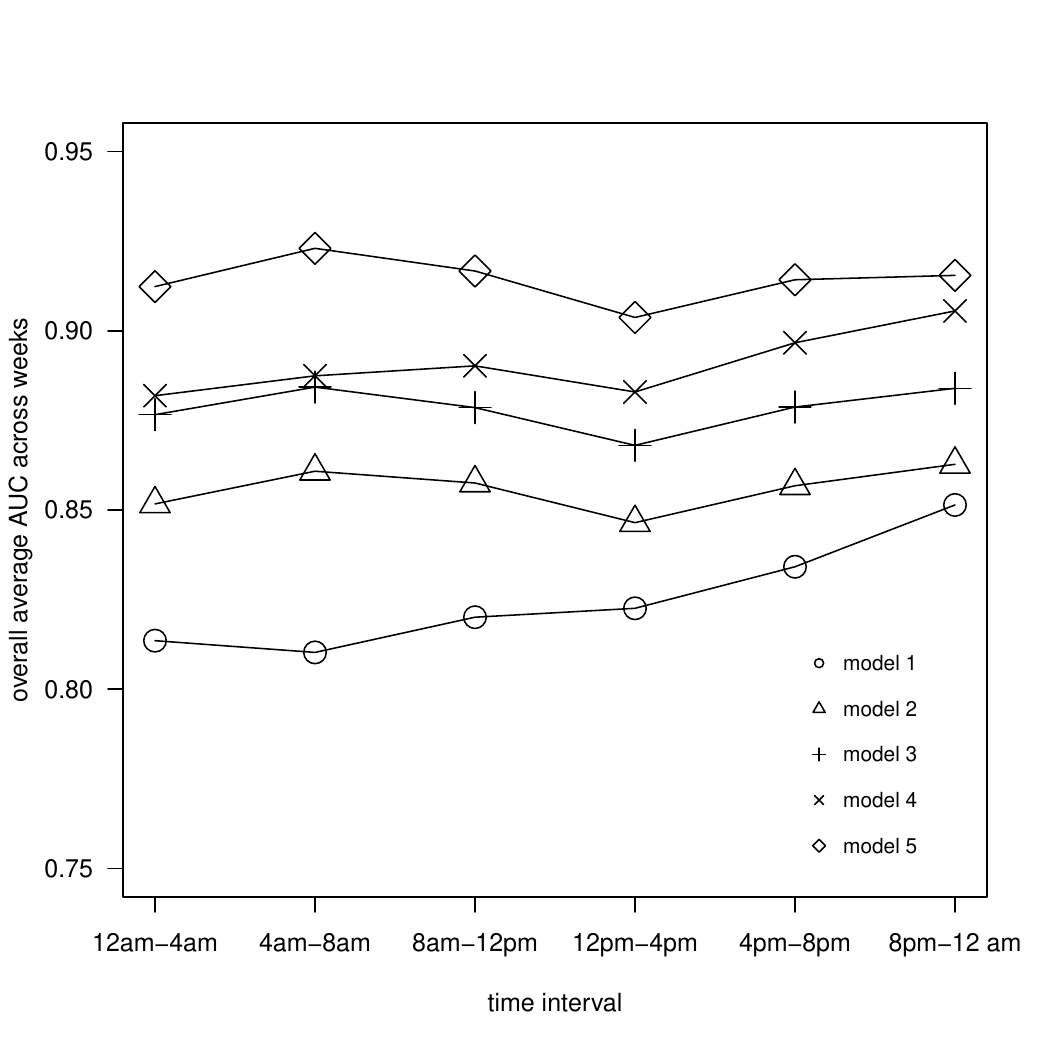}
\vspace{-4mm}
\caption{Overall average AUC over the 25 weeks [week of 4 October 2020 up to week of  March 21 2021]}
\label{F:AUCoverall}
\end{figure}

\begin{table}[h]
\caption{\label{T:top2040pct} \textcolor{black}{Summary of the `percentage of actual events captured for any upcoming week', in the top 20\% and top 40\% monitored areas in different time intervals of the day, for the 25 upcoming weeks (i.e. week of 4 October 2020 up to week of 21 March 2021). Improvement in terms of additional percentage of actual events captured by Model 5 over Model 1 is provided in brackets for each case. 
}}
\centering
\begin{subtable}{\textwidth}
\caption{Top 20 \% monitored area}
\resizebox{\textwidth}{!}{
\begin{tabular}{lcccccccccc}
  \toprule
 ~ &\multicolumn{5}{c}{Average value over the 25 upcoming weeks}&\multicolumn{5}{c}{Standard deviation over the 25 upcoming weeks}\\ \cmidrule(l){2-6} \cmidrule(l){7-11}
 & Model 1 & Model 2 & Model 3 & Model 4 & Model 5 (Impr.) & Model 1 & Model 2& Model 3 & Model 4 & Model 5 \\ 
  \midrule
12 am-4 am & 0.499 & 0.592 & 0.639 & 0.683 & \textbf{0.770} (27.1\%) & 0.081 & 0.088 & 0.066 & 0.056 & \textbf{0.047}\\ 
  4 am-8 am & 0.561 & 0.638 & 0.673 & 0.720 & \textbf{0.792} (23.1\%) & 0.110 & 0.104 & 0.081 & 0.083 & \textbf{0.080}\\ 
  8 am-12 am & 0.549 & 0.635 & 0.654 & 0.709 & \textbf{0.803} (25.4\%) & 0.097 & 0.076 & 0.076 & 0.054 & \textbf{0.061}\\ 
  12 am-4pm & 0.563 & 0.584 & 0.634 & 0.691 & \textbf{0.756} (19.3\%) & 0.057 & 0.062 & 0.070 & 0.052 & \textbf{0.045}\\ 
  4 pm-8 pm& 0.581 & 0.629 & 0.668 & 0.723 & \textbf{0.766} (18.5\%) & 0.059 & 0.049 & 0.038 & 0.037 & \textbf{0.035}\\ 
  8 pm-12 am & 0.611 & 0.637 & 0.695 & 0.746 & \textbf{0.764} (15.3\%) & 0.042 & 0.042 & 0.035 & 0.035 & \textbf{0.031}\\ 
   \bottomrule
\end{tabular}}
\end{subtable}\vspace{3mm}
\begin{subtable}{\textwidth}
\caption{Top 40 \% monitored area}
\resizebox{\textwidth}{!}{
\begin{tabular}{lcccccccccc}
  \toprule
 ~ &\multicolumn{5}{c}{Average value over the 25 upcoming weeks}&\multicolumn{5}{c}{Standard deviation over the 25 upcoming weeks}\\ \cmidrule(l){2-6} \cmidrule(l){7-11}
 & Model 1 & Model 2 & Model 3 & Model 4 & Model 5 (Impr.) & Model 1 & Model 2& Model 3 & Model 4 & Model 5 \\
  \midrule
12 am-4 am & 0.811 & 0.878 & 0.903 & 0.897 & \textbf{0.940} (12.9\%) & 0.068 & 0.049 & 0.038 & 0.045 & \textbf{0.030}\\ 
  4 am-8 pm & 0.809 & 0.876 & 0.902 & 0.908 & \textbf{0.951} (14.2\%) & 0.071 & 0.074 & 0.054 & 0.064 & \textbf{0.050}\\ 
  8 pm-12 am & 0.827 & 0.890 & 0.894 & 0.921 & \textbf{0.962} (13.5\%) & 0.074 & 0.052 & 0.052 & 0.035 & \textbf{0.026}\\ 
  12 am-4 pm& 0.826 & 0.873 & 0.898 & 0.913 & \textbf{0.939} (11.3\%) & 0.046 & 0.039 & 0.037 & 0.031 & \textbf{0.026}\\ 
  4 pm-8 pm & 0.849 & 0.885 & 0.907 & 0.930 & \textbf{0.949} (10.0\%) & 0.041 & 0.030 & 0.028 & 0.026 & \textbf{0.020}\\ 
  8 pm-12 am & 0.878 & 0.896 & 0.924 & 0.943 & \textbf{0.952} (7.4\%) & 0.025 & 0.019 & \textbf{0.016} & 0.019 & 0.017\\ 
   \bottomrule\\
\end{tabular}}
\end{subtable}
{\color{black}* Average and standard deviation of AUC values have been calculated over a 25 week period}
\end{table}

\begin{figure}[ht]
\begin{subfigure}{.48\textwidth}
\center
\includegraphics[width=7cm, height=7cm]{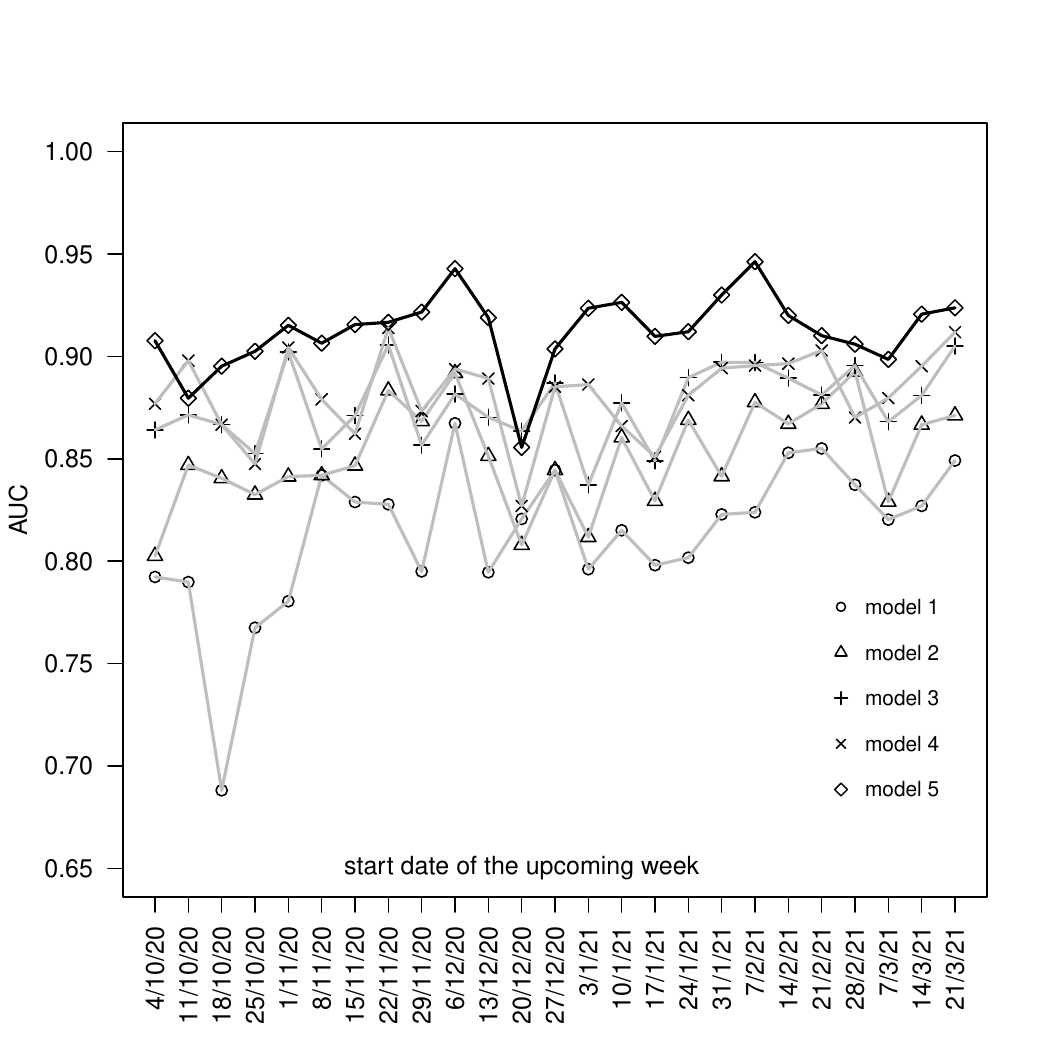}
\caption{ 12 am to 4 am}
\end{subfigure}
\begin{subfigure}{.48\textwidth}
\center
\includegraphics[width=7cm, height=7cm]{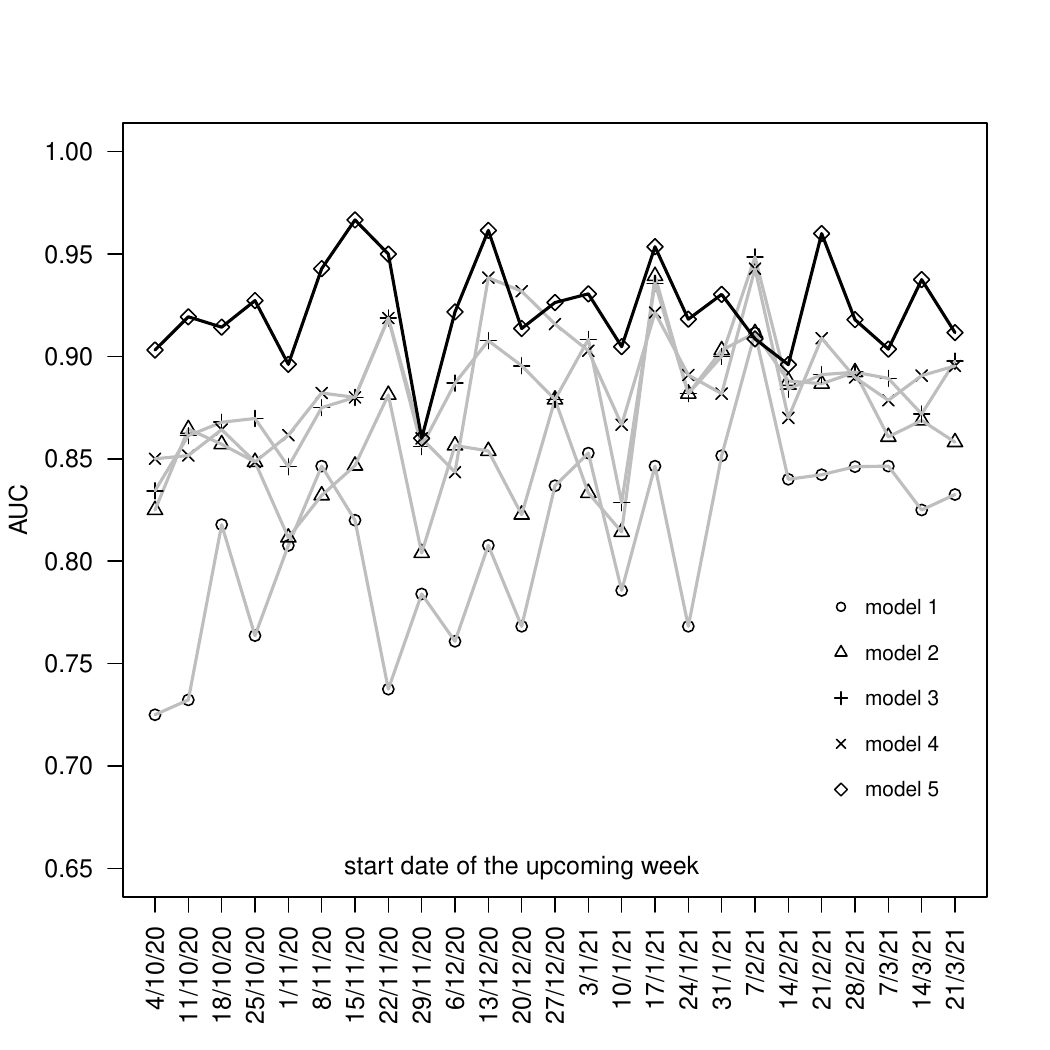}
\caption{4 am to 8 am}
\end{subfigure}

\begin{subfigure}{.48\textwidth}
\center
\includegraphics[width=7cm, height=7cm]{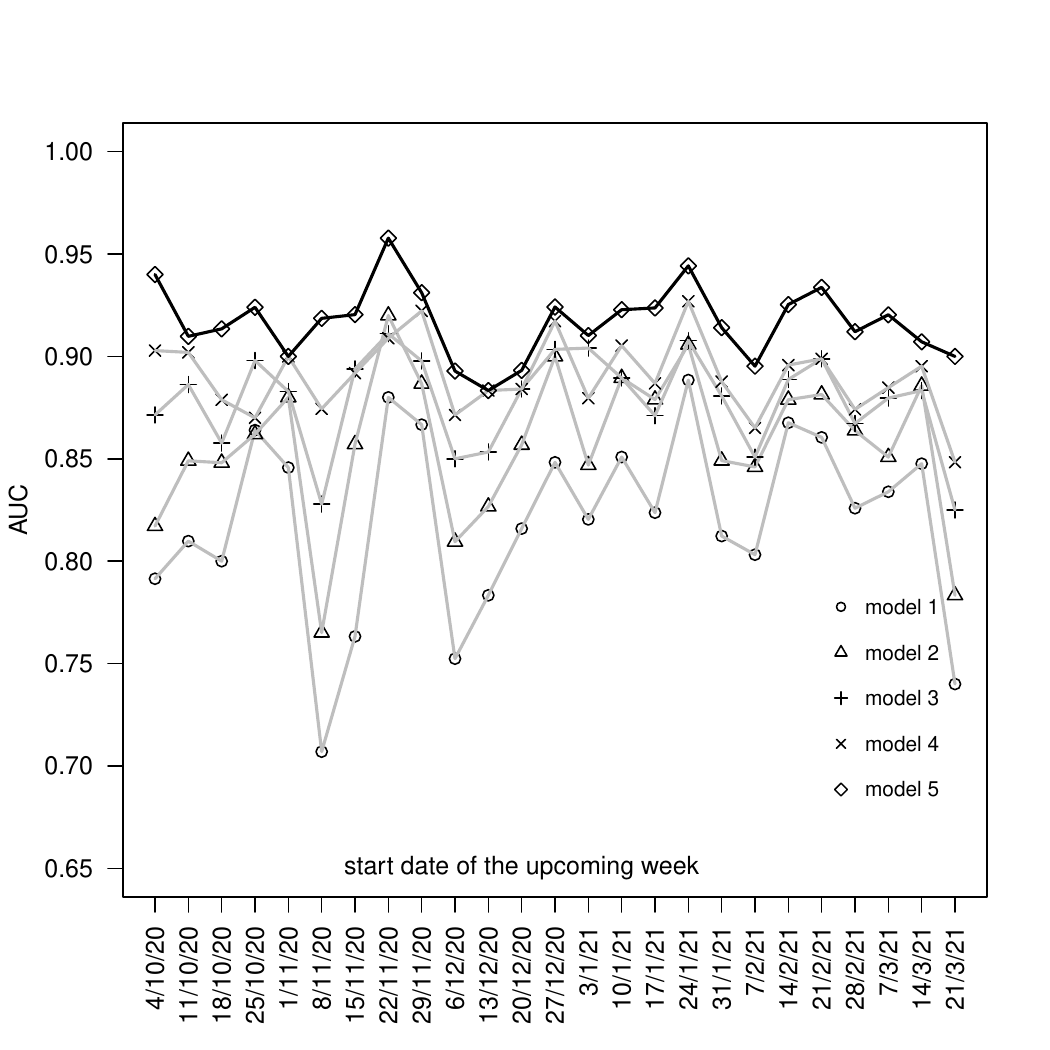}
\caption{8 am to 12 pm}
\end{subfigure}
\begin{subfigure}{.48\textwidth}
\center
\includegraphics[width=7cm, height=7cm]{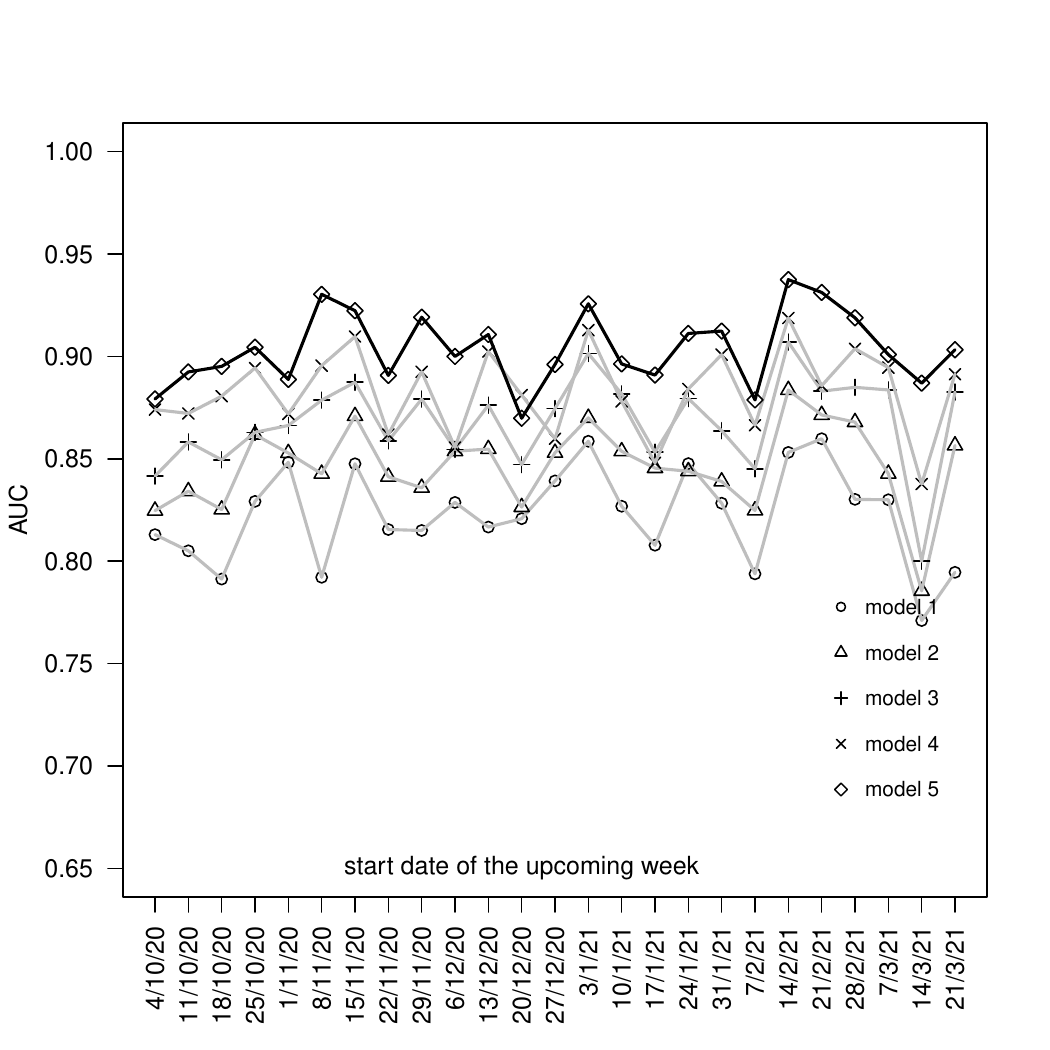}
\caption{12 pm to 4 pm}
\end{subfigure}

\begin{subfigure}{.48\textwidth}
\center
\includegraphics[width=7cm, height=7cm]{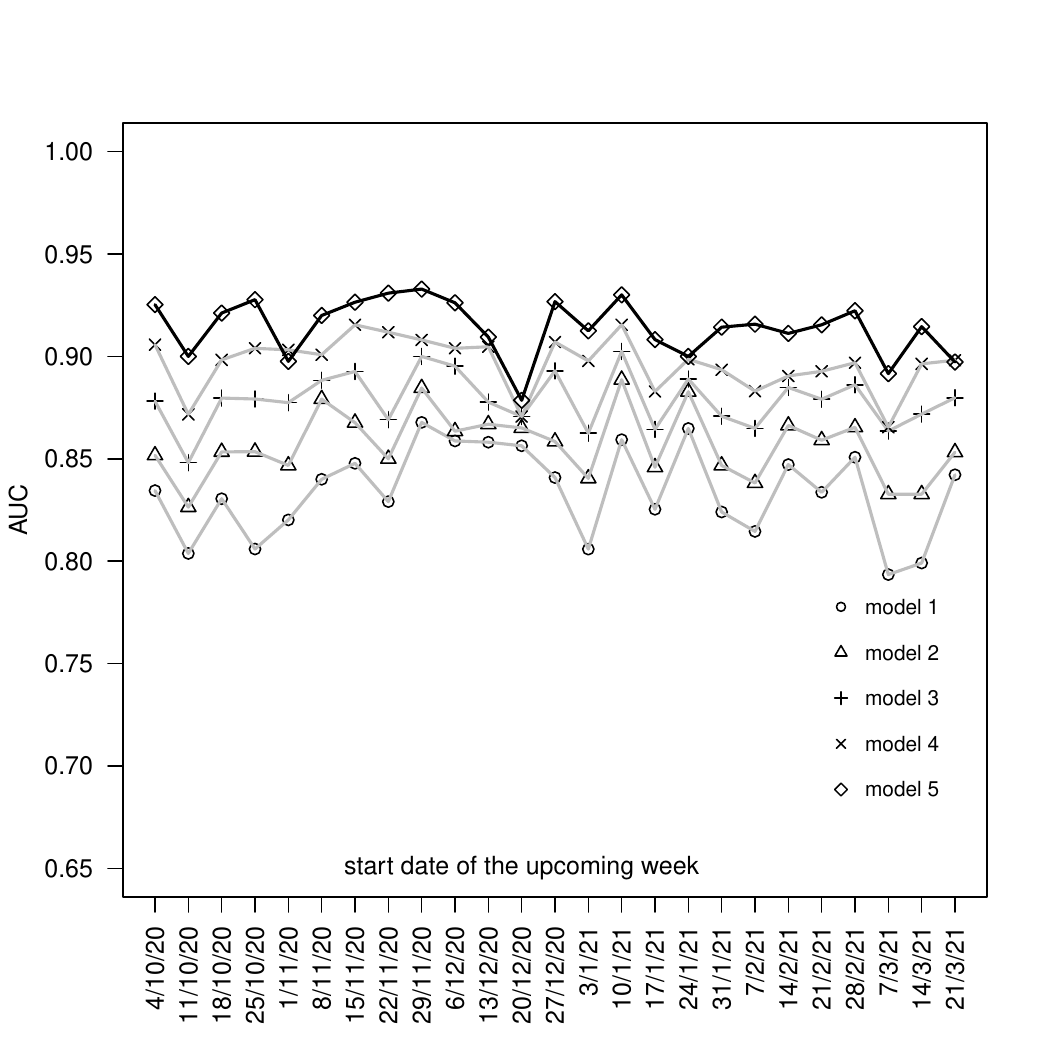}
\caption{4 pm to 8 pm}
\end{subfigure}
\begin{subfigure}{.48\textwidth}
\center
\includegraphics[width=7cm, height=7cm]{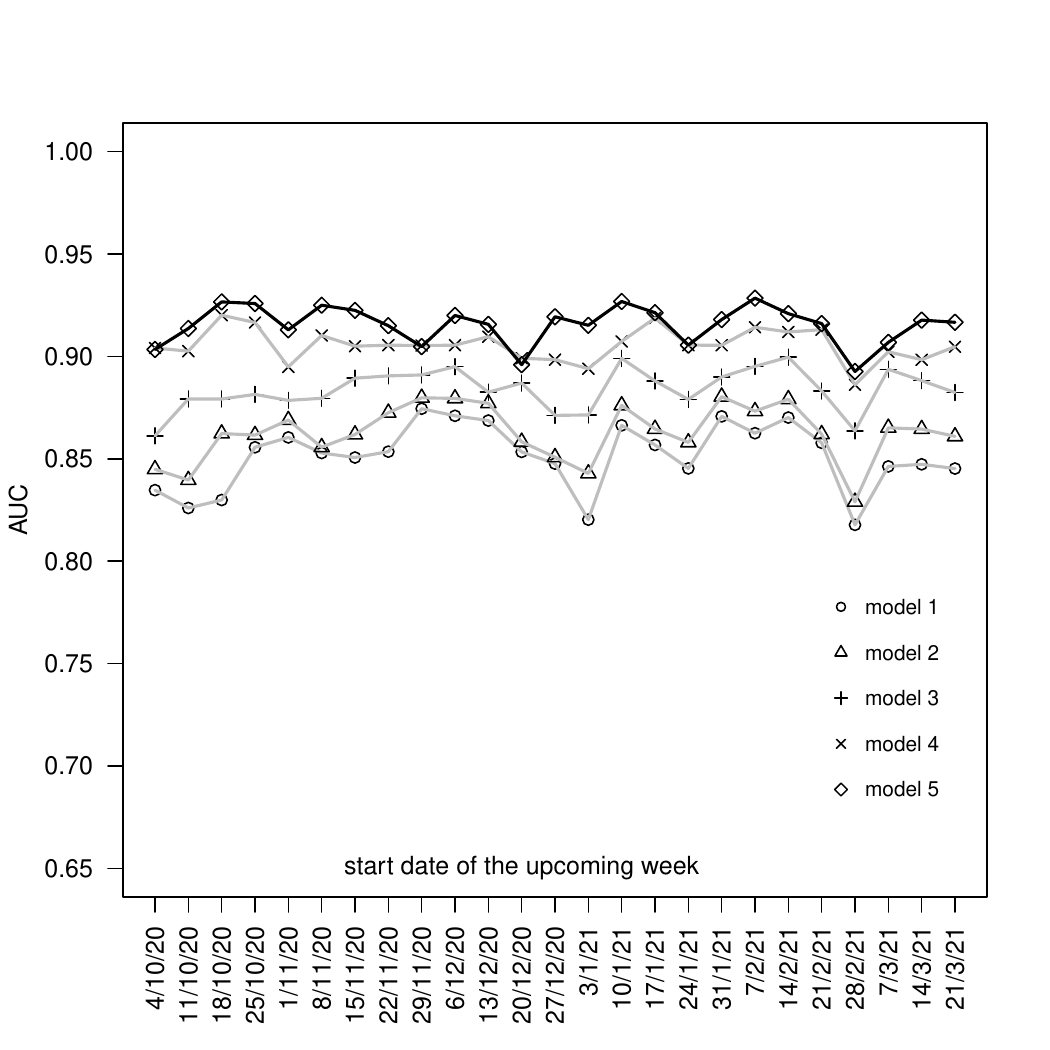}
\caption{8 pm to 12 am}
\end{subfigure}
\caption{AUC by week [week of 4 October 2020 up to week of March 21 2021] for different time intervals in the day based on different models}
\label{F:AUCweekwise}
\end{figure}
\FloatBarrier

\subsection{Predictive accuracy with expert inputs}
In this section, we design our experiments to measure the impact of expert information varying in terms of amount and quality. The expert inputs are simulated in our study as this information was not available through the Delhi police. However, their intention is to have an approach that can incorporate expert inputs when available. As a baseline for comparison we use Model 5 from the previous section that does not contain any expert input, but relies on the historical data alone. We refer to this model as a `No Intel' model in our results. The value of expert input is estimated in terms of gain in AUC when different amount of information with varying quality is provided by the user. 
To measure the impact of amount of information, we consider that the user provides information about some percentage of the actual events that happen in the next temporal block, while to measure the impact of quality of information, we consider that the information provided by the user about incidents to happen in the next temporal block are within an accuracy radius. We call the first dimension as the extent of expert information, and we call the second dimension as the accuracy radius of expert information. For instance, if the actual number of events that happen in the next temporal block are 100, then a 20\% extent of information would mean that the expert has insights on 20 out of the actual 100 events that would occur, and a 100 meter accuracy would mean that the expert is able to predict the crime locations with an accuracy radius of 100 meter. In our experimentation we vary the extent of information as $10\%, 20\%, 30\%$ and $50\%$, while the accuracy radius is varied as $100$ meter, $500$ meter and $1000$ meter. The results are expected to be better when the extent of information is larger and the accuracy radius is smaller. In our interactions with the Delhi police department, we identified that the extent of information and accuracy radius of expert inputs available every week typically fall in the lower side of the ranges chosen by us in our experimentation. 

{\color{black}The steps for simulating expert input has been outlined below. Suppose $p$ represents proportion of events on which the experts are able to provide information
(i.e. indicating extent of information) and $d$ represents radius within which the experts are able to suggest event locations (i.e. accuracy radius of expert information).
To simulate expert inputs for a given week $k$, across specific time-windows—12am-4am, 4am-8am, 8am-12pm, 12pm-4pm, 4pm-8pm, and 8pm-12am—the following procedure is adopted:
\begin{enumerate}
    \item \textbf{Identify Actual Event Locations:} Consider the actual event locations, denoted as $L_1, L_2, \ldots, L_n$, within the specified time-window during week $k$.
    \item \textbf{Random Selection of Events:} Randomly select $m = \left\lfloor n \cdot p \right\rfloor$ events from the set $L_1, \ldots, L_n$, where $p$ represents the proportion of events for which experts can provide input.
    \item \textbf{Simulate Expert Inputs:} For each event location selected in Step 2, generate a random location within a disc of radius $d$ around the original event location. This new location represents the simulated expert input, capturing the accuracy of expert predictions.
    \item \textbf{Assign Temporal Information:} For each simulated expert input, assign the mid-point of the corresponding time-window as the time associated with that expert input.
\end{enumerate}
}

\begin{table}[htbp]
  \centering
  \caption{\textcolor{black}{AUC values with varying extent of information and accuracy radius of expert inputs. Improvement percentage is the difference in AUC as compared to the `No Intel' case expressed in terms of percentage.}}
    \begin{tabularx}{\textwidth}{c *{7}{Y}}
    \toprule
          &       & \multicolumn{3}{c}{Intel on 10\% incidents within} & \multicolumn{3}{c}{Intel on 20\% incidents within} \\
\cmidrule(l){3-5} \cmidrule(l){6-8}    Time  & No Intel & 100m  & 500m  & 1000m & 100m  & 500m  & 1000m \\
\cmidrule(l){1-2} \cmidrule(l){3-5} \cmidrule(l){6-8}
     12 am - 4 am  &                             0.924  &                             0.922  &                             0.922  &                             0.912  &                             0.925  &                             0.937  &                             0.920  \\
     4 am - 8 am  &                             0.912  &                             0.923  &                             0.928  &                             0.916  &                             0.928  &                             0.914  &                             0.907  \\
     8 am-12 pm  &                             0.900  &                             0.912  &                             0.908  &                             0.887  &                             0.917  &                             0.910  &                             0.898  \\
     12 pm -4 pm  &                             0.903  &                             0.917  &                             0.914  &                             0.896  &                             0.914  &                             0.909  &                             0.908  \\
     4 pm- 8 pm  &                             0.897  &                             0.907  &                             0.905  &                             0.903  &                             0.905  &                             0.906  &                             0.906  \\
     8 pm- 12 am  &                             0.917  &                             0.920  &                             0.918  &                             0.910  &                             0.930  &                             0.929  &                             0.922  \\
    \midrule
Overall average  & 0.909 & 0.917 & 0.916 & 0.904 & 0.920 & 0.917 & 0.910 \\ 
Improvement percent                 &       & 0.8\% & 0.7\% & -0.5\% & 1.1\% & 0.8\% & 0.1\%\\
        \midrule
          &       &       &       &       &       &       &  \\
    \midrule
          &       & \multicolumn{3}{c}{Intel on 30\% incidents within} & \multicolumn{3}{c}{Intel on 50\% incidents within} \\
\cmidrule(l){3-5} \cmidrule(l){6-8}    Time  & No Intel & 100m  & 500m  & 1000m & 100m  & 500m  & 1000m \\
\cmidrule(l){1-2} \cmidrule(l){3-5} \cmidrule(l){6-8}
     12 am - 4 am  &                             0.924  &                             0.939  &                             0.931  &                             0.936  &                             0.958  &                             0.949  &                             0.932  \\
     4 am - 8 am  &                             0.912  &                             0.930  &                             0.921  &                             0.923  &                             0.953  &                             0.944  &                             0.942  \\
     8 am-12 pm  &                             0.900  &                             0.925  &                             0.922  &                             0.915  &                             0.952  &                             0.938  &                             0.913  \\
     12 pm -4 pm  &                             0.903  &                             0.936  &                             0.930  &                             0.914  &                             0.949  &                             0.939  &                             0.916  \\
     4 pm- 8 pm  &                             0.897  &                             0.914  &                             0.917  &                             0.906  &                             0.925  &                             0.930  &                             0.914  \\
     8 pm- 12 am  &                             0.917  &                             0.939  &                             0.938  &                             0.926  &                             0.950  &                             0.948  &                             0.933  \\
    \midrule
  Overall average  & 0.909 & 0.930 & 0.926 & 0.920 & 0.948 & 0.942 & 0.925 \\ 
  Improvement percent               &       & 2.1\% & 1.7\% & 1.1\% & 3.9\% & 3.3\% & 1.6\% \\
    \bottomrule
    \end{tabularx}%
  \label{tab:expert_input}%
\end{table}%

Table \ref{tab:expert_input} shows the change in AUC when the extent of information and accuracy raduis of the expert information made available to the model is varied. Clearly, with a smaller accuracy radius of 100 meter, we see a gain of approximately $1\%-4\%$ in AUC, and with an accuracy radius of 500 meter, we see a gain of approximately $1\%-3\%$ in AUC. A large accuracy radius of 1000 meter leads to only minor gains in AUC, especially for cases where the amount of intelligence is on the higher side. Information about some street crime expected to happen in a particular week within a radius of $1000$ meter is perceived by the Delhi police department as little information, and therefore our observation of a lower gain in AUC for an accuracy radius of $1000$ meter is in agreement with their perception. A 1\% increment in AUC approximately amounts to 10-12 incidents out of 500-600 incidents that happen in a week. Therefore, the expert inputs are helpful in identifying a good number of incidents that cannot be captured by historical data. Our results in this section also highlight the extent of useful information that we are able to extract from the historical data and the quality of prediction that we are able to perform without any expert inputs. For instance, if information on $50\%$ of the incidents to happen in a future week with an accuracy radius of $100$ meter leads to only $4\%$ increase in AUC, it suggests that our proposed model has already been able to capture a large amount of information from the historical data alone. Therefore, the expert information is valuable to the extent of only $4\%$ AUC gain. The proposed predictive hotspot mapping approach not only  reduces the human effort significantly with the help of a machine that learns from past data, but also provides a way for the machine to incorporate new insights that might be available in expert inputs. The system also gives confidence to the decision makers that both past information and new insights are incorporated in the predictive hotspots generated by the proposed approach. {\color{black}In Section~\ref{S:additional_results} we analyse an expert who suggests only key locations, like famous temples, flyovers, markets and metro stations, as possible locations and we observe that such an expert does not add much gain in information as this information is already contained in historical data.}

 \section{Implications of the proposed model for patrolling decisions }
 \label{S:discussion}
To our best knowledge, the hotspot mapping model developed in this paper is the first of its kind that seamlessly incorporates  multiple real-world complexities in a single model. It of course is the first hotspot model to be applied in the context of street crime in Delhi.  Here, we discuss the application of our proposed model (Model 5 in Section \ref{S:results}) and its implications for patrolling decisions. The main utility of the model is to provide a hotspot map for an upcoming week, for any given time interval of the day. As an illustration, Parts (a) and (b) of Figure \ref{F:illustration_549_5_6} show the predicted hotspot map for the week of 21 March 2021, for the time intervals `4 pm to 8 pm' and `8 pm to 12 am', respectively. The red part of the map marks the top 20\% likely spots for street crime and the yellow part marks the next 20\% likely spots, thus red and yellow cover the top 40\% areas to be monitored in Delhi, in the respective time intervals during the week starting 21 March 2021. 
The police department can use such a hotspot map to decide on where to deploy their resources to deter street crime at different times of the day for an upcoming week.

\subsection{Dynamic patrolling allocations depending on the predictions}
It is important to note that the hotspot maps change every week and also at different times during the day. It is beyond the cognitive ability of humans to quickly incorporate insights from past and new data and make modifications in patrolling decisions quickly. Our approach suggests that with significant temporal changes in the hotspot maps the patrolling resources should be deployed dynamically. To help visualize such changes, we provide two illustrations below.
\begin{enumerate}
\item First, we consider changes in hotspot maps during the same upcoming week but for successive time intervals in the day.   Part (a) of Figure \ref{F:illustration_549_5_6}  shows the predicted hotspot map for the week of 21 March 2021,  for the time interval, `4 pm to 8pm'. Part (b) of the figure  shows the predicted hotspot map for the same week but the next time interval, `8 pm to 12 am'. The hotspot maps for successive time intervals shown in Parts (a) and (b) of the figure appear deceivingly similar.  However, there are several changes in the map, which are highlighted in  Part (c) of the figure using additional color coding.  In Part (c), the blue spots indicate locations that were neither yellow nor red  for `4 pm to 8pm'  but are marked as red or yellow for  `8 pm to 12 am'. These would be new spots that require patrolling in the time interval `8 pm to 12 am' in comparison to `4 pm to 8 pm'.  Similarly, the green spots in the map denote those points that were either red or yellow for `4 pm to 8 pm' but are neither red nor yellow for `8 pm to 12 am'. So, patrolling resources from these spots could be withdrawn and redeployed. {\color{black}There are 1.4\% green
and 1.4\% blue dots in Part (c). Since the number of highlighted locations is fixed, the number of new locations added (green)
will equal the number of old locations removed (blue).}
We observe that on an average about 650 new locations per week get flagged as a red or yellow spot requiring patrolling.

\item Second, we consider changes in hotspot maps for two successive weeks but for the same time interval in the day. Part (a) of Figure \ref{F:illustration_542_549_6}  shows the predicted hotspot map for the week of 14 March 2021,  for the time interval `8 pm to 12 am'. Part (b) of the figure  shows the predicted hotspot map for its next week (i.e. week of 21 March 2021) but the same time interval `8 pm to 12 am'. In Part (b) of the figure, we use additional color coding to highlight the changes, where  blue spots are those that were neither yellow nor red  for the  week of 14 March 2021,  but are marked as red or yellow for  the week of 21 March 2021. These would be the new spots that require patrolling for the week of 21 March 2021 in comparison to the week of 14 March 2021.  Similarly, the green spots in the map denote those points that used to be either red or yellow for the week of 14 March 2021 but are neither red nor yellow for the week of 21 March 2021. So, patrolling resources from these spots could be withdrawn and redeployed. {\color{black}There are about 3.1\% green and similar number of blue dots.}
\end{enumerate}

\begin{figure}[ht]
\begin{subfigure}{.48\textwidth}
\center
\includegraphics[width=8cm]{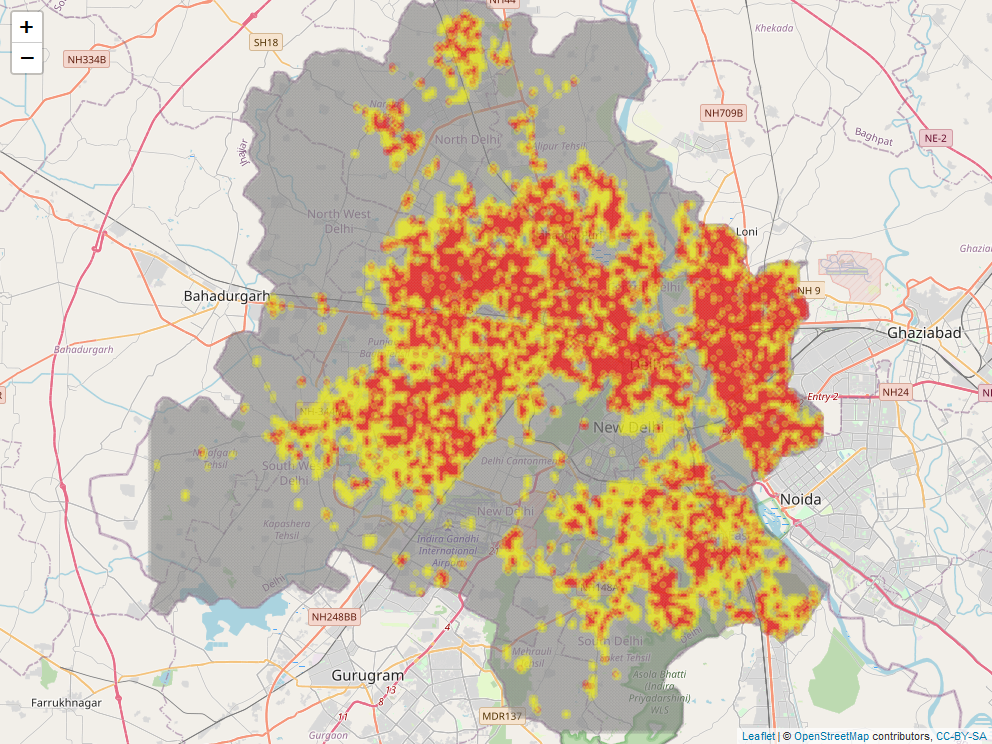}
\caption{4 pm to  8 pm}
\end{subfigure}\hfill
\begin{subfigure}{.48\textwidth}
\center
\includegraphics[width=8cm]{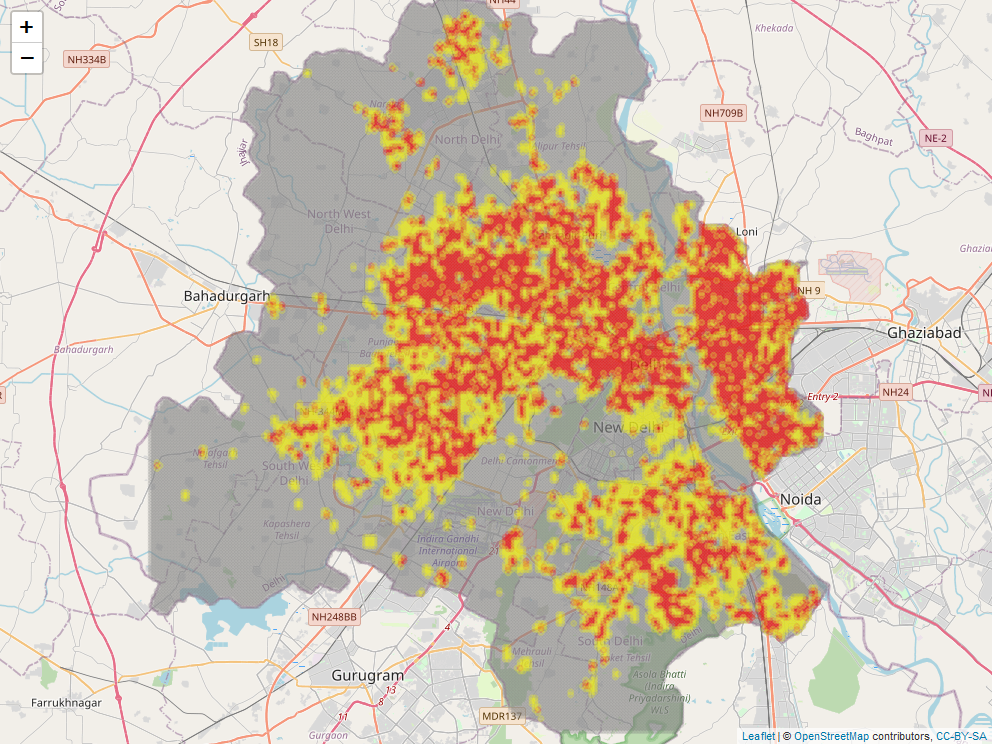}
\caption{8 pm to  12 am}
\end{subfigure}

\begin{subfigure}{\textwidth}
\center
\includegraphics[width=10cm]{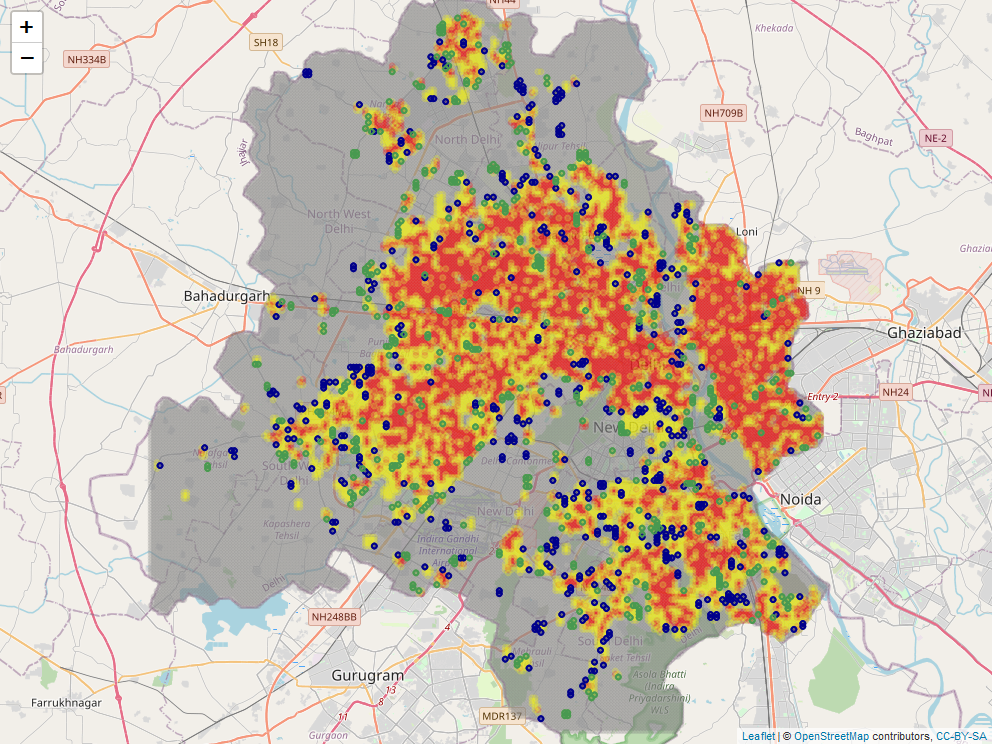}
\caption{8 pm to 12 am (changes from `4pm to 8 pm'  highlighted)}
\end{subfigure}
\caption{Predicted hotspot maps for two successive time intervals for the week of 21 March  2021, based on Model 5. Red marks the top 20\% likely spots and and yellow the  next 20\% likely spots. Part (c) highlights differences between Parts (a) and (b): blue indicates spots that were neither red nor yellow in Part (a) but are marked as red or yellow in Part (b), green indicates spots that were red or yellow in Part (a) but are neither red nor yellow in Part (b). \textcolor{black}{There are 1.4\% green
and 1.4\% blue dots in Part (c).}}
\label{F:illustration_549_5_6}
\end{figure}

The dynamic nature of the problem emphasizes regular updates in the hotspot maps as new data arrive every week. As observed in Figures \ref{F:illustration_549_5_6} and \ref{F:illustration_542_549_6}, the overall locations of incidents might seem deceivingly stable, which might suggest constancy in allocation of resources; however, not updating the posterior may lead to new locations not getting flagged. We provide an elaborate discussion on this in Section~\ref{S:additional_results}. Not using an updated model may result in a loss of AUC of about 0.5\%  which could amount to not predicting the correct locations for around 5 events in a week or 20 events in a month, which is important from a policing point of view.

\begin{figure}[ht]
\begin{subfigure}{.48\textwidth}
\center
\includegraphics[width=8cm]{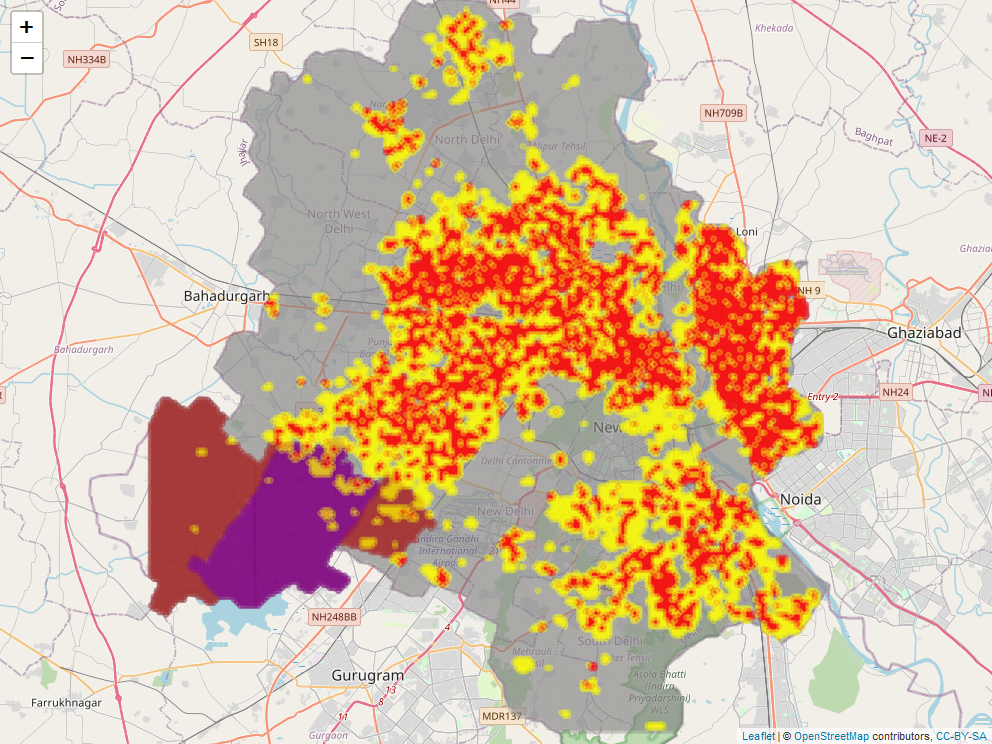}
\caption{week of 14 March  2021 (8 pm to  12 am)}
\end{subfigure}\hfill
\begin{subfigure}{.48\textwidth}
\center
\includegraphics[width=8cm]{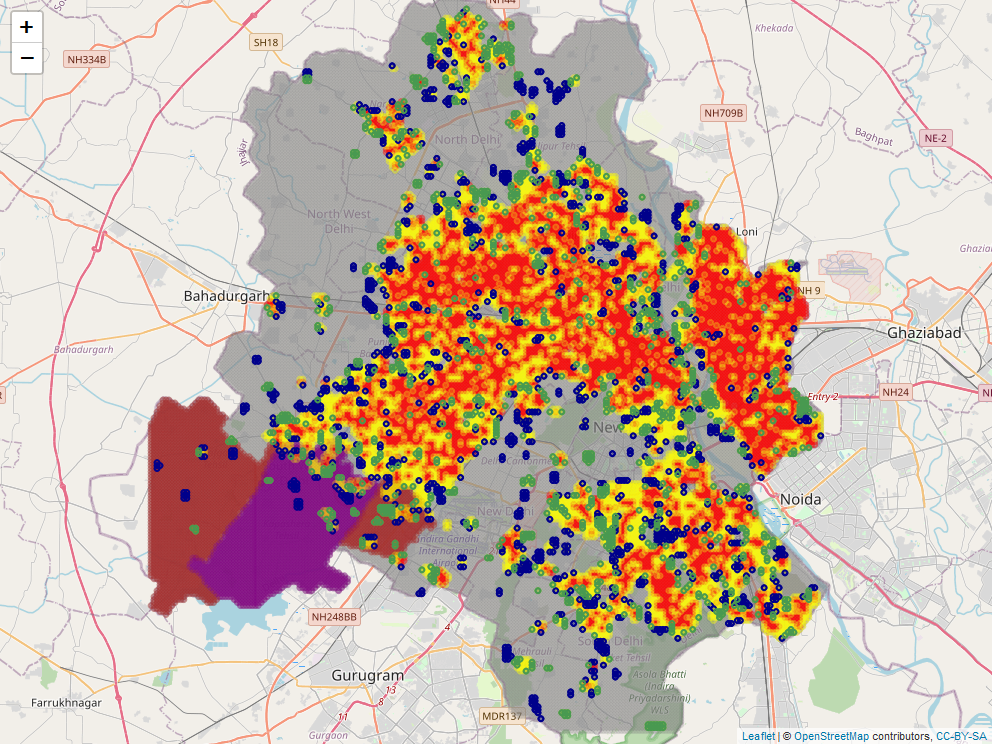}
\caption{week of 21 March  2021 (8 pm to  12 am)}
\end{subfigure}
\caption{\label{F:illustration_542_549_6} Predicted hotspot maps for two successive weeks, for the week of 14 March 2021 and the week of 21 March 2021, based on the proposed Model 5. Red marks the top 20\% likely spots and yellow the  next 20\% likely spots. Part (b) highlights some spots in relation to Part (a): blue indicates spots that were neither red nor yellow in Part (a) but should be marked as red or yellow in Part (b), green indicates spots that were red or yellow in Part (a) but should neither be red nor yellow in Part (b). {\color{black}There are about 3.1\% green and 3.1\% blue dots.}}
\end{figure}

\begin{figure}[ht]
\begin{subfigure}{.49\textwidth}
\center
\includegraphics[width=8cm]{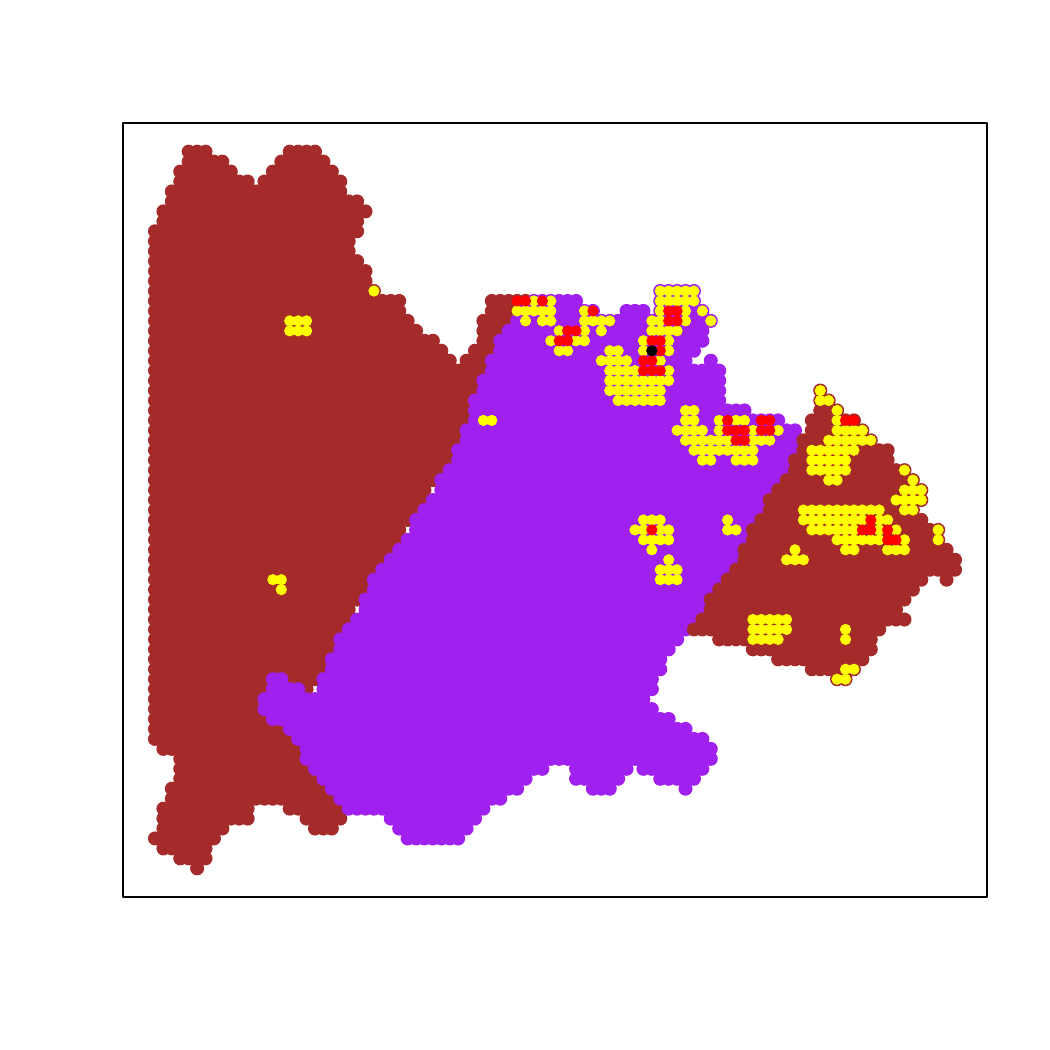}
\vspace{-12mm}
\caption{week of 14 March  2021 (8 pm to  12 am)  }
\end{subfigure}
\begin{subfigure}{.49\textwidth}
\center
\includegraphics[width=8cm]{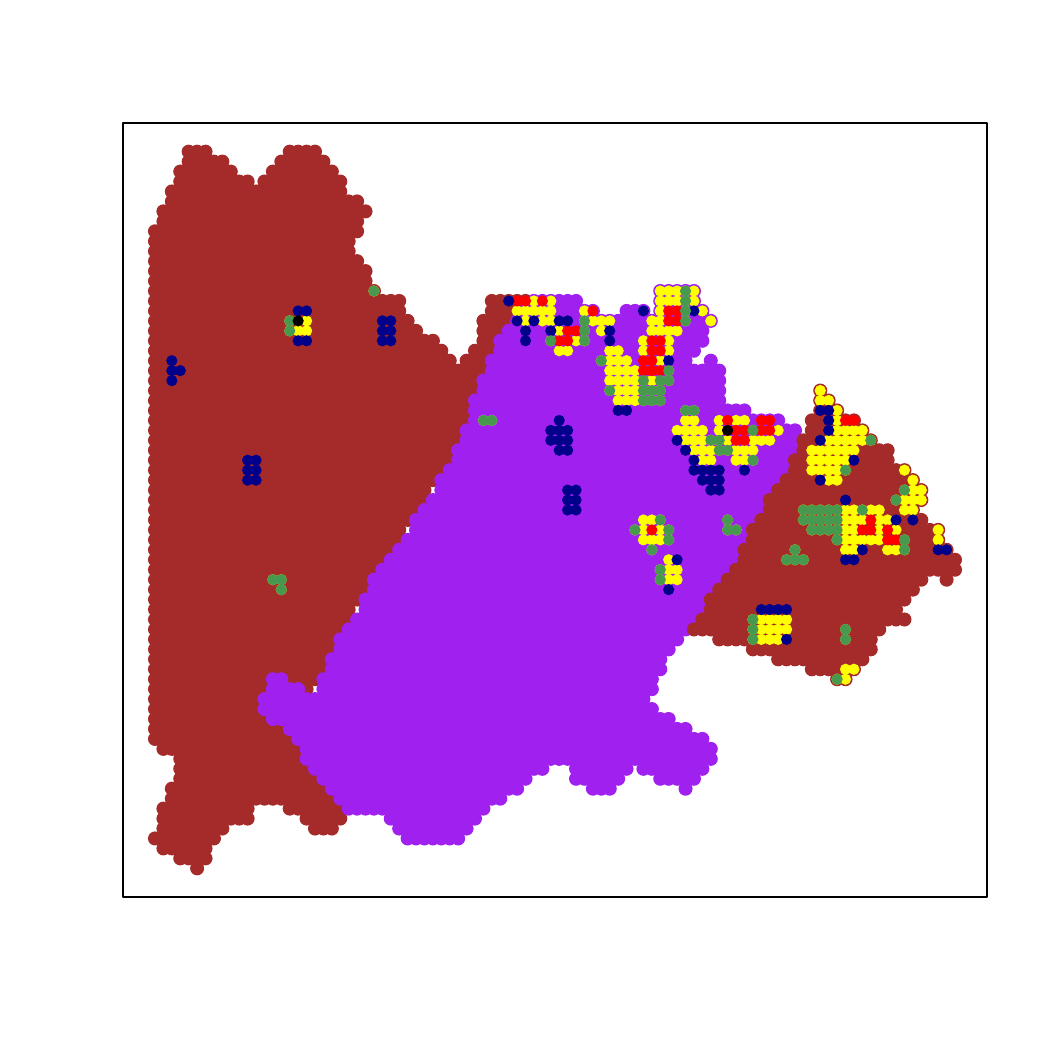}
\vspace{-12mm}
\caption{week of 21 March  2021 (8 pm to  12 am) }
\end{subfigure}
\caption{\label{F:illustration_542_549_6zoom} Map focused on three police jurisdictions within Delhi, viz. `Jaffarpur Kalan', `Chhawla' and `Dwarka Sector 3'. Predicted hotspot maps are shown  for two successive weeks, for the week of 14 March  2021 and the week of March 21 2021, based on the proposed Model 5. In the afore mentioned order, the three police station jurisdictions are marked in `brown', `purple' and again `brown'.  Red marks the top 20\% likely spots and yellow the next 20\% likely spots. Part (b) highlights some spots in relation to Part (a): blue indicates spots that were neither red nor yellow in Part (a) but should be marked as red or yellow in Part (b), green indicates spots that were red or yellow in Part (a) but should neither be red nor yellow in Part (b).}
\end{figure}

\subsection{Variability within police station jurisdictions and insights on key locations}
Patrolling allocation decisions in Delhi are usually taken at a police jurisdiction level. In the current process, we have learned from the police department officials that such allocations are relatively stable for several weeks. This means that changes in allocation of vehicles are made only when a large number of observed events suggest a shift in the pattern. However, we emphasize based on the findings from our study that the allocations must be changed continuously and dynamically with emerging data for an effective patrolling. Our proposed model helps make such decisions at a police station level and ensures an early capture of the pattern shifts. As an illustration,   Figure \ref{F:illustration_542_549_6zoom} shows the map focused on three police jurisdictions within Delhi, viz. `Jaffarpur Kalan', `Chhawla' and `Dwarka sector 3'. Predicted hotspot maps (for the same time interval in the day) are shown  for two successive weeks, for the week of 14 March  2021 and week of March 21 2021, based on our proposed model. The blue spots in Part (b) are those that were neither yellow nor red  for the  week of 14 March 2021, but are marked as red or yellow for  the week of 21 March 2021. These are new spots requiring  patrolling in these jurisdictions. Similarly, the green spots in the map denote those points that used to be either red or yellow for the week of 14 March 2021 but are neither red nor yellow for the week of 21 March 2021. These are spots from where patrolling could be withdrawn and deployed elsewhere. The number of events by police station jurisdiction can vary widely, as can be seen in Table \ref{T:eventbyPS}, which gets factored into the dynamically estimated predictive hotspot maps.

\begin{table}[ht]
\centering
\caption{\label{T:eventbyPS} Number of actual events by three police station jurisdictions, over 10 weeks}
\begin{tabular}{lccc}
  \toprule
 week start date & Jaffarpur Kalan & Chhawla& Dwarka sector 3 \\ 
  \midrule
 17/1/21 & 1 & 1 & 1 \\ 
   24/1/21 & 0 & 3 & 3 \\ 
  31/1/21 & 0 & 4 & 1 \\ 
   7/2/21 & 0 & 1 & 1 \\ 
  14/2/21 & 0 & 7 & 2 \\ 
  21/2/21 & 0 & 3 & 2 \\ 
 28/2/21 & 1 & 6 & 0 \\ 
   7/3/21 & 1 & 1 & 2 \\ 
   14/3/21 & 0 & 5 & 4 \\ 
   21/3/21 & 0 & 1 & 2 \\  
   \bottomrule
\end{tabular}
\end{table}

The hotspot maps  provide useful insights on key locations that are usually expected to be prone to street crime incidents. Here, we consider four types of such key locations, viz. metro stations,  famous temples, markets and flyovers (i.e. bridges built across busy roads to ease traffic congestion). Metro stations, temples and markets tend to attract crowds and so can be prone to street crime events. On the other hand, some areas beneath the flyovers tend to be isolated spots, which can be conducive for street crime activity.  Figure \ref{F:pred_map_549_6prone} shows the predictive hotspot map for the week starting 21 March 2021 for the time interval 8 pm - 12 am, along with the key locations. The spots marked with triangles indicate metro stations and circles indicate either famous temples, markets or flyovers. We can see that most of the key locations are part of the predicted red and yellow hotspots, so they are predicted as prone to street crime by our  model. However, not all key locations are marked as hotspots. For example, 20 of the 191 metro stations are not marked as hotspots and  the method helps identify the specific  key metro stations requiring patrolling.  Table \ref{T:keylocations_549_6} shows the  number of key locations predicted as hotspots (i.e. red and yellow) for the week starting 21 March 2021 for the time interval 8 pm to 12 am. Another important aspect to note here is that key locations will not account for all the main hotspots, as there are several others. So, it would not suffice to focus attention on mainly the key locations and model helps in identifying a more complete set of locations for patrolling.

\begin{table}[ht]
\centering
\caption{\label{T:keylocations_549_6} Number of key locations predicted as hotspots (i.e. red and yellow) for the week starting 21 March 2021 for the time interval 8 pm to 12 am }
\resizebox{.5\textwidth}{!}{\begin{tabular}{lcccc}
\toprule
& \multicolumn{2}{p{2cm}}{\hspace{3mm}Hotspots}&\multicolumn{2}{p{2cm}}{~}\\
  \midrule
Key location type & Red & Yellow & Other & Total \\ 
  \midrule
famous temples &  13 &   4 &   3 &  20 \\ 
flyovers &  50 &   5 &   1 &  56 \\ 
  markets &  22 &   6 &   2 &  30 \\ 
  metro stations$^{\star}$ & 146 &  25 &  20 & 191 \\  
   \bottomrule
\end{tabular}}
~\\
\small{$^{\star}$only metro stations within Delhi are counted}
\end{table}

\subsection{Prescriptive analytics}
{\color{black}
Given a predictive density estimate of crime, determining optimal deployment strategies is still a complex task due to the dynamic, spatio-temporal nature of urban crime. 
The kernel density estimate obtained from our proposed approach can be used to systematically determine patrolling positions given the availability of resources.
Two complementary approaches can be employed to address this challenge. The first leverages the Markov Decision Process (MDP) framework, allowing for sequential decision-making under uncertainty and enabling adaptive patrol strategies that respond to evolving risk patterns. We discuss this approach briefly in Section~\ref{sec:mdp}. The second approach utilizes network allocation techniques to optimize resource distribution across a city's spatial grid, ensuring maximum coverage and efficient utilization of limited patrol units. For instance, \citet{brandtetal2022} use an integer programming formulation for such allocations. However, both methodologies offer robust tools for designing data-driven, responsive policing strategies.
}

Without delving into the advanced approaches for prescriptive analytics, which we keep for a future work, we comment on the number of resources that may be required in our context and how it compares to the existing numbers deployed. As seen in Table \ref{T:top2040pct} top 20\% areas that are marked for monitoring by our model can be expected to capture 75-80\% of actual street crime events. Patrolling resources to tackle street crime can include police on vans or bikes or foot. If the police wants to ensure the presence of 1 resource in a 2 sq km area to deter street crime then to cover 20\% of the area of Delhi ($\approx$ 1480 sq km), we would need about 145 resources. The number of Prakhar vans which help in anti-street-crime has increased to about 88 vans \citep{nayudu2021delhi}, and to 118 more recently (see \ref{sec:rev_interview}), thus indicating that additional resources including patrol bikes and on-foot personnel should complement the vans. An emerging technology that can considerably improve the effectiveness and the area coverage of the existing patrol resources is the use of drones. Equipping patrol vehicles with drone monitoring ability in hotspot areas can potentially be effective while covering a larger area than what a patrol vehicle could otherwise achieve. The use of drones for crime monitoring is in active consideration by the Delhi police \citep{pti2022delhi}. 

\begin{figure}[t]
\center
\includegraphics[height=7cm]{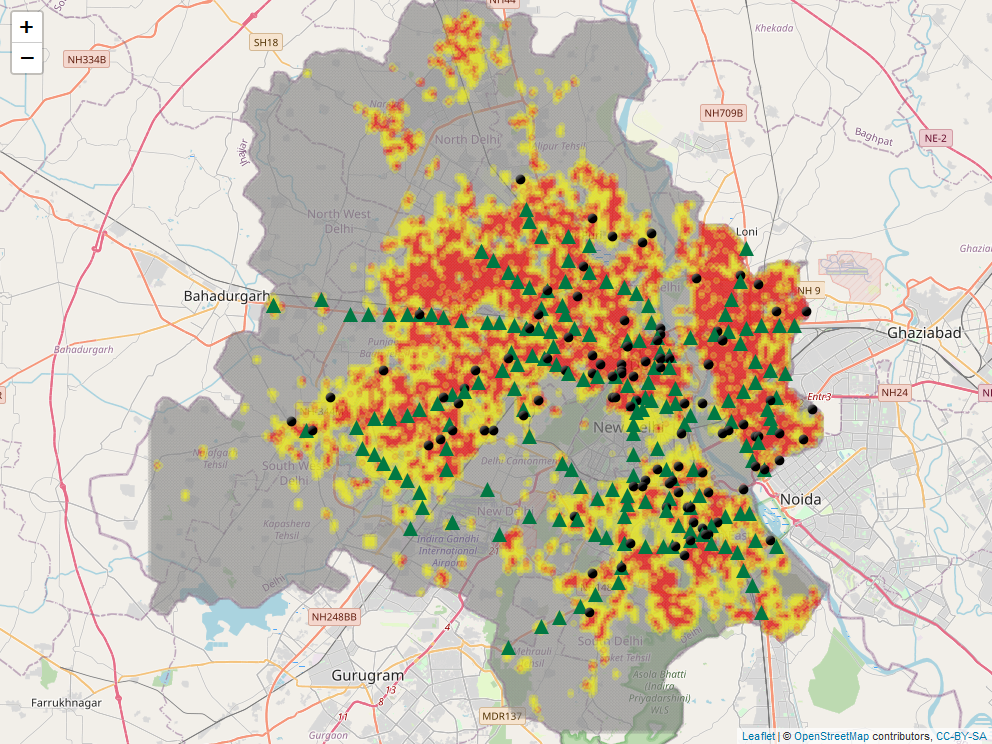}
\caption{\label{F:pred_map_549_6prone} Predictive hotspot map for the week starting 21 March 2021 for the time interval 8 pm - 12 am, along with the key locations. As before, red spots mark the top 20\% likely spots and yellow spots mark the next 20\% likely spots for street crime as predicted by Model 5. In addition, the spots marked with triangles indicate metro stations and circles indicate either famous temples, markets or flyovers.}
\end{figure}

\section{Conclusions}
\label{S:conclusion}
Street crime prediction and prevention is an important problem that is faced by law enforcement agencies worldwide. With emphasis on automating operations and relying on data-driven decisions, agencies across the world are in need of efficient hotspot maps that can help them make effective patrolling decisions. Many of the existing approaches, rely on the creation of hotspot maps based on historical data alone, that makes the decision maker not involved in the process. This also leads to decision makers having lack of confidence in the model as they feel that there is no way they can incorporate crucial inputs in the model that they know cannot be learned from past data. The paper addresses this gap by suggesting an approach that is able to incorporate both historical and expert data to create hotspot maps. To our best knowledge, such an approach has not been studied in the context of street crime prediction. Additionally, we make a number of other methodological contributions to the area of hotspot map creation by coming up with a spatio-temporal block-weighted adaptive kernel density estimation approach that handles several estimation challenges effectively. Our proposed approach has been applied to the city of Delhi that has the highest crime rate among 19 Indian metropolitan cities with population larger than 2 million. The proposed approach has been developed in collaboration with the Delhi police department that aims to exploit data and technology to manage crime in the city. In this context, our proposed spatio-temporal block-weighted adaptive kernel density model can be easily implemented to generate hotspot maps on a regular basis for predicting locations prone to street crime in Delhi at different times of the day based on historical data and expert inputs.

The results of our proposed approach have been very promising as it achieves excellent accuracy consistently across many weeks of prediction. By monitoring $20\%$ of the locations proposed by our model one can capture about $80\%$ of the street crimes and by monitoring $40\%$ of the locations one can capture about $95\%$ of the street crimes. In the absence of such a model, many of the important decisions are taken in an ad hoc manner based on certain rules of thumb. Our work generated a number of key insights that have important implications for allocating the usually limited resources available with the law enforcement agencies. We highlighted the importance of dynamically modifying patrolling decisions over time with new data emerging by showing that hotspot maps on the same day can be significantly different at different times of the day, and also there is variability withing the same time interval across weeks. We also showed that patterns that are perceived by the law enforcement agency to be stable or slowly changing are actually deceiving in nature. Additionally, rules like patrolling key location types, such as, famous temples, flyovers, markets and metro stations often get over generalized. Our model is able to differentiate among the locations within these location types by identifying spots that are more/less prone to crimes. Therefore, our model has also been able to nullify a number of assumptions that the law enforcement agencies often make in patrolling allocations. The utility of such models is likely to go up in the future as many of the major cities across the world plan to utilize drone-based surveillance systems. 

{\color{black}
Our work in this paper can be extended in a variety of ways. The most straightforward extension would be to utilize the predicted hotspot maps for more efficient resource allocation. Currently, while our model identifies high-risk areas, the next logical step would involve developing strategies for optimal deployment of law enforcement resources, such as patrol units, based on these hotspots. This could include dynamic resource allocation where patrol units are re-deployed in real-time as new data on crime patterns emerges throughout the day. Because of unavailability of data the past patrol vehicle locations are endogenous in our implementation and not incorporated separately in the model. Since the location of patrol vehicles are likely to change the crime dynamics, it would be important auxiliary information that should be considered at the modeling stage followed by a dynamic resource allocation approach. Exploring the impact of other information, such as, population density, socio-economic indicators, weather conditions, holidays, and special events would be an interesting direction to pursue.}
{\color{black}
Other extensions can focus on further enhancing the model by taking the kind of crime into consideration and then making allocation decisions. Additionally, it will be interesting to also understand the public behavior on what kinds of crime are reported or not reported; in other situations, formal complaints are not filed after reporting of the crime over a call. Such issues of underreporting or not following up with a formal complaint may stem from a lack of trust or perceived inconvenience that requires attention of the law enforcement agencies. While our current study excludes prank calls and focuses on verified incidents, understanding the factors that influence the decision to report a crime over a call (or formally) could lead to more comprehensive and nuanced crime prevention strategies. Methodologies such as Poisson regression with spatial effects \cite{liu2024quantifying}, adapted from related fields, could be employed to model reporting delays, or multiple calls on the same incident may be modeled to indicate the urgency and genuineness of the call. These issues represent a significant opportunity for future research in the area, which would enhance the accuracy and reliability of crime prediction models by integrating more nuanced behavioral insights.
}

\section{Acknowledgements}
The authors of the study would like to acknowledge the support provided by Brij Disa Centre for Data Science and Artificial Intelligence and Krishnamurthy Tandon School of Artificial Intelligence at Indian Institute of Management Ahmedabad.







\bibliographystyle{pomsref}
\bibliography{Ref_PCR}

\ECSwitch 

\ECHead{E-Companion: Data preparation, Gibbs Sampling Steps, Additional Results, Interview Findings and Extensions}

 
 \section{Data preparation and description}\label{S:data_prep}
 
 Before using the data from DPD, we performed a number of steps to prepare the data and bring it in a form conducive for our analysis. Some of the data preparation steps are as follows:
 
 {\bf Removing duplicates.}  We observed that sometimes, there were multiple calls related to the same event, which could be identified by comparing the text description given by the caller. If the descriptions had a large verbatim overlap of text then they were clearly duplicate calls and only one of them was retained. We accomplished this using standard text analysis tools in the  R package (\citealt{cran}). After removing duplicate records, the data contained {\color{black}51230 calls} related to street crime for the time period under consideration.

{\bf Geocoding event locations.}  More recently, the Delhi police has started recording the longitude and latitude coordinates of the event locations in their system for most events when the patrol vehicle reaches the location. However, it is still not 100\% populated and historically this information is even less readily available. We note that the latitude and longitude information were recorded for about 60\% of the events in our data.  For the remaining records, we used geocoding based on Google API to populate the location coordinates based on the text description of the locations recorded during the PCR call, which is typically a text written using  a combination of English and Hindi words.  

{\bf Geocoding police station locations.} We manually looked up the geographical coordinates of the police stations, in whose jurisdiction the events were recorded.  There are over 200 different police stations in the city.  This information turns out to be useful for some sanity checks on the data for analysis of the results from the model.

{\bf Data cleaning.}  As a sanity check on the street crime event location coordinates, we flagged as unusable, any records where the coordinates of the event were either not found due to lack of sufficient information in the text description, or turned out to be more than 5 kilometers away (radially) from the police station to whose jurisdiction the event had been assigned in DPD. {\color{black}Delhi police records latitude and longitude of crimes, and assigns the nearest police station in whose jurisdiction the crime happens. Since, location recording and police station assignment tasks happen manually, it provides a two-step check on the accuracy of the data. If the crime coordinates and police station are very far away from each other, the data point entered manually is likely to be erroneous. Therefore we removed such records from the database choosing a radial distance of 5 kilometers as a cut-off, as the jurisdiction regions for the police stations in Delhi are much smaller. After this clearning process, we were able to finally obtain usable geographical coordinates for 90\% of the events.}

{\color{black}
Further details on dataset preparation are provided in Table~\ref{tab:data_preprocessing}. Our proposed model requires 1 year of historical data for prediction of an upcoming week, therefore our predictions are generated from the week of 4 October 2020. Interestingly, during the model learning phase there are structural shifts in the number of calls and crimes happening in the region. However, the model performance (as per the results in Section~4.1 in the paper) is robust to such structural shifts in the past. Figure~\ref{fig:rev_no_of_calls} shows the period for which the data was available along with the number of calls received on a daily basis. It also shows the major COVID-19 lockdown that happened in the country and time from which the predictive hotspots are generated.
\begin{figure}[htp]
\centering
\includegraphics[width=10cm]{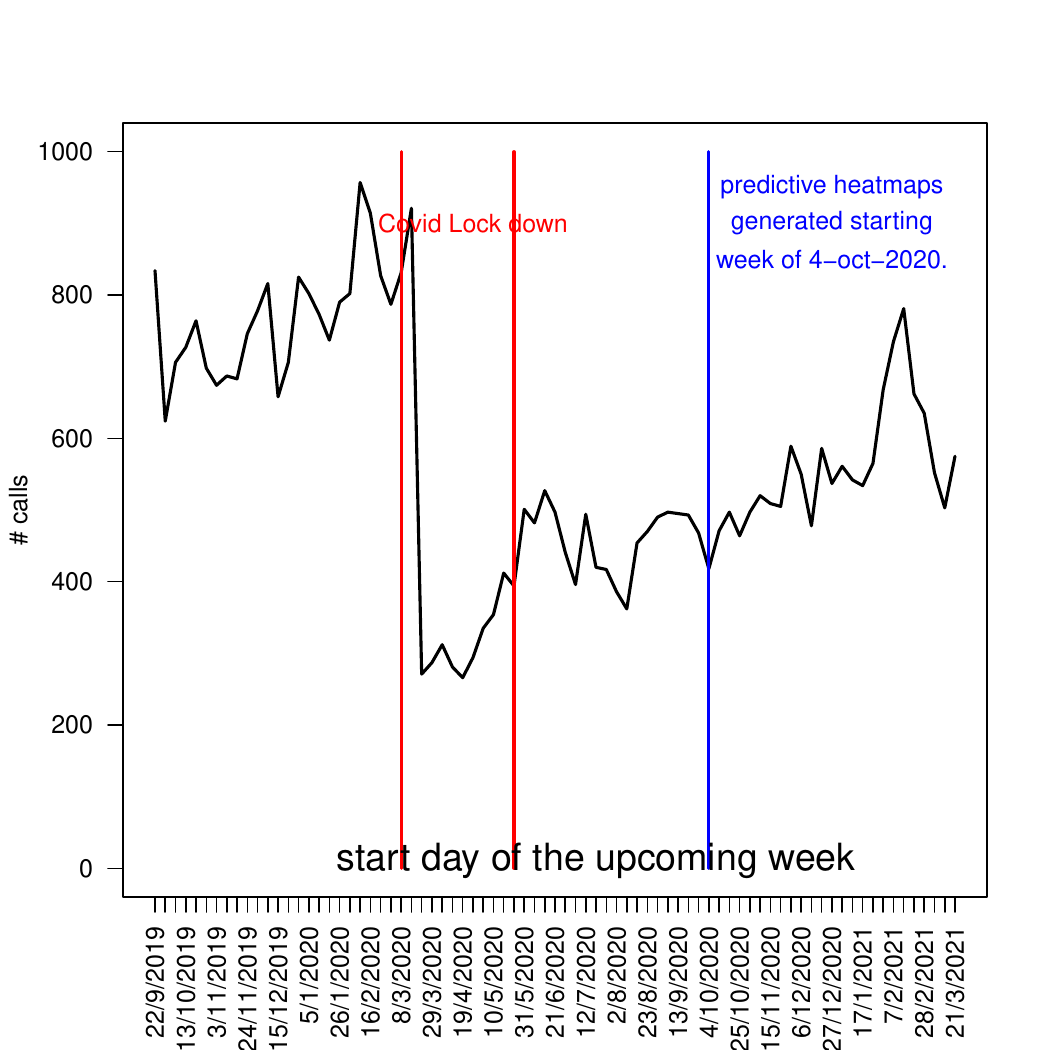}
\caption{\textcolor{black}{Number of calls received per day.}}
\label{fig:rev_no_of_calls}
\end{figure}
\begin{figure}[htp]
\center
\includegraphics[width=10cm]{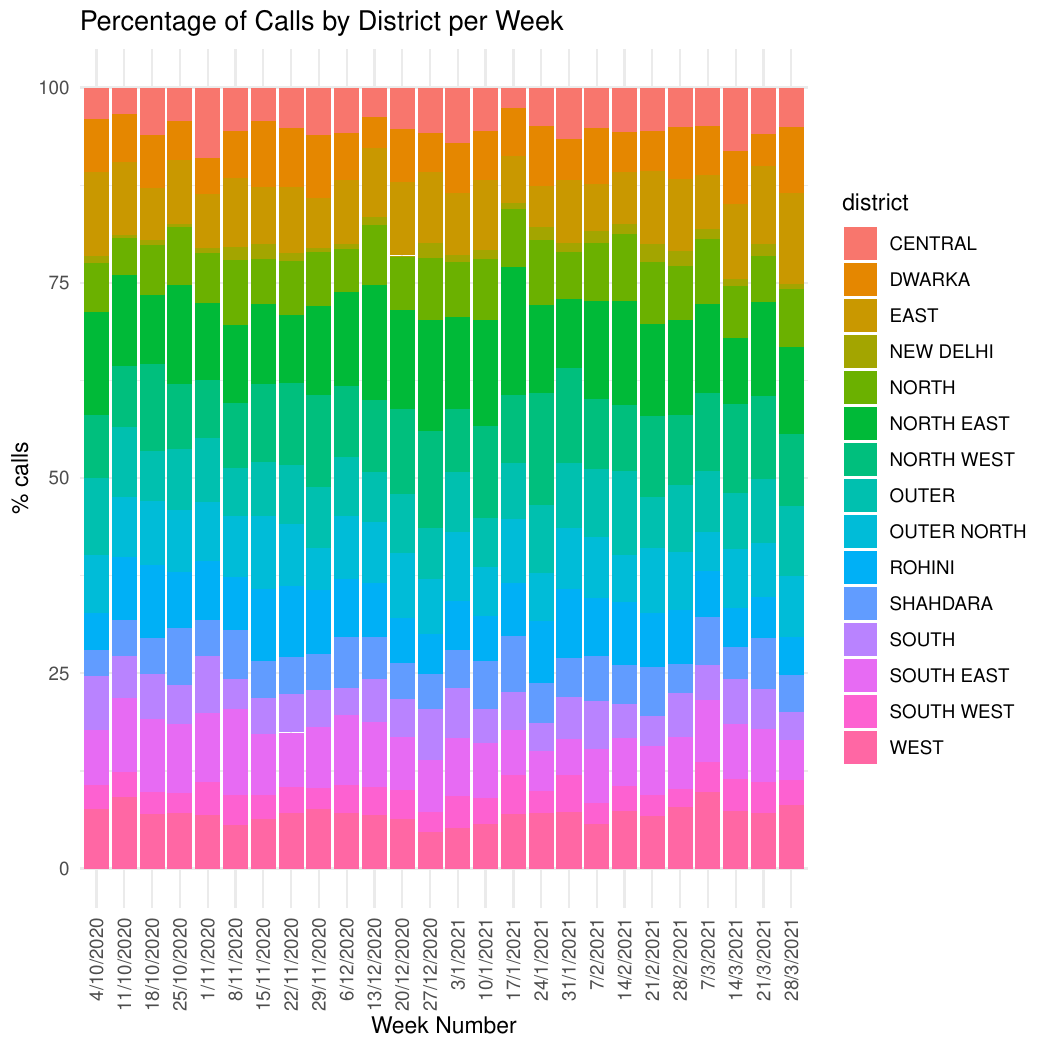}
\caption{\textcolor{black}{Percentage calls received by various districts of Delhi per week.}}
\label{fig:rev_district_calls}
\end{figure}
\begin{figure}[htp]
\center
\includegraphics[width=10cm]{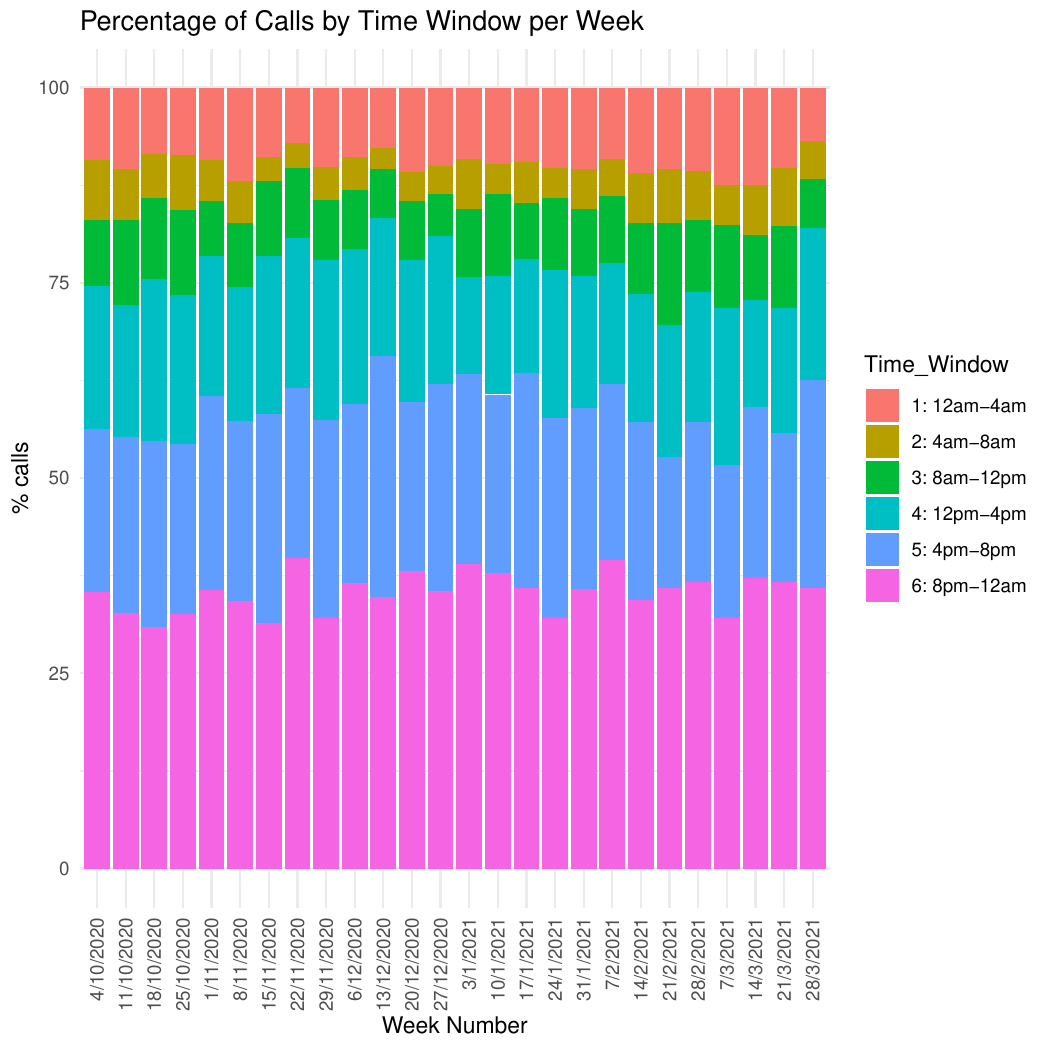}
\caption{\textcolor{black}{Percentage calls received by 4-hour time window per week.}}
\label{fig:rev_timewindow_calls}
\end{figure}
We provide further insights on the data through Figures~\ref{fig:rev_district_calls} and~\ref{fig:rev_timewindow_calls}, which indicate the percentage of calls received from various districts of Delhi per week and percentage of calls received by time window per week, respectively.
}

\begin{table}[h!]
\centering
\begin{tabular}{p{0.7\textwidth}r}
\toprule
\textbf{Description} & \textbf{Number of Records} \\ 
\midrule
Total records received from police database (street crime) & 65,057 \\ 
Total after removing duplicate records (i.e. multiple calls on same event) & 51,230 \\ 
Number of records with invalid latitude/longitude coordinates & 5,295 \\
Final usable records & 45,935 (~90\%) \\ 
\bottomrule
\end{tabular}
\caption{\textcolor{black}{Summary of records after data preprocessing}}
\label{tab:data_preprocessing}
\end{table}

 
  \section{Gibbs sampling steps}
  \label{S:gibbs}
  
 In order to generate simulations from the joint posterior distribution given in Equation~(16) in the paper, we use Gibbs sampling. Here, we describe the steps of Gibbs sampling.
 ~\\
 {\bf Step 0.} Assign starting values for the parameters in the set $$\boldsymbol{\Theta}=\{\alpha_1,\beta_1, \alpha_2, \beta_2, \alpha_3, \beta_3, (w_1, \ldots, w_B), (I_1, J_1), (I_2, J_2), \ldots, (I_n, J_n)\}.$$
 ~\\
 {\bf Step 1.} Gibbs step for $\alpha_1$ involves simulating from the pdf with
 \begin{eqnarray}
 \Pi\left( \alpha_1 \big\vert (\boldsymbol{s}_x, \boldsymbol{s}_y, \boldsymbol{t}), \boldsymbol{\Theta}\backslash \{\alpha_1\} \right)\propto \alpha_1^n e^{ -\frac{\sum_{l=1}^n\left(s_{xl}-x_{I_lJ_l} \right)^2\alpha^2_1 A^{2\beta_1}_{I_lJ_l}}{2}}, ~\alpha_1\in(0,\infty),
 \end{eqnarray}
 which is same as simulating from 
  $$\alpha_1\sim Gamma\left(shape=n/2+1, rate= \frac{\sum_{l=1}^n\left(s_{xl}-x_{I_lJ_l} \right)^2 A^{2\beta_1}_{I_lJ_l}}{2} \right).$$
   ~\\
 {\bf Step 2.} Gibbs step for $\beta_1$ involves simulating from the pdf with
 \begin{eqnarray}
 \Pi\left( \beta_1 \big\vert (\boldsymbol{s}_x, \boldsymbol{s}_y, \boldsymbol{t}), \boldsymbol{\Theta}\backslash \{\beta_1\} \right)\propto e^{\beta_1\sum_{l=1}^n\log(A_{I_lJ_l}) -\frac{\sum_{l=1}^n\left(s_{xl}-x_{I_lJ_l} \right)^2\alpha^2_1 A^{2\beta_1}_{I_lJ_l}}{2}}, ~\beta_1\in(0,1).
 \end{eqnarray}
We simulate $\beta_1$ from an approximate discrete distribution by restricting the support to a suitable grid, viz $\beta_1\in \{0,0.01, 0.02,\ldots, 0.99\}$.
~\\
{\bf Step 3.} Gibbs step for $\alpha_2$ involves simulating from the pdf with
 \begin{eqnarray}
 \Pi\left( \alpha_2 \big\vert (\boldsymbol{s}_x, \boldsymbol{s}_y, \boldsymbol{t}), \boldsymbol{\Theta}\backslash \{\alpha_2\} \right)\propto \alpha_2^n e^{ -\frac{\sum_{l=1}^n\left(s_{yl}-y_{I_lJ_l} \right)^2\alpha^2_2 A^{2\beta_2}_{I_lJ_l}}{2}}, ~\alpha_2\in(0,\infty),
 \end{eqnarray}
 which is same as simulating from 
  $$\alpha_2\sim Gamma\left(shape=n/2+1, rate= \frac{\sum_{l=1}^n\left(s_{yl}-y_{I_lJ_l} \right)^2 A^{2\beta_2}_{I_lJ_l}}{2} \right).$$
   ~\\
{\bf Step 4.} Gibbs step for $\beta_2$ involves simulating from the pdf with
 \begin{eqnarray}
 \Pi\left( \beta_2 \big\vert (\boldsymbol{s}_x, \boldsymbol{s}_y, \boldsymbol{t}), \boldsymbol{\Theta}\backslash \{\beta_2\} \right)\propto e^{\beta_2\sum_{l=1}^n\log(A_{I_lJ_l}) -\frac{\sum_{l=1}^n\left(s_{yl}-y_{I_lJ_l} \right)^2\alpha^2_2 A^{2\beta_2}_{I_lJ_l}}{2}}, ~\beta_2\in(0,1).
 \end{eqnarray}
We simulate $\beta_2$ from an approximate discrete distribution by restricting the support to a suitable grid, viz $\beta_2\in \{0,0.01, 0.02,\ldots, 0.99\}$.
~\\
{\bf Step 5.} Gibbs step for $\alpha_3$ involves simulating from the pdf with
 \begin{eqnarray}
  \Pi\left( \alpha_3\big\vert (\boldsymbol{s}_x, \boldsymbol{s}_y, \boldsymbol{t}), \boldsymbol{\Theta}\backslash \{\alpha_3\} \right)\propto
  e^{\alpha^2_3\sum_{l=1}^n A^{2\beta_3}_{I_lJ_l}\cos\left( (t_l-t_{I_lJ_l})\pi/12\right) - \log(I_{0}(\alpha^2_3 A^{2\beta_3}_{I_lJ_l}))}, \alpha
  _3\in (0, \infty).
 \end{eqnarray}
We simulate $\alpha_3$ from an approximate discrete distribution by restricting the support to a suitable grid, viz $\alpha_3\in \{0,0.01, 0.02,\ldots, 10\}$.
~\\
{\bf Step 6.} Gibbs step for $\beta_3$ involves simulating from the pdf with
 \begin{eqnarray}
  \Pi\left( \beta_3\big\vert (\boldsymbol{s}_x, \boldsymbol{s}_y, \boldsymbol{t}), \boldsymbol{\Theta}\backslash \{\beta_3\} \right)\propto
  e^{\alpha^2_3\sum_{l=1}^n A^{2\beta_3}_{I_lJ_l}\cos\left( (t_l-t_{I_lJ_l})\pi/12\right) - \log(I_{0}(\alpha^2_3 A^{2\beta_3}_{I_lJ_l}))}, \beta
  _3\in (0, 1).
 \end{eqnarray}
We simulate $\beta_3$ from an approximate discrete distribution by restricting the support to a suitable grid, viz $\beta_3\in \{0,0.01, 0.02,\ldots, 0.99\}$.
~\\
{\bf Step 7.} Gibbs step for $(w_1,w_2,\ldots, w_{B}$ involves simulating from the pdf with
\begin{eqnarray}
  \Pi\left( (w_1,w_2,\ldots,w_B)\big\vert (\boldsymbol{s}_x, \boldsymbol{s}_y, \boldsymbol{t}), \boldsymbol{\Theta}\backslash \{w_1,w_2,\ldots,w_B\} \right) \propto \prod_{l=1}^n w_{I_l},
 \end{eqnarray}
which is same as simulating from
$$ (w_1,w_2,\ldots,w_B) \sim \mbox{Dirichlet}( 1+f_1, 1+f_2,\ldots, 1+f_B), $$
where $f_j$=number of $l\in\{1,2,\ldots,n\}$ with  $I_l=j$.
~\\
{\bf Step 8.} Gibbs step for $(I_l, J_l)$ for each $l\in \{1,2,\ldots, n\}$ involves simulating from the pdf with
\begin{eqnarray}
  &&\Pi\left( (I_l, J_l)\big\vert (\boldsymbol{s}_x, \boldsymbol{s}_y, \boldsymbol{t}), \boldsymbol{\Theta}\backslash \{(I_l, J_l)\} \right)\nonumber\\
&&\propto \alpha_1 A^{\beta_1}_{I_lJ_l}e^{ -\frac{\left(s_{xl}-x_{I_lJ_l} \right)^2\alpha^2_1 A^{2\beta_1}_{I_lJ_l}}{2}} \cdot  \alpha_2A^{\beta_2}_{I_lJ_l}e^{ -\frac{\left(s_{yl}-y_{I_lJ_l} \right)^2\alpha^2_2 A^{2\beta_2}_{I_lJ_l}}{2}} \cdot \frac{e^{\alpha_3^2A^{2\beta_3}_{I_lJ_l}\cos\left( (t_l-t_{I_lJ_l})\pi/12\right)}}{24I_{0}(\alpha_3^2 A^{2\beta_3}_{I_lJ_l})}\cdot \frac{w_{I_l}}{n_{I_l}},\nonumber\\
 &&~~~~ I_l \in \{1,2,\ldots, B\}, ~ J_l\in\{1,2,\ldots, n_{I_l} \}.
 \end{eqnarray}
We repeat the steps 1 to 8 many times and discard the first half of the simulations as warm-up. The posterior summaries are calculated based on the simulations post warm-up.

{\color{black}
Over multiple experiments with two chains of MCMC with 1000 samples, we
found that the MCMC tended to stabilize after 100 warm up samples. Being a computation
intensive exercise, where the model had to be repeatedly run and updated for every week
over one year of historical data, we chose a warm up sample size of 100 followed by 100
simulations for the purpose of parameter estimation and computing predictions for each
upcoming week. As an illustration, Figure~\ref{fig:rev_mcmc} shows the
trace plots based on 1000 MCMC simulations (with 2 chains) for the latest model update
in our data. The figure shows the trace plots for the parameters
$\alpha_1, \alpha_2, \beta_1, \beta_2, \beta_3$, and some of the weight parameters: $w_1, w_{11},
w_{20}, w_{52}$. We see that the two chains mix and stabilize after 100 simulations.
}
\begin{figure}[htp]
\center
\includegraphics{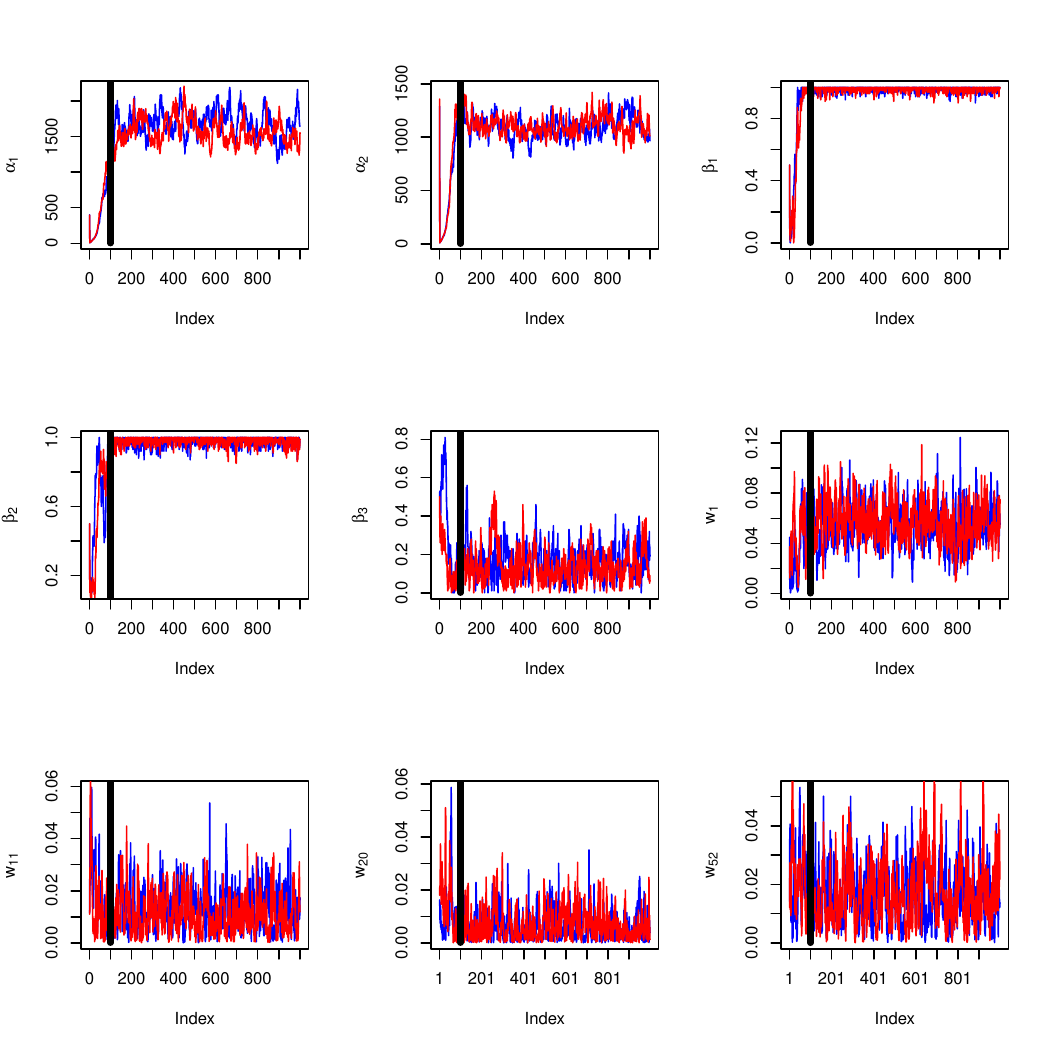}
\caption{\textcolor{black}{Trace plots for the parameters,
$\alpha_1, \alpha_2, \beta_1, \beta_2, \beta_3$, and weight parameters, $w_1, w_{11},
w_{20}, w_{52}$.}}
\label{fig:rev_mcmc}
\end{figure}

\section{Additional results}
\label{S:additional_results}
{\color{black}
In this section, we perform two analyses. First, we analyse an expert who suggests only key locations, like famous temples, flyovers, markets and metro stations, as possible locations of future crime. Based on such an expert, we add an expert block that contains latitude and longitudes of the key locations, and then create the predictive hotspot. The results of the analysis with and without such an expert is provided through Table~\ref{tab:red_auc_comparison}, which clearly shows that there is not much gain in information as this information is already present in the historical data. We observe that based on such an input only about 1\% of actual events would be within 100m, only about 5\% of actual events would be within 200m and about 10\% of actual events would be within
1000m.

\begin{table}[h]
\centering
\begin{tabular}{ccc}
\toprule
\textbf{Time} & \textbf{AUC with Expert} & \textbf{AUC without Expert} \\ \midrule
12am-4am & 0.920 & 0.924 \\ 
4am-8am & 0.904 & 0.912 \\ 
8am-12pm & 0.895 & 0.900 \\ 
12pm-4pm & 0.907 & 0.903 \\ 
4pm-8pm & 0.896 & 0.897 \\ 
8pm-12am & 0.915 & 0.917\\ \midrule
Average & 0.906 & .909\\
\bottomrule
\end{tabular}
\caption{\textcolor{black}{Comparison of AUC with and without expert input across different time windows.}}
\label{tab:red_auc_comparison}
\end{table}
}

Second, we analyse the model parameters of Model 5. Figure \ref{F:par_dataA} gives posterior summaries for the parameters of Model 5 with adaptive bandwidths used to predict the locations of each of the 25 upcoming weeks, starting with the week of 4 October 2020 up to the week of 21 March 2021. In the figure, the posterior means of the parameters are shown as a solid line and the 95\% credible interval is marked with dotted lines. The key point to note here is that the parameter estimates relating to bandwidths, i.e. $(\alpha_1, \beta_1, \alpha_2, \beta_2, \alpha_3, \beta_3)$ change significantly over the weeks (see Parts (a)-(d) of the figure). This indicates that the spatio-temporal patterns do change over time and hence it is important to dynamically estimate these parameters with every week of newly available data. Parts (g)-(h) of the figure show how the weightings of the historical temporal blocks change. For brevity, we only show the posterior summaries for weights $w_{25}$ and $w_{52}$ as an illustration.
{\color{black}Note that had the weights been uniform, the weight for each of the weeks would be
$\frac{1}{52}$. However, when the weights are estimated we observe that they are ranging
between .01 and .03. If we check the weights ($w_1$–$w_{52}$) estimated for generating predictions for a particular week, we can see that different lags become prominent. For example, in the week of 4 October 2020, $w_{25}$ happens to be more prominent than $w_{52}$, while in the week starting 21 March 2021, $w_{52}$ is more prominent than $w_{25}$. As to why some lags may become more prominent while predicting a particular week cannot be ascertained easily, but it suffices to say that the proposed method tries to capture such dependencies dynamically.}


\begin{figure}[htp]
\begin{subfigure}{.48\textwidth}
\center
\includegraphics[width=7cm, height=5cm]{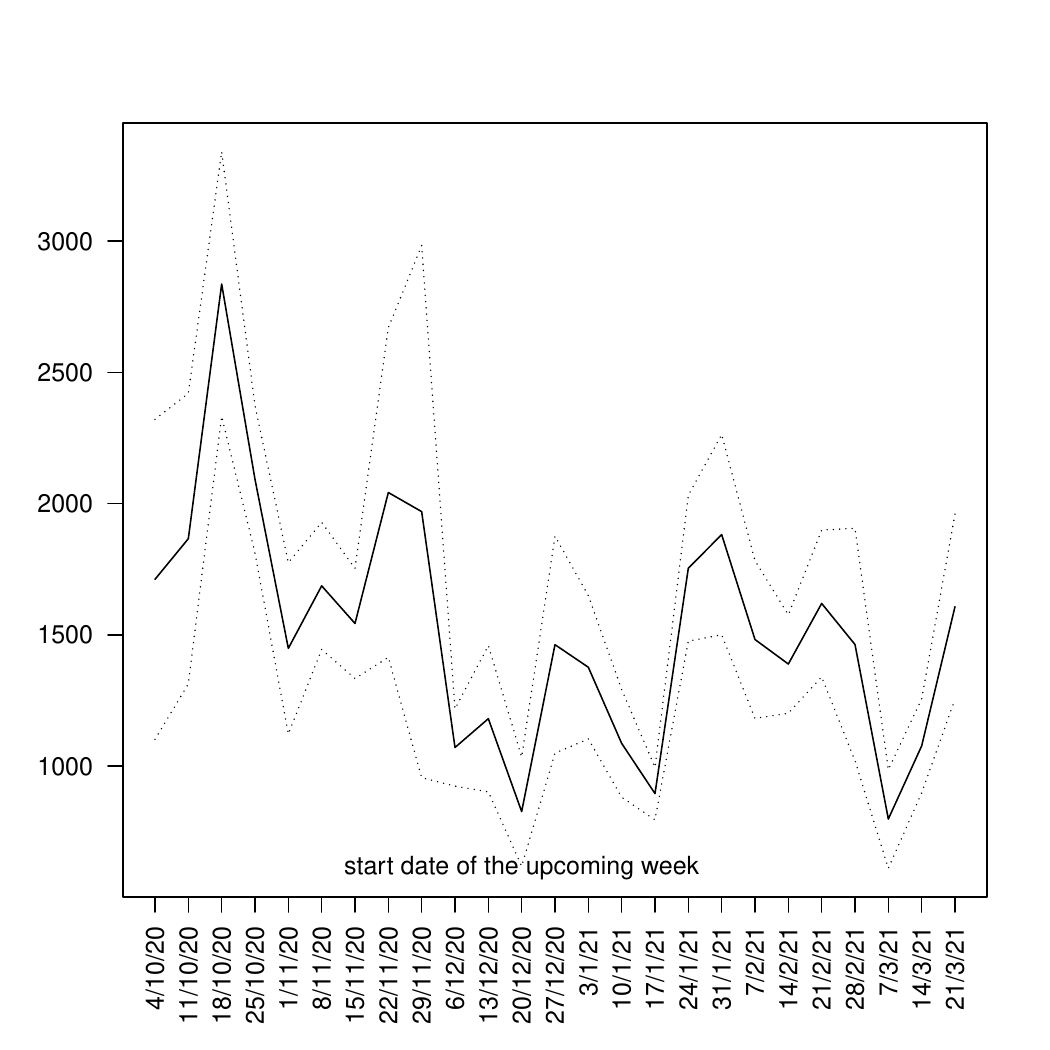}
\caption{ $\alpha_1$}
\end{subfigure}
\begin{subfigure}{.48\textwidth}
\center
\includegraphics[width=7cm, height=5cm]{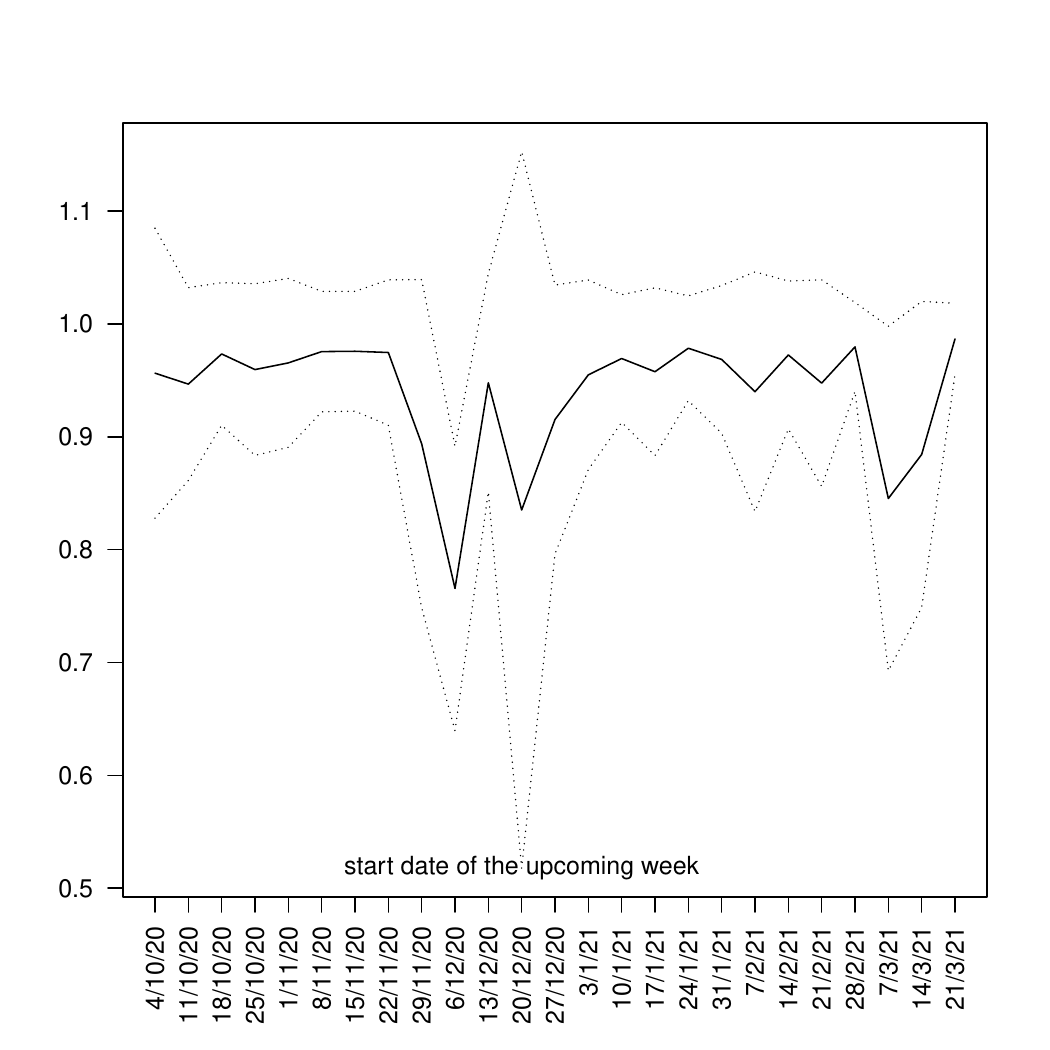}
\caption{$\beta_1$}
\end{subfigure}

\begin{subfigure}{.48\textwidth}
\center
\includegraphics[width=7cm, height=5cm]{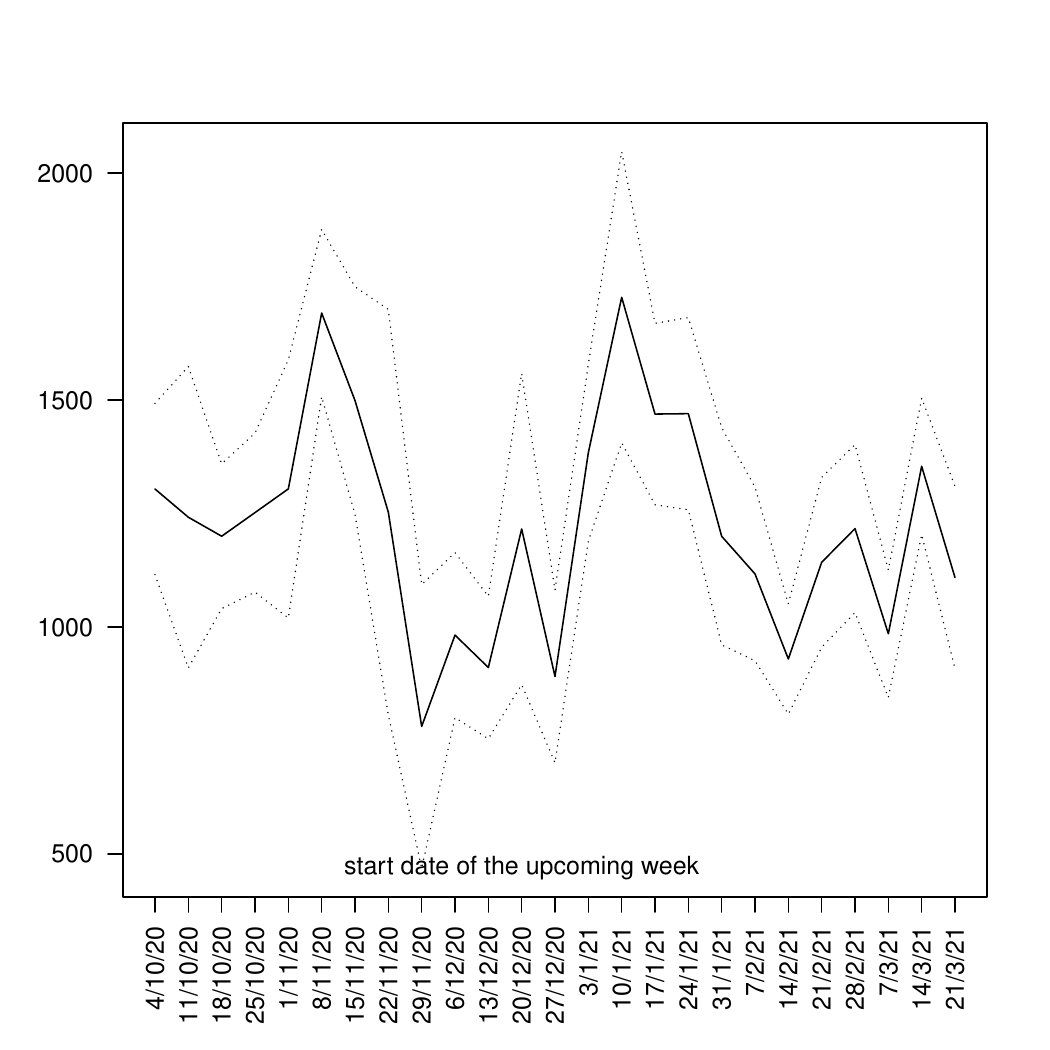}
\caption{ $\alpha_2$}
\end{subfigure}
\begin{subfigure}{.48\textwidth}
\center
\includegraphics[width=7cm, height=5cm]{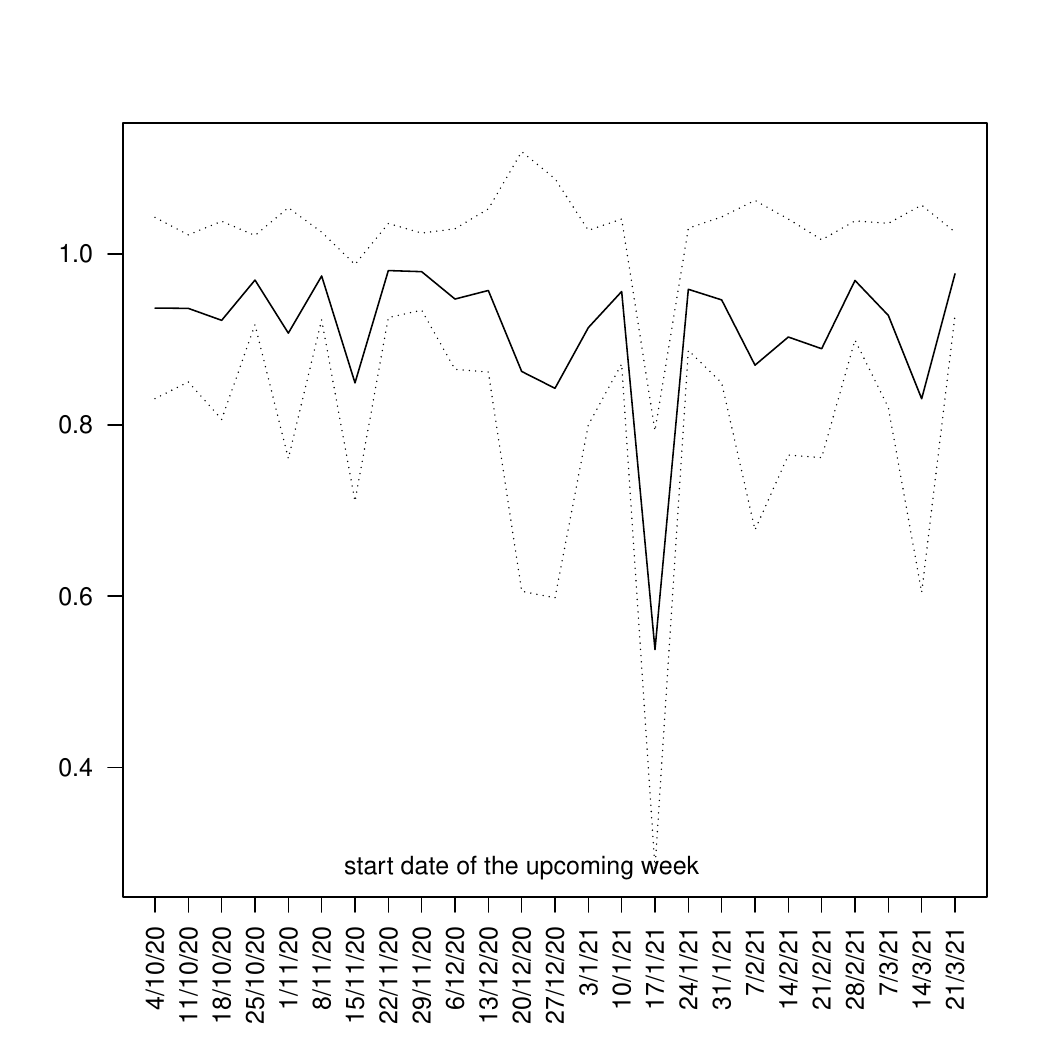}
\caption{$\beta_2$}
\end{subfigure}

\begin{subfigure}{.48\textwidth}
\center
\includegraphics[width=7cm, height=5cm]{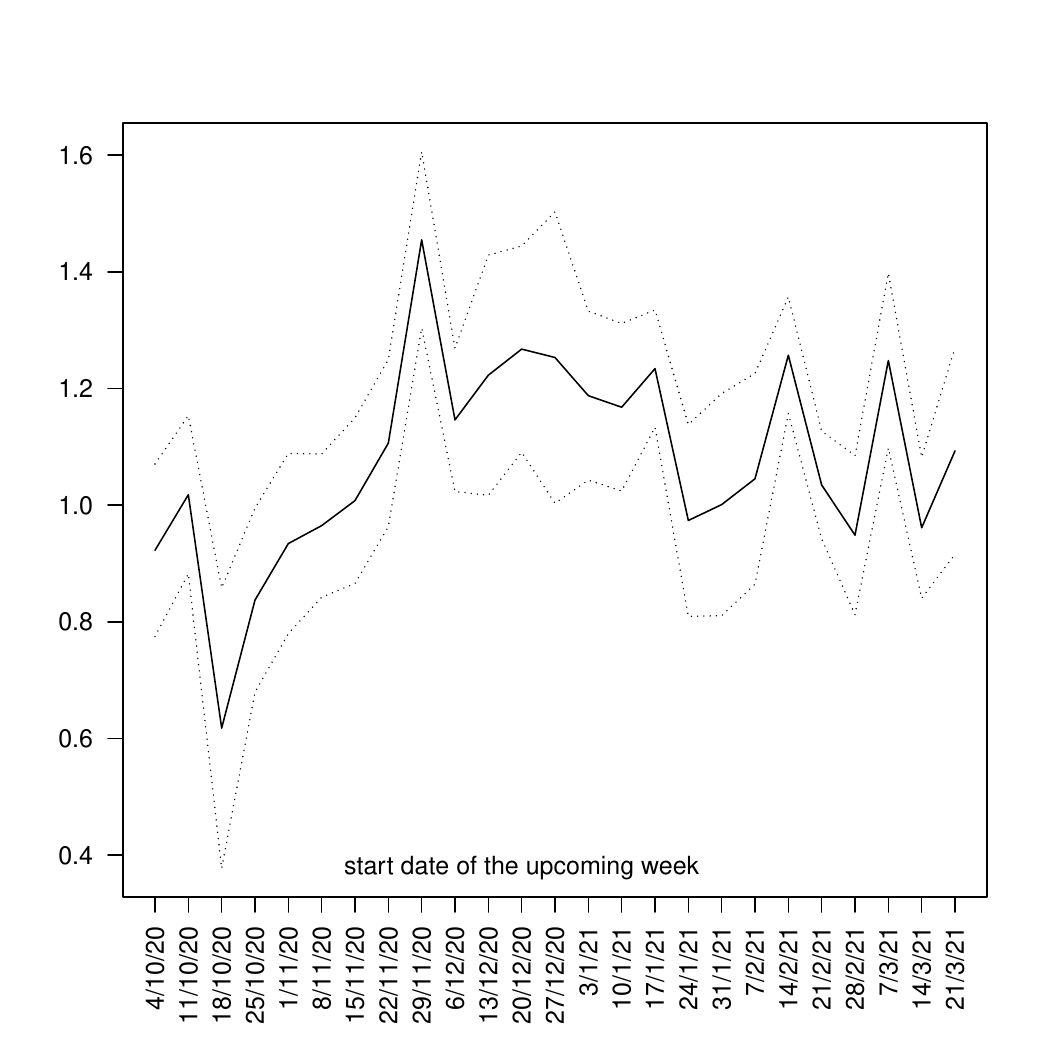}
\caption{ $\alpha_3$}
\end{subfigure}
\begin{subfigure}{.48\textwidth}
\center
\includegraphics[width=7cm, height=5cm]{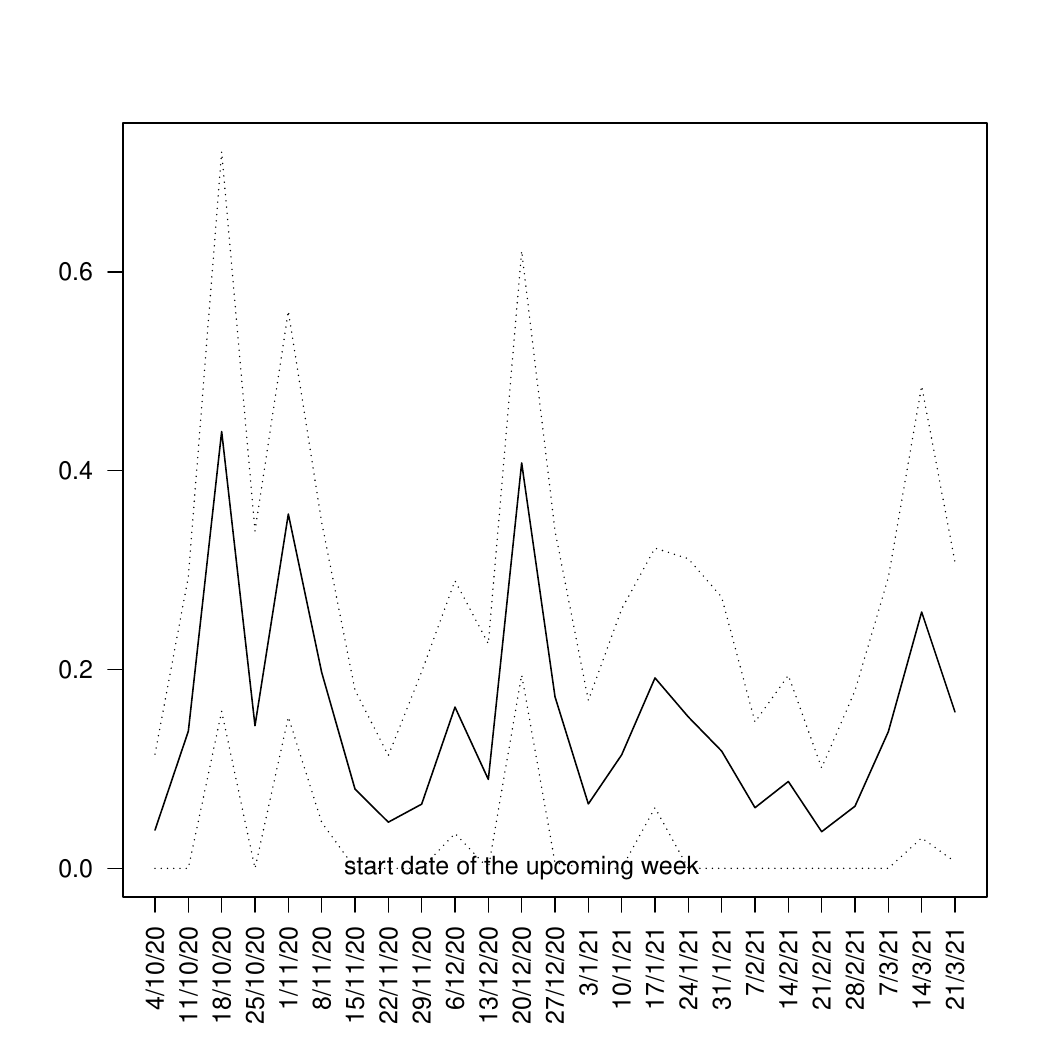}
\caption{$\beta_3$}
\end{subfigure}

\begin{subfigure}{.48\textwidth}
\center
\includegraphics[width=7cm, height=5cm]{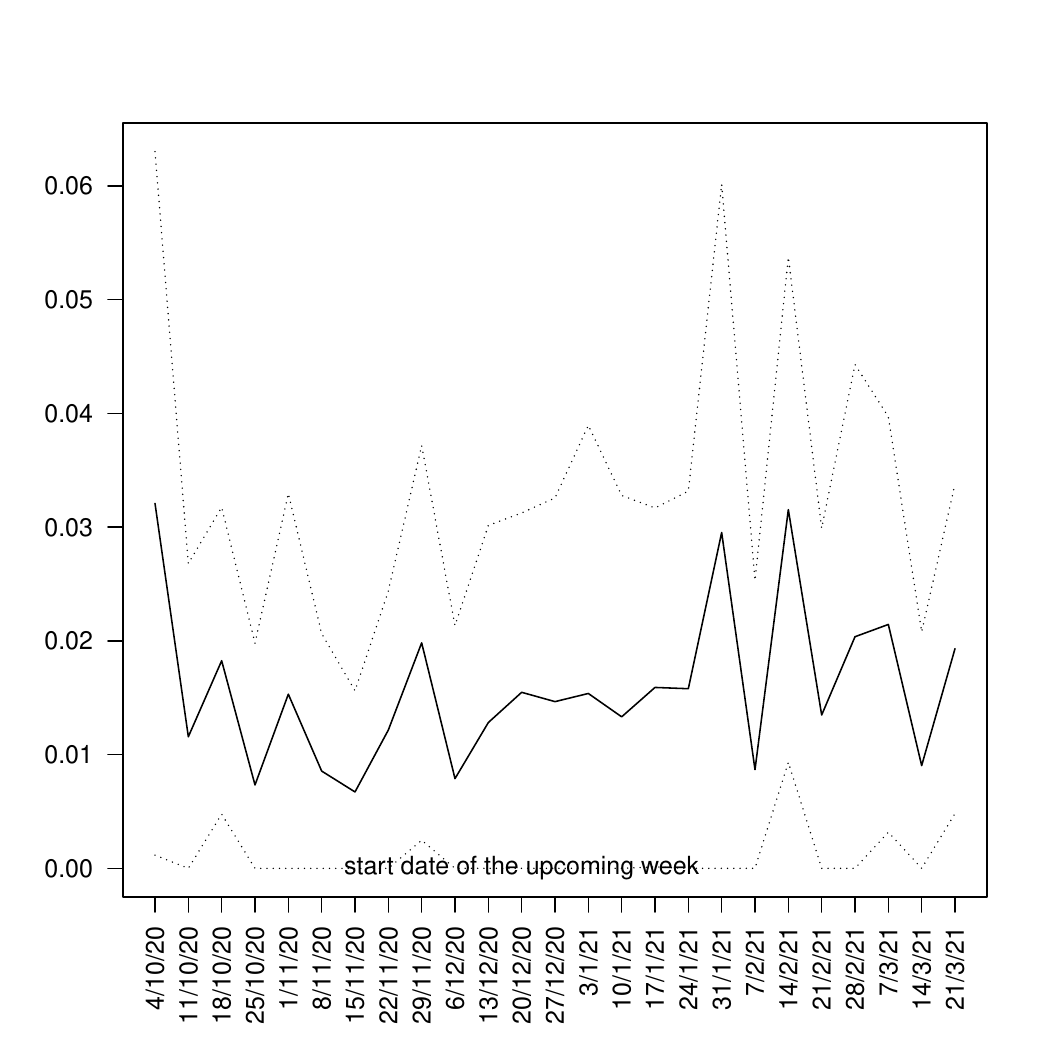}
\caption{ $w_{25}$}
\end{subfigure}
\begin{subfigure}{.48\textwidth}
\center
\includegraphics[width=7cm, height=5cm]{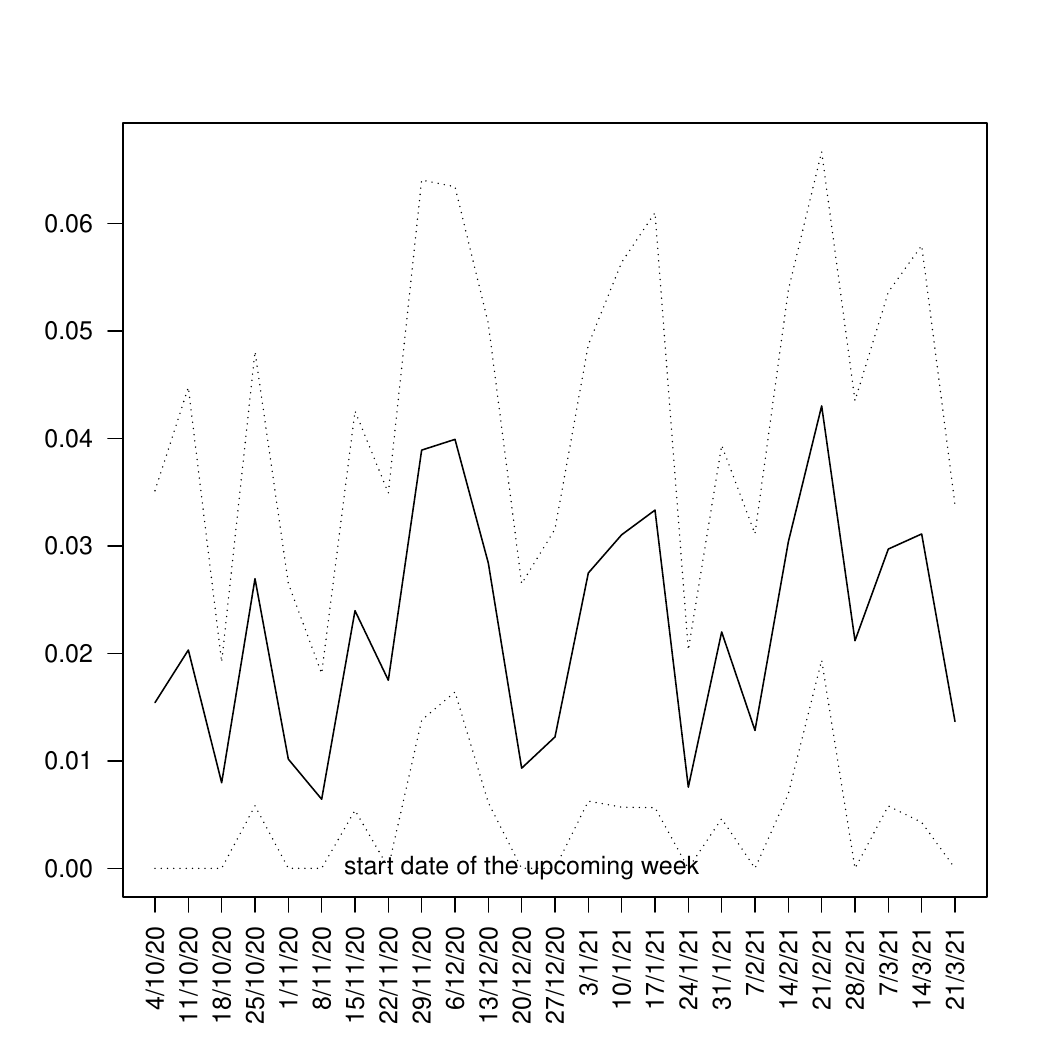}
\caption{$w_{52}$}
\end{subfigure}
\caption{Posterior summaries of parameters for Model 5  (solid line is the posterior mean and dotted lines mark the 95\% credible interval)}
\label{F:par_dataA}
\end{figure}

The analysis in Section~5 in the paper illustrated that the predictive hotspot maps change each time the map gets updated for a given temporal block and time interval, which may then require re-positioning of resources. To further understand the magnitude of these changes, we compute the transition probability matrices to track the changes in (red, yellow, other) spots from one update of the hotspot map to the next, starting with the week of 4 October 2020 up to the week of 21 March 2021, and within each week for  the six successive daily time intervals, viz `12am to 4 am', `4 am to 8am',..., `8 pm to 12 pm'.  Here, `other' just means `neither red nor yellow'. For any two consecutive updates of the hotspot map, the transition probability matrix would give the percentage of red spots in the current map that remained red, became yellow, or became other, in the next updated hotspot map. Similarly, it also gives the percentage of yellow in the current map that became red, remained yellow or became other, in the next updated map; and the percentage of spots that were `other' in the current map, but became red, or became yellow, or remained other, in the next updated map. Table \ref{T:trans_prob} Part (a) shows the element-wise average across all the transitional probability matrices so computed, and Part (b) gives the element-wise standard deviations. During successive updates of the hotspot map, on an average, about 90\% red spots remain as red spots and about 80\% of yellow spots remain as yellow spots, but about 10\% red move to yellow. Further, about 3\% and 0.03\% of `other' locations become yellow and red respectively. To understand the significance of these changes, note that the `other' consists of about 60\% of 36263 locations, which means that over 650 locations become yellow and over 6 locations become red, from other.

Next, we would like to further emphasize the importance of regular updates to the hotspot maps with updated data every week.  We noted earlier, using Parts (a) and (b)  of Figure~4 in the paper, that hotspot maps for two different time intervals look deceivingly similar, but show many differences when looked closely in Part (c).  Therefore, although the distribution of street crime locations in Delhi do not appear to change drastically, they do change. The pertinent question is, what if we do not update the hotspot map predictions regularly and skip a few weeks of data update?  Figure \ref{F:AUC_ifnotlatest} shows the impact on AUC if the hotspot map update is skipped to not include a few weeks of new data. Specifically, for any upcoming week, we compute the AUC with respect to the hotspot prediction as given by our model updated using the latest available event data, as well as what the AUC would have been had we skipped the model update by 1 week, 2 weeks and so on. The figure shows the average AUC, over the different weeks we considered for this analysis along with the 10th and 90th percentiles. We can see from the figure that, on an average, the AUC is highest when the model is updated with latest information. The impact on AUC is lesser if the model is not updated up to 8 weeks, as compared to not updating the model for more than 8 weeks. It is to be noted that reduction of even a 0.5\% AUC would mean 5 events in a week or 20 events in a month, which is important from a policing point of view.

\begin{table}[ht]
\caption{\label{T:trans_prob} Summaries of transitional probability matrices computed to track the changes in (Red, Yellow, other) spots from one update of the map to the next, starting with the week of 4 October 2020 up to the week of 21 March 2021, and within each week, for six successive time intervals, viz. `12am to 4 am', `4 am to 8am',..., `8 pm to 12 pm'.  }
\centering
\begin{subtable}{.48\textwidth}
\begin{center}
\caption{Element-wise average of the matrices }
\begin{tabular}{rrrr}
  \toprule
 & Red& Yellow& Other\\ 
  \midrule
Red& 0.8958 & 0.1035 & 0.0007 \\ 
  Yellow & 0.1033 & 0.8080 & 0.0887 \\ 
  Other & 0.0003 & 0.0295 & 0.9702 \\ 
   \bottomrule
\end{tabular}
\end{center}
\end{subtable}
\begin{subtable}{.48\textwidth}
\begin{center}
\caption{Element-wise sd of the matrices }
\begin{tabular}{rrrr}
  \toprule
 & Red& Yellow& Other\\ 
  \midrule
Red & 0.0384 & 0.0371 & 0.0020 \\ 
  Yellow & 0.0369 & 0.0767 & 0.0407 \\ 
  Other & 0.0007 & 0.0135 & 0.0140 \\ 
   \bottomrule
\end{tabular}
\end{center}
\end{subtable}
\end{table}

\begin{figure}[htb]
\center
\includegraphics[width=8cm, height=8cm]{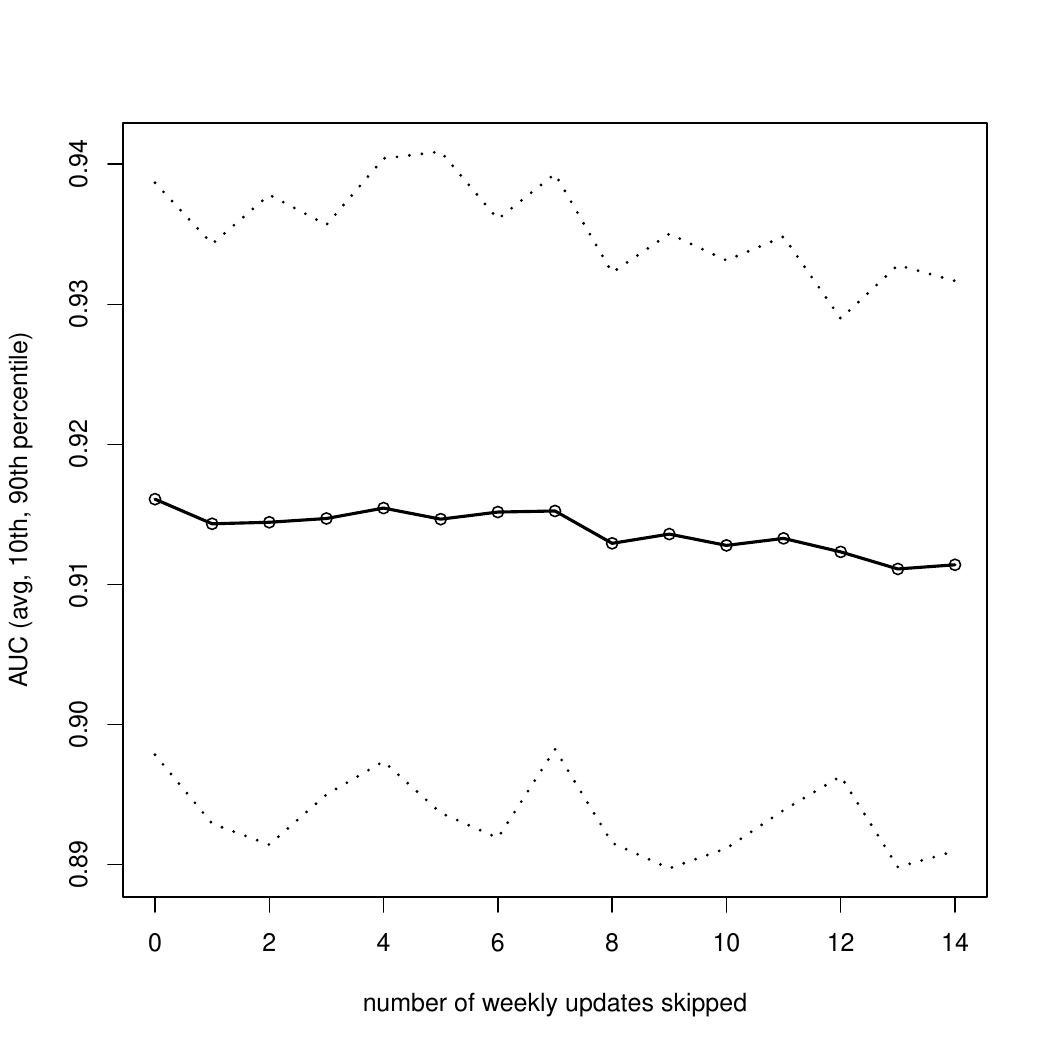}
\caption{\label{F:AUC_ifnotlatest} Impact on AUC if hotspot map is not updated with a few weeks of emerging data }
\end{figure}

{\color{black}
\subsection{Further discussion on parameters}
The parameters $(\alpha_1, \beta_1)$ together relate to the adaptive bandwidth of the KDE along the longitude dimension. Similarly, $(\alpha_2, \beta_2)$ correspond to the latitude dimension, and $(\alpha_3, \beta_3)$ to the time during the day. Equation (24) of the paper provides the exact relations. These parameters are estimated on a weekly basis, and Figure~\ref{F:par_dataA} shows the posterior mean and credible intervals for the parameters over 25 weeks, starting from the week of 10 October 2020, to the week of 21 March 2021.

Approximately, we can interpret $\left(\frac{1}{\alpha_1}\right)$ and $\left(\frac{1}{\alpha_2}\right)$ as the average bandwidth radius for the longitude and latitude dimensions, respectively, although the specific bandwidth used around each data point would vary at different locations/times in the adaptive setup. The parameters $\beta_1$ and $\beta_2$ determine the adaptive nature of the bandwidth for the longitude and latitude dimensions. If $\beta_i = 0$, then the bandwidth used for the KDE would be constant across locations or times of the day. 

For example, in Figure~\ref{F:par_dataA}, we observe that while $\beta_2$ usually stayed around 0.9, it dipped close to 0.6 during the week of January 17, 2021, implying that the bandwidth required for KDE along the latitude direction was relatively more uniform during that week compared to others. Figure~\ref{F:par_dataA}, which displays the summaries for 25 weeks, suggests that these parameters change significantly week over week and therefore need to be updated regularly.

Table~\ref{tab:rev_table_param} shows the summaries of all the parameters for the most recent week (starting 21 March 2021) in our data. We note that the overall average estimated ranges for the bandwidth along longitude and latitude are 49.9–78.0 meters and 85.0–122.6 meters, respectively. For the time during the day, the overall average estimated range for bandwidth is 47.2–65.5 minutes. The $\beta$ parameters further adjust the bandwidths depending on the location and time during the day, as per Equation (24). 

Additionally, Figure~\ref{fig:rev_fig_weight} displays the estimated weights corresponding to each of the 52 lags. We observe that some lags have magnitudes greater than $\frac{1}{52}$ (e.g., lags 1, 21, 32, 49, 50), indicating that they play a significant role in predicting events for the week of 21 March 2021. It is particularly noteworthy that, apart from the most recent week (lag = 1 week), a lag close to one year (lag = 50 weeks) also plays an important role in the predictions for the week of 21 March 2021.

\begin{table}[h!]
\centering
\caption{\textcolor{black}{Parameter estimates and corresponding 95\% credible intervals for the week starting March 21, 2021.}}
\begin{tabular}{crll}
\toprule
\textbf{Parameter} & \textbf{Estimate} & \textbf{95\% CI} & \textbf{Corresponding range} \\
\midrule
$\alpha_1$ & 1607.36 & [1254.8, 1959.9] & [49.9, 78.0] meters \\
$\beta_1$ & 0.99 & [0.955, 1.018] &  \\
$\alpha_2$ & 1109.13 & [908, 1310.3] & [85.0, 122.6] meters \\
$\beta_2$ & 0.98 & [0.928, 1.025] &  \\
$\alpha_3$ & 1.09 & [0.916, 1.27] & [47.2, 65.5] minutes \\
$\beta_3$ & 0.16 & [0.007, 0.308] &  \\
\bottomrule
\end{tabular}
\label{tab:rev_table_param}
\end{table}

\begin{figure}[htp]
\center
\includegraphics[width=10cm]{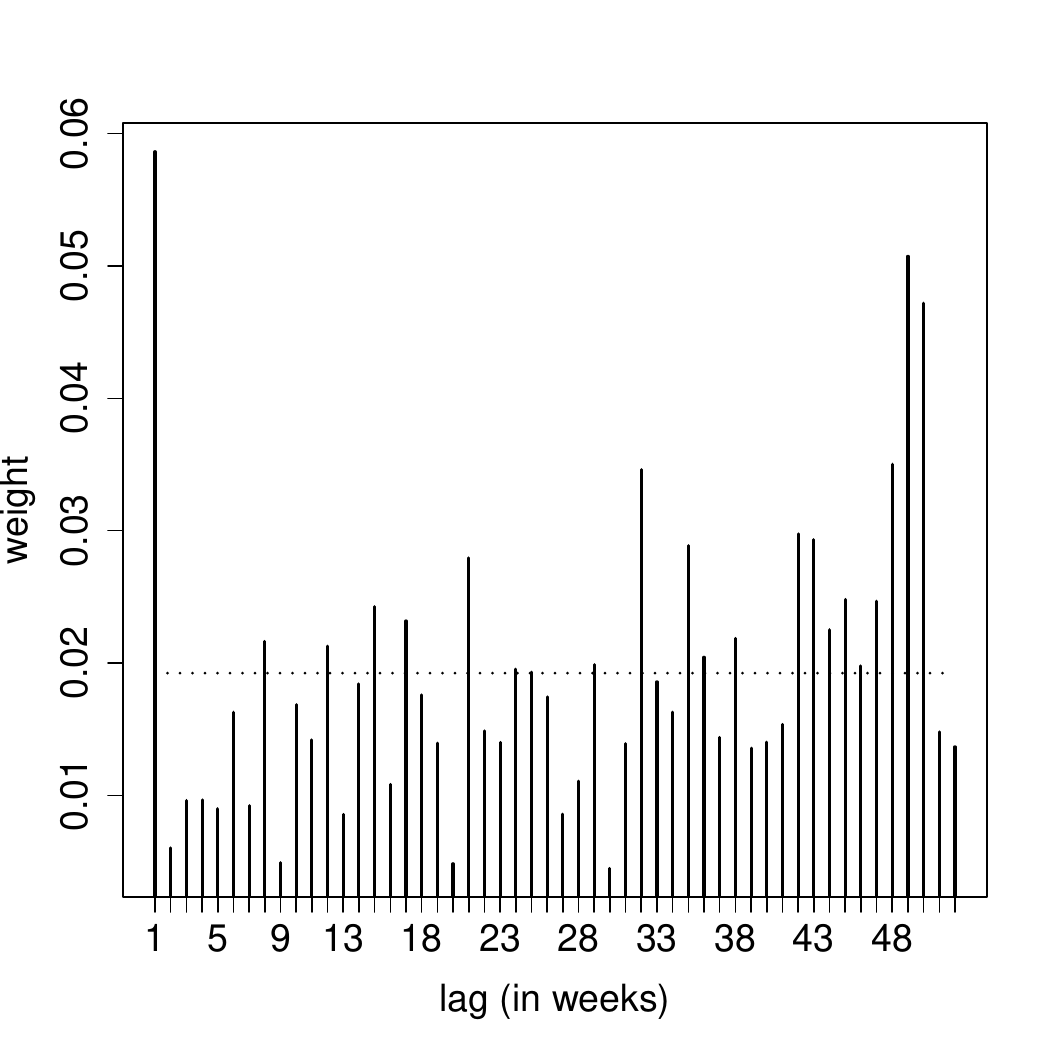}
\caption{\textcolor{black}{Estimated weights corresponding to the 52 lags.}}
\label{fig:rev_fig_weight}
\end{figure}

}

{\color{black}
\section{Interview findings and integration of expert inputs}
\label{S:interview_findings}
We conducted a detailed interview with the Deputy Commissioner of Police (DCP), Delhi PCR, who is the head of the control room and responsible for managing all the patrol vehicles and their allocation. The important findings are provided in Section~\ref{sec:rev_interview}. An interaction on patrolling strategy and incorporation of expert inputs, was conducted with multiple patrol officers to understand how past data and human inputs are currently integrated and what is their response in using a predictive approach that integrates human inputs. We summarize the findings from this study in Section~\ref{sec:rev_pilot}. Sections~\ref{sec:rev_interview} and~\ref{sec:rev_pilot} provide a global as well as a local view of the patrolling operations in Delhi.

\subsection{Interview with DCP, Delhi PCR}\label{sec:rev_interview}
In an effort to gain a deeper understanding of the operational processes and potential for resource allocation within the PCR vehicle management system, we conducted a detailed interview that focused on several key aspects, including PCR vehicle allocation, operational strategies, and the feasibility of incorporating predictive models for PCR management. The following questions were explored during the in-person interview:
\begin{itemize}
    \item What is the usual process of PCR allocation and functioning?
    \item How many PCR vans are operational in Delhi?
    \item How frequently are crime hotspot locations updated?
    \item How are specific issues like street crime considered in PCR allocation?
    \item What is the potential utility of predictive models for PCR allocation?
    \item How do PCR vans choose patrol spots, and is their patrolling randomized or systematic?
    \item Are the PCR vans mobile or static. If it is a combination of the two then what is the suggested logic.
    \item Is it feasible to share past PCR van locations?
\end{itemize}
The responses to the above questions have been provided below categorized into segments with the answers to the questions addressed in one or more of the segments.

\textbf{Recent Structural Changes:} The DCP shared that a recent policy change in Delhi shifted the responsibility for PCR operations from the central control room to individual police stations (213 in number with jurisdiction of 7 km$^2$ on average). Therefore, the Delhi police has explored both policies, a centralized approach and decentralized approach to resource allocation. However, the current DCP has been instrumental in reverting operational control back to the central control room as they found the approach to be more favourable. 
A centralized approach is advantageous as it allows them to capitalize on past data comprehensively, but at times makes it difficult to incorporate local expert inputs. According to the DCP, a seamless approach to integrating inputs from individual police stations with a central database of past crimes will be helpful, as it will support the creation of predictive models that account for macro patterns in the data as well as dynamic micro developments.

\textbf{Determination of PCR Van Locations:}
PCR van locations are largely determined by a software tool called CMap, which provides call density data based on the number of emergency calls in different areas. The steps involved are:
\begin{enumerate}
    \item The PCR uses CMap to generate a list of high-call-density locations.
    \item This list is shared with the district DCPs, who review it based on their jurisdictional knowledge and suggest adjustments.
    \item Final PCR locations are a combination of CMap data and district DCP inputs.
    \item Patrol areas are similarly determined using CMap and DCP inputs, with adjustments for sudden law and order situations or major events like Independence Day.
    \item For major events or VIP movements, the district DCPs provide specific instructions on patrol vehicle placement.
    \item Although CMap inputs remain mostly static, significant changes can occur due to district DCPs recommendations.
\end{enumerate}

\textbf{Number of PCR Vehicles:} Delhi has a total of 802 PCR vehicles, of which around 775 are usually operational, while others may be under maintenance. These include:
\begin{itemize}
    \item 30 “Parakram” anti-terrorist vehicles
    \item 20 vehicles deployed at tourist locations
    \item 15 all-women PCR vans for marketplaces and women’s colleges
    \item 118 “Prakhar” vehicles which are always mobile and primarily focused on street crime
    \item 37 fully static vehicles are permanently stationed near sensitive areas, like Prime Minister's residence
    \item 350 PCR vehicles are static but moves to attend to calls and returns back to the same location
    \item Remaining vehicles are semi-static, i.e. they have one base point and 2-3 halting points. These vehicles attend calls and then return to eaither a base point or one of the halt points.
\end{itemize}

\textbf{Call Density Data:} The CMap system generates call density map based on past 1 year data which is used to monitor crime hotspots. These calls are often related to crimes like snatching, robbery, motor vehicle theft, and other thefts. The call density data is updated monthly to identify any changes, though the perceived fluctuations are typically minimal. For serious crimes, the police does not rely on call data.

\textbf{Functioning of Prakhar Vehicles for Street Crime:} Prakhar vehicles, specifically designed to handle street crime, are constantly in motion, especially in densely populated or market areas. Their primary role is to attend to PCR calls, although they may also conduct vehicle checks at picket points. These vehicles enhance police visibility and have successfully aided in apprehending criminals. As part of the “Safe City” initiative, Prakhar vehicles are equipped with cameras to provide live monitoring from the control room, with a coverage radius of 25 meters.

\textbf{Incorporating Predictive Models:} 
The idea of using predictive models is new to the Delhi police, but offers potential for more efficient PCR vehicle deployment given that they now actively want to pursue a centralized approach. Their current operations will be complemented by predictive analytics to identify future high-density crime locations. Integrating real-time data from patrol vehicles, along with expert inputs and past data, could lead to a more dynamic and responsive system for patrolling and crime prevention. Currently, the past PCR vehicle location data is not recorded and stored, but it can be done in the future if it is beneficial for the predictive approach.

\subsection{Insights from patrol officers on patrolling strategy}\label{sec:rev_pilot}

We interviewed three patrol officers to understand their patrol routines and gain insights into integration of past information with new information. The primary focus was to learn about their approach to patrolling and assess the role of technology, if any, in their daily operations. The project objective, which centers on predicting future street crime locations based on historical data, was also explained to them.

\textbf{Beat Structure and Management:}
Each police station is typically divided into 10 beats (areas), with two beats usually managed by an Assistant Sub-Inspector (ASI). The police station is overseen by the Station House Officer (SHO), who plays a supervisory role. The call density data is analyzed at the control room on a monthly basis, and the necessary information is communicated to the SHO. Based on their observations from the past data and the insights from patrol officers, certain local patrolling decisions are taken at the level of the police station, which includes, deciding the positioning/movement of the resources, deployment of on-foot officers, etc.

\textbf{Patrolling Strategy within a Beat:}
The officers highlighted that streets with free-flowing traffic require constant monitoring, particularly for crimes such as snatching, where perpetrators often use two-wheelers for quick getaways. Early mornings, when streets are less crowded, were identified as a high-risk time for such crimes. Each beat is kept relatively small to facilitate fast and frequent patrols. In crowded areas, snatchers often operate on foot, using the crowd for concealment and making quick escapes. Dimly lit or busy roads also create favorable conditions for street crime, especially after dark. For instance, bus stops and locations where people tend to be distracted by mobile phones were cited as common hotspots.
%
While the officers rely heavily on past data to anticipate where crimes might occur, they also take several other factors into account. Events that draw large crowds, such as local gatherings or festivals, can temporarily turn an otherwise low-risk area into a potential crime hotspot. They also monitor areas where criminals recently released on bail are known to operate and often look for suspicious individuals. They also monitor suspicious vehicles, such as vehicles with fake or no number plates. 



\textbf{Use of the ``eBeat Book'' application:}
The ``eBeat Book'' is a mobile app used by patrol officers during patrolling. It provides the following functionalities:
\begin{itemize} \item Access to vehicle registration data for background checks. \item A facial recognition feature (currently not fully functional). \item The ability to mark “Important Places” (new liquor shops, areas with frequent gatherings, etc.)
\item Location tagging, which records latitude and longitude coordinates for specific places. \item Information on recently released criminals, available for immediate reference. \end{itemize}
The app is fully integrated with the police control room, enhancing coordination between patrol teams and the central unit. Snapshots of the app that we obtained from the Delhi police are provided through Figure~\ref{fig:rev_eBeat}. The possibility to enter new places that require monitoring is already available on the app. This feature enables a seamless collection of expert inputs for integration in the proposed predictive modeling approach.

\begin{figure}[ht]
\begin{subfigure}{.45\textwidth}
\center
\includegraphics[width=7cm]{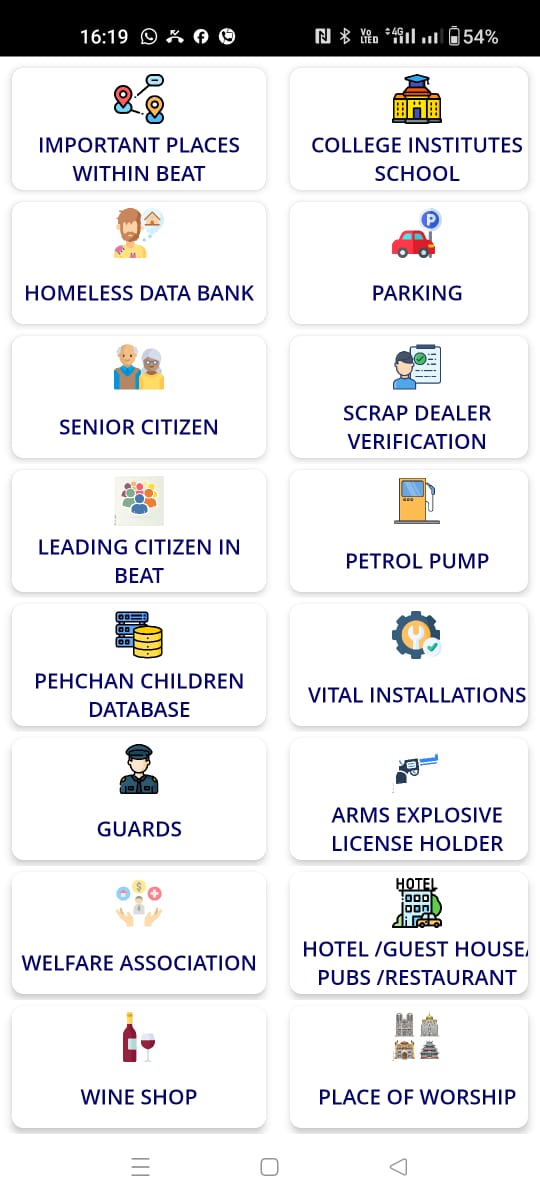}
\caption{Various features available in the application for the patrol officers.}
\end{subfigure}
\hfill
\begin{subfigure}{.45\textwidth}
\center
\includegraphics[width=7cm]{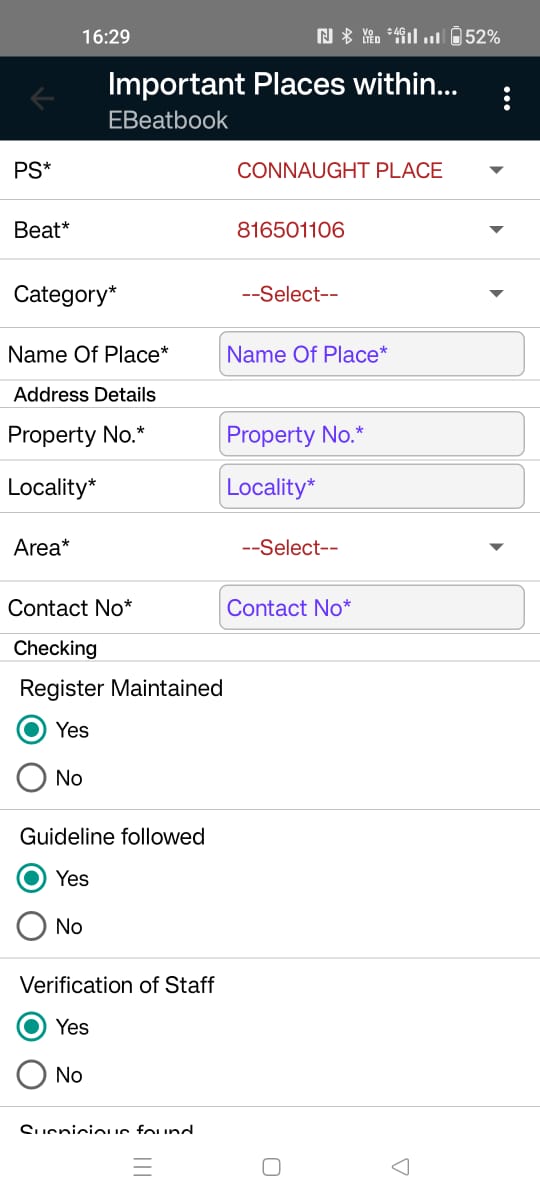}
\caption{Feature that allows patrol officers to enter new places for monitoring.}
\end{subfigure}
\caption{\textcolor{black}{``eBeat Book'' application}}
\label{fig:rev_eBeat}
\end{figure}

\textbf{Impact of Police Presence on Crime:}
A strong police presence usually deters street crime, as criminals are unlikely to act when they know police can easily intercept them. However, crimes may happen close to police presence if the spot is not easily and immediately accessible by officers. Geography plays a role in how effective the police presence is; for example: at metro stations with multiple exits across busy roads, criminals might take advantage of the time it takes for police to cross to the other side; even if police vehicles are nearby, obstacles like road dividers or heavy traffic can also slow the police response. To counter such difficulties, the officers strategically position surprise checkpoints at entry and exit routes to locations where they anticipate crime.
}

\textbf{Perspective on Predictive Tools:}
Although no predictive tools are currently in use, the officers believe that around 70-75\% of street crime locations remain consistent over time. {\color{black}This belief is not entirely true as the data shows dynamically changing crime patterns in different parts of the city from one time period to the other. While the officers found it challenging to fully grasp the concept of predictive modeling, they expressed enthusiasm about the idea of combining their on-ground intuition with macro-level data from predictive models to enhance policing efforts.}


{\color{black}
\section{Incorporating covariates in the proposed model}\label{sec:covariates}
To describe the kde formulation that contains additional covariates, such as, population density, socio-economic indicators, weather, or patrol vehicle location, we organize the data as shown in Table~\ref{T:blockdata_cov}. Table~\ref{T:blockdata_cov} is similar to Table~2 in the paper, but contains an additional row of covariate. The ideas discussed in this section can be extended to multiple covariates; however, for simplicity we assume a single additional covariate.
\begin{table}
\centering
\caption{Data with an additional covariate organised for block-weighted adaptive kde }
\label{T:blockdata_cov}
\renewcommand{\arraystretch}{1.2} 
\resizebox{\textwidth}{!}{
\begin{tabular}{|l|c|c|c|c|c|}
\hline
~& \multicolumn{4}{c|}{Historical blocks}& Latest block  \\
\hline
 Temporal block & $b-B$ & $b-B+1$& $\ldots$ & $b-1$ & $b$ \\
\hline
Block index $i$ & $1$ & $2$& $\ldots$ & $B$ & $~$  \\
\hline
Lon. coordinates& $x_{11},x_{12},\ldots,x_{1n_1}$ & $x_{21},x_{22},\ldots,x_{2n_2}$& $\ldots$ & $x_{B1},x_{B2},\ldots,x_{Bn_B}$& $s_{x1},s_{x2},\ldots,s_{xn}$  \\
\hline
Lat. coordinates& $y_{11},y_{12},\ldots,y_{1n_1}$ & $y_{21},y_{22},\ldots,y_{2n_2}$& $\ldots$ & $y_{B1},y_{B2},\ldots,y_{Bn_B}$& $s_{y1},s_{y2},\ldots,s_{yn}$  \\
\hline
Event time (in hours)& $t_{11},t_{12},\ldots,t_{1n_1}$ & $t_{21},t_{22},\ldots,t_{2n_2}$& $\ldots$ & $t_{B1},t_{B2},\ldots,t_{Bn_B}$& $t_{1},t_{2},\ldots,t_{n}$ \\
\hline
Covariate& $z_{11},z_{12},\ldots,z_{1n_1}$ & $z_{21},z_{22},\ldots,z_{2n_2}$& $\ldots$ & $z_{B1},z_{B2},\ldots,z_{Bn_B}$& $z_{1},z_{2},\ldots,z_{n}$ \\
\hline
Block weight& $w_1$ & $w_2$& $\ldots$ & $w_B$ & $~$ \\
\hline
\end{tabular}}\\
\end{table}
With an additional covariate, $z$, Equation~(7) in the paper can be updated as follows:
\begin{equation}
f(s_x, s_y, t, z) = \sum_{i=1}^{B} \frac{w_i}{n_i} \sum_{j=1}^{n_i} 
\left\{ \frac{1}{h_{1ij}} 
k\left( \frac{s_x - x_{ij}}{h_{1ij}} \right)
\frac{1}{h_{2ij}}
k\left( \frac{s_y - y_{ij}}{h_{2ij}} \right)
k\left( t - t_{ij} \right)
\frac{1}{\theta_{ij}} k\left( \frac{z - z_{ij}}{\theta_{ij}} \right) \right\}. 
\end{equation}
Estimation of $\theta$ may be done similar to other bandwidths, $h_1, h_2$ and $\tau$. The above approach allows incorporation of additional covariates or auxiliary information in the model. We note that appropriate estimation techniques can be adapted for the case when covariate $z$ may be binary or discrete  (for example, \cite{chu2015plug}).
A simple implementation on incorporating holidays as covariates was performed in the kde approach, the results for which are provided through Table~\ref{tab:cov}. Every record in the data was populated with information about whether the day of crime corresponded to a holiday (weekend or public holidays) or not. The average performance changed only slightly but we believe that other more informative covariates may lead to improvements in performance.
}

\begin{table}[h!]
\centering
\caption{AUC while predicting street crime events for the week of 21 March 2021.}
\begin{tabular}{ccc}
\toprule
Time Group & Model without Covariate & Model with Covariate (Holiday or Not) \\
\midrule
12 am -- 4 am   & 0.924 & 0.941 \\
4 am -- 8 am    & 0.912 & 0.910 \\
8 am -- 12 pm   & 0.900 & 0.899 \\
12 pm -- 4 pm   & 0.903 & 0.902 \\
4 pm -- 8 pm    & 0.897 & 0.893 \\
8 pm -- 12 am   & 0.917 & 0.916 \\ \midrule
Overall       & 0.909 & 0.910 \\
\bottomrule
\end{tabular}
\label{tab:cov}
\end{table}

{\color{black}
\section{A Markov Decision Process inspired Framework for Patrol Vehicle Allocation}\label{sec:mdp}
Effective allocation of police patrol vehicles is a critical component of proactive crime prevention and rapid response. Given a predictive density estimate, it is essential to determine optimal deployment strategies that ensure maximum coverage of areas with high crime risk. The dynamic nature of urban crime, influenced by both spatial and temporal patterns, makes this problem particularly challenging.

To address this challenge, a framework inspired by the Markov Decision Process (MDP) paradigm can be adopted. The key idea is to use crime predictions to inform decision making for future patrol vehicle deployments, optimizing for objectives such as minimum expected response time or maximum deterrence. The framework may build upon a spatio-temporal crime forecasting model where data is represented in the form \((s_x, s_y, t, z)\), indicating a spatial location \((s_x, s_y)\), time \(t\), and a contextual covariate \(z\). Here, the variable \(t\) captures the temporal granularity (e.g., hourly, daily, weekly), and \(z\) includes auxiliary information such as population density, socio-economic indicators, weather, or patrol vehicle location. For prediction, this data must be available for all spatial locations within the time range of interest. 
For practical implementation, we may assume \(z\) to remain constant at a given location over a time range of specified granularity. Complex scenarios can be modeled with multivariate  \(Z\).

In multivariate cases involving \(Z\), one may use a term for each $z$ or may reduce the dimensionality via principal components analysis. When \(Z\) is binary or discrete, one may use appropriate approaches for estimation.
A particularly relevant instance of \(Z\) is the set of patrol vehicle locations. Incorporating this information can significantly improve the prediction and lead to an effective allocation strategy. Currently, in our work, patrol vehicle positions for historical time blocks \(b - B, \ldots, b\) are not recorded, limiting the direct applicability of such models. However, we recommend that future deployments should ensure that this data is systematically collected and archived.

Assuming access to historical and real-time patrol vehicle data, we propose a formulation where the patrol vehicle deployment at future time \(t\) is represented by an action \(Z_t\).
\[
Z_t(s_x, s_y) = 
\begin{cases}
1 & \text{if a patrol van is assigned to } (s_x, s_y, t), \\
0 & \text{otherwise}.
\end{cases}
\]
The predictive model estimates the density of crime locations conditional on patrol vehicle deployment, denoted by \(\hat{f}(s_x, s_y, t, Z_t)\). A new crime location at time \(t\), denoted \((s'_x, s'_y)\), is assumed to be drawn from this density. The reward associated with a deployment strategy \(Z_t\) is quantified by a function \(R((s'_x, s'_y), Z_t)\), which captures the effectiveness of the deployment in addressing crime. For example, one can define the reward as the negative of the minimum distance between the predicted crime location and the nearest patrol van:
\[
R((s'_x, s'_y), Z_t) = -\min_{(s''_x, s''_y)} \left( \text{dist}((s'_x, s'_y), (s''_x, s''_y)) \right)
\quad \text{subject to } Z_t(s''_x, s''_y) = 1.
\]
The objective is to find the deployment vector \(Z_t\) that maximizes the expected reward, which corresponds to minimizing the average distance between predicted crime events and patrol van locations. The optimal deployment is therefore given by:

\[
Z_t^{\text{opt}} = \arg\max_{Z_t \in \mathcal{Z}} \int R((s'_x, s'_y), Z_t) \cdot \hat{f}(s'_x, s'_y, t, Z_t) \, ds'_x ds'_y,
\]
where $\mathcal{Z}$ is a set of feasible allocations. This MDP-inspired framework provides a promising direction for integrating predictive analytics with operational decision-making in urban policing. A full development and implementation of this approach is left for future work.
}

\end{document}